\documentclass[onefignum,onetabnum]{siamart220329}

\usepackage{lipsum}
\usepackage{amsfonts}
\usepackage{graphicx}
\usepackage{epstopdf}
\usepackage{algorithmic}
\usepackage{bm}
\usepackage{amsmath}
\usepackage{amssymb}
\usepackage{mathtools}
\usepackage{physics}
\usepackage{overpic}
\usepackage{epstopdf}
\usepackage{graphicx}
\usepackage{bm}
\usepackage[font = small]{caption}
\usepackage[font = small]{subcaption}
\usepackage[normalem]{ulem}
\usepackage{cleveref}

 \ifpdf
  \DeclareGraphicsExtensions{.eps,.pdf,.png,.jpg}
\else
  \DeclareGraphicsExtensions{.eps}
\fi

\newcommand{\pd}[2]{\frac{\partial #1}{\partial #2}}
\newcommand{\pdd}[2]{\frac{\partial^2 #1}{\partial #2^2}}

\newcommand{\Ai}{\mathrm{Ai}}

\newcommand{\tmath}[1]{\texorpdfstring{$#1$}{#1}}

\newcommand{\beq}{\begin{equation}\begin{aligned}}
\newcommand{\eeq}{\end{aligned}\end{equation}}

\newcommand{\beqs}{\begin{equation*}\begin{aligned}}
\newcommand{\eeqs}{\end{aligned}\end{equation*}}

\usepackage[normalem]{ulem}

\newcommand \bk{\color{black}}
\newcommand \rd{\color{red}}

\newcommand{\eps}{\ensuremath{\varepsilon}}
\newcommand{\Ord}[1]{\ensuremath{\mathcal{O}\left( #1\right)}}

\newsiamremark{remark}{Remark}
\newsiamremark{hypothesis}{Hypothesis}
\crefname{hypothesis}{Hypothesis}{Hypotheses}
\newsiamthm{claim}{Claim}

\headers{Pattern Localisation in Swift-Hohenberg Equation}{A. L. Krause, V. Klika, E. Villar-Sep\,{u}lveda, A. R. Champneys, E. A. Gaffney}

\title{Pattern Localisation in the Swift-Hohenberg Equation via Slowly Varying Spatial Heterogeneity\thanks{
\funding{E.~V-S. has received PhD funding from ANID, Beca Chile Doctorado en el extranjero, number 72210071.}}}

\author{
Andrew L. Krause\thanks{Mathematical Sciences Department, Durham University,
Upper Mountjoy Campus, Stockton Rd,
Durham DH1 3LE, United Kingdom (\email{andrew.krause@durham.ac.uk}).}
\and V\'{a}clav Klika\thanks{Department of Mathematics, FNSPE, Czech Technical University in Prague, Trojanova 13, 120 00 Praha, Czech Republic (\email{vaclav.klika@fjfi.cvut.cz})}
\and
Edgardo Villar-Sep\'{u}lveda\thanks{{Engineering Mathematics}, {University of Bristol}, {{Ada Lovelace Building, Tankard's Cl, University Walk}, {Bristol}, {BS8 1TW}, United Kingdom (\email{edgardo.villar-sepulveda@bristol.ac.uk}, \email{a.r.Champneys@bristol.ac.uk}).}}
\and
Alan R. Champneys\footnotemark[4]
\and Eamonn A. Gaffney\thanks{{Mathematical Institute}, {University of Oxford}, {{Andrew Wiles Building}, {Oxford}, {OX2 6GG}, United Kingdom} (\email{gaffney@maths.ox.ac.uk}).}
}

\usepackage{amsopn}

\begin{document}

\maketitle

\begin{abstract}
Theories of localised pattern formation are important to understand a broad range of natural patterns, but are less well-understood than more established mechanisms of domain-filling pattern formation. Here, we extend recent work on pattern localisation via slow spatial heterogeneity in reaction-diffusion systems to the Swift-Hohenberg equation. We use a WKB asymptotic approach to show that, in the limit of a large domain and slowly varying heterogeneity, conditions for Turing-type linear instability localise in a simple way, with the spatial variable playing the role of a parameter. For nonlinearities locally corresponding to supercritical bifurcations in the spatially homogeneous system, this analysis asymptotically predicts regions where patterned states are confined, which we confirm numerically. We resolve the inner region of this asymptotic approach, finding excellent agreement with the tails of these confined pattern regions. In the locally subcritical case, however, this theory is insufficient to fully predict such confined regions, and so we propose an approach based on numerical continuation of a local homogeneous analog system. Pattern localisation in the heterogeneous system can then be determined based on the Maxwell point of this system, with the spatial variable parameterizing this point. We  compare  this theory of localisation via spatial heterogeneity to localised patterns arising from homoclinic snaking, and suggest a way to distinguish between different localisation mechanisms in natural systems based on how these structures decay to the background state (i.e.~how their tails decay). We also explore cases where both of these local theories of pattern formation fail to capture the interaction between spatial heterogeneity and underlying pattern-forming mechanisms, suggesting that more work needs to be done to fully disentangle exogenous and intrinsic heterogeneity. 
\end{abstract}



\section{Introduction}
A contemporary question in many areas of science is to understand the origin of natural spatial structures \cite{rohani1997spatial, meron2012pattern, painter2021systems, krause_near_2021}.
In particular, given an observed spatial distribution (henceforth, pattern), it is relevant to understand if the mechanisms underlying its formation are due to intrinsic self-organisation (e.g.~from Turing-like pattern forming mechanisms \cite{turing1952chemical, krause_near_2021}) or to exogenous factors, such as environmental heterogeneity in the context of ecosystems or developing tissue structures.
Examples can be found in plant-root initiation \cite{brena2014mathematical, brena2015stripe},  chemical system self-organization \cite{lee1994} and hierarchical patterning in embryology \cite{raspopovic2014digit, onimaru2016fin}, as well as in fluid dynamics \cite{soward2016}, vegetation patterning and neuroscience \cite{patterson2023spatial}. Questions of intrinsic or extrinsic factors underlying pattern formation become even more intricate with regards to localised pattern formation, whereby oscillatory spatial structures are confined to distinct spatial regions, falling away to background states that would not be classified as structured or patterned regions \cite{knobloch2015spatial}. In this paper, we consider a spatially-heterogeneous Swift-Hohenberg equation as a prototype model for understanding the interplay between nonlinearity and spatial heterogeneity in giving rise to localised patterns. 

Spatially heterogeneous systems are likely more realistic models than their simpler homogeneous counterparts, particularly for embryological and ecological phenomena. Turing himself was aware of this, noting that most biological structures likely ``evolve from one pattern into another, rather than from homogeneity into a pattern" \cite{turing1952chemical}. A major reason for emphasizing simpler homogeneous models is due to how much more difficult even relatively simple techniques, such as linear stability analysis, become in the heterogeneous case \cite{krause_WKB, krause_near_2021}. Nevertheless, there is a growing body of work exploring such heterogeneous systems numerically \cite{herschkowitz1972,herschkowitz1975,benson1993diffusion, barrio1999two, page2003pattern, page2005complex,kao2014spatial, krause2018heterogeneity, vandenberg2023turing} and in asymptotic regimes \cite{herschkowitz1972,iron2001spike, kolokolnikov2018pattern,krause_WKB, gaffney2023}, among other approaches \cite{benson1998unravelling, scheel2018wavenumber, van2021pattern, calderon2022turing}. Spatial and spatiotemporal heterogeneity has been used to design Turing spaces matching complex prepatterns and pattern-forming regions \cite{woolley2021bespoke}, as well as in orienting stripes \cite{hiscock2015orientation, coelho2021stripe}. 

Recent work \cite{krause_WKB,gaffney2023,dalwadi2023universal,krause2018heterogeneity,patterson2023spatial}
has explored how bifurcations seen in spatially homogeneous {settings  } play out in spatially heterogeneous systems, where the heterogeneity passes through values around these bifurcation points. The qualitative features here, in the case of Turing-type bifurcations, are a localisation of classical Turing conditions leading to a {type } of confined pattern formation distinct from the localisation observed in spatially homogeneous systems. An important lesson arising from this work is that a naive local theory of heterogeneity (essentially treating spatial variables as parameters) can successfully explain observed behaviours in heterogeneous systems in some cases (e.g.~\cite{krause_WKB,gaffney2023}, and even for spatiotemporally forced systems \cite{dalwadi2023universal}), but critically fails to explain some emergent dynamics (e.g.~\cite{krause2018heterogeneity,patterson2023spatial}). Clarifying when the intuitive local picture accurately captures the dynamics, and when it does not, is the main goal of this paper, {with a secondary aim of investigating how the decay of localised pattern back to baseline  (i.e.~the patterned solution's tail) may indicate the underlying pattern formation mechanism.}

To address these issues, we will study a particularly simple model of slowly varying heterogeneity. Namely, we consider a heterogeneous Swift-Hohenberg equation of the form,
\beq\label{main_sh0}
    \pd{u}{t} = r(x)u-\left(1+\eps^2 \pdd{}{x}\right)^2u + N(u), \quad \quad \quad x\in[0,1],
\eeq
where we assume that $N(0)=N_u(0)=0$ and $0 < \eps \ll 1$. To represent a closed system so that any pattern formation is an emergent property of the system rather than due to external forcing at the boundary, the associated boundary conditions are taken to be the generalised Neumann conditions 
\beq\label{bcs}
   u_x(t,0) = u_{xxx}(t,0) = 0 = u_x(t,1) = u_{xxx}(t,1). 
\eeq
{We always enforce the condition 
\begin{equation}\label{bcsr} r_x(0)=r_{xxx}(0)=r_x(1)=r_{xxx}(1)=0
\end{equation}
so that $$(ru)_x(0)=(ru)_x(1)=(ru)_{xxx}(0)=(ru)_{xxx}(1)=0.$$ This  ensures the variation in $r(x)$ adjacent to the boundaries does not give different behaviors from the system with constant $r$, to avoid  the prospect of the heterogeneity introducing a boundary layer at the domain edges that would not be present in the homogeneous system. Finally, we} 
{also consider initial conditions that are small random perturbations across the domain, so that the perturbations are homogeneous {\it on average}, as further detailed in Appendix \ref{appendix_numerics}. In particular, this implicitly entails that we analyse the impact of the heterogeneity in $r(x)$, rather than any potential interactions between heterogeneities in $r(x)$ and heterogeneities in the  initial conditions, with such interactions out of scope of the current study.}  

This model has the corresponding energy functional \cite{kao2014spatial},
\beq\label{energy}
    E(u) = - \int_0^1 \frac{1}{2}\left(r(x) u^2-\left(u+\eps^2\pdd{u}{x}\right)^2\right)+F(u)dx, \quad F(u) = \int_0^u N(v)dv,
\eeq
from which we see that all stable states for asymptotically large times must be stationary (ruling out heterogeneity-induced spatiotemporal dynamics, as in \cite{page2005complex, krause2018heterogeneity, kolokolnikov2018pattern}). We assume that all functions are sufficiently smooth, and in particular that $|r'(x)| = o(1/\eps)$, i.e.~the heterogeneity varies slowly. This model is essentially equivalent to a spatial dynamics formulation (as in \cite{beck2009snakes} and elsewhere) with a slowly varying heterogeneity relative to any other length scales in the problem. We will also consider a homogeneous analog of \cref{main_sh0} given by,
\begin{equation}\label{homog_sh}
    \pd{u}{t} = r_hu-\left(1+\eps^2 \pdd{}{x}\right)^2u + N(u), \quad \quad \quad x\in[0,1],
\end{equation}
assuming the same boundary conditions \cref{bcs}. This model also has an energy functional of the form of \cref{energy}. Our goal is then to understand when the dynamics of \cref{main_sh0} can be understood by looking at the dynamics of \cref{homog_sh} with $r(x) = r_h$ locally in $x$ (i.e.~when a `quasi-static' approach in space can be justified, {including at the bifurcation point corresponding to $r_h=0$}). 

We illustrate this in \cref{fig:homog_het_comparison}, showing a long-time solution of the homogeneous system in panel (a), and a corresponding heterogeneous system in panel (b) for the same value of $\eps$, with panels (c) and (d) showing simulations with smaller values of $\eps$. 
The red lines correspond to $r(x)=0$, and hence to where a naive local theory would predict pattern confinement. Following asymptotic analyses of heterogeneous reaction-diffusion \cite{krause_WKB} and reaction-cross-diffusion systems \cite{gaffney2023}, we will justify this intuitive picture in the limit of small $\eps$ by showing that the linear stability problem leads to locally supported solutions within these regions.
{The extent of the bleed of pattern beyond the red lines will also be briefly considered in \cref{appc} that documents more subtle aspects of the asymptotic analysis.}   We also proceed to fill an important gap in these previous papers by carrying out a boundary-layer analysis at the bifurcation crossing, to approximate how the tails of the solutions behave near points where $r(x)=0$. Importantly, these ideas from linear theory will be shown to only work when the local picture is supercritical. 

{We will use the term `locally' throughout to indicate that we mean the nature of the Turing bifurcation corresponding to \cref{homog_sh} at the point $r(x)=0$ in \cref{main_sh0}.} In the locally subcritical case, we will develop an alternative prediction for the confinement region based on the idea of a local Maxwell point of the energy functional \cref{energy}. We will {also} numerically explore cases where neither approach successfully predicts heterogeneity-induced pattern localisation, raising important questions about how to understand such systems in general. Overall these ideas will give a partial answer to what we can learn about a system's underlying patterning mechanisms based on observed patterned states. {We note that this perspective of slow passage through a bifurcation in space is somewhat different from the perspective of inhomogeneous problems arising, e.g., in the fluid dynamics context \cite{harris2000inhomogeneous, soward2016} where a weakly nonlinear theory is developed for small amplitude solutions, with an emphasis on phase mixing. In contrast, we will consider large amplitude forcing that is sufficiently slow to allow for a `local' picture near the transition region to instability.}

We remark {that the form of localisation from locally supercritical bifurcations in the presence of heterogeneity differs from  that of} localised solutions arising in spatially homogeneous models, such as \cref{homog_sh}, due to homoclinic snaking \cite{woods1999,burke2006,beck2009snakes,kozyreff2009physicaD,knobloch2015spatial,al2021localized,villar2025beyond}. We give examples of {the latter type of solution in \cref{fig:snaking_example}, where we have simulated the homogeneous model using initial data constructed from the simulation in \cref{fig:homog_het_comparison}(a) by setting some parts of this solution to zero, and then evolving forward in time.  This leads to a {fundamentally different kind of} localized solution. We remark that there are several ways to numerically find such states once a good parameter regime is known, such as via numerical continuation \cite{uecker2019pattern}. There are several differences to the simulations shown in \cref{fig:homog_het_comparison} with these localised states, both in their qualitative properties (e.g.~sharper tails, and a degree of translation invariance away from the boundaries) and in the details of the nonlinearities in determining their existence (e.g.~they generally arise in the bistable regime of a subcritical Turing bifurcation). {In particular, the tails in the case of heterogeneity across a {(locally)} supercritical bifurcation appear algebraic, while the localised solutions arising from snaking are exponential.} Motivated by these distinctions, we will return to compare and contrast such localised states with those arising in spatially heterogeneous models later. {Importantly, we will  not consider the interplay of these distinct mechanisms (i.e.~the impact of heterogeneity on localised solutions arising from snaking), which was instead studied in \cite{kao2014spatial}.}

\begin{figure}
    \centering
    \begin{subfigure}{.495\textwidth}\includegraphics[width=1\textwidth]{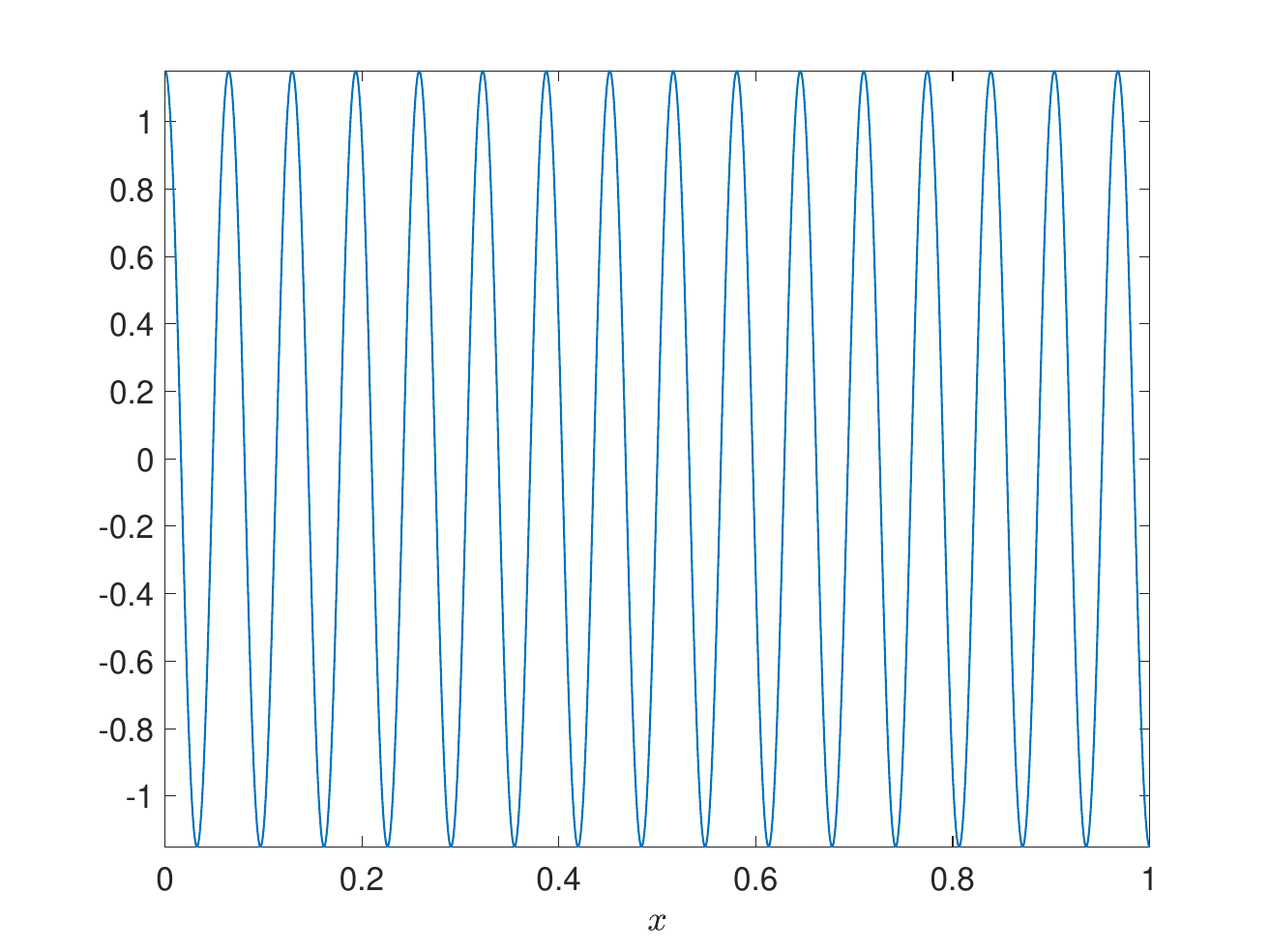}
    \caption{$r = 1$, $\eps = 0.01$}
    \end{subfigure}
    \begin{subfigure}{.495\textwidth}\includegraphics[width=1\textwidth]{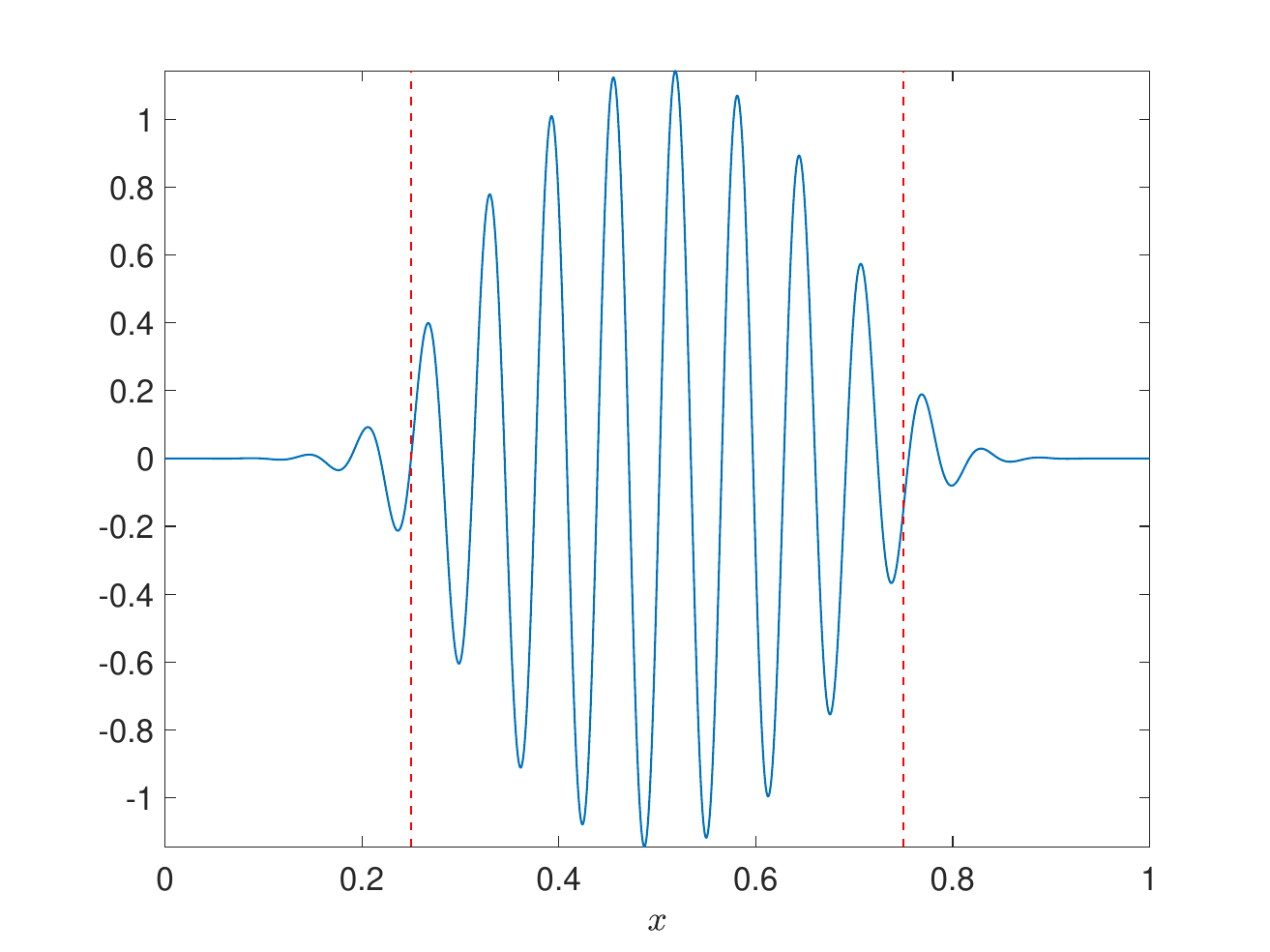}
    \caption{$r = -\cos(2\pi x)$, $\eps = 0.01$}
\end{subfigure}

\begin{subfigure}{.495\textwidth}\includegraphics[width=1\textwidth]{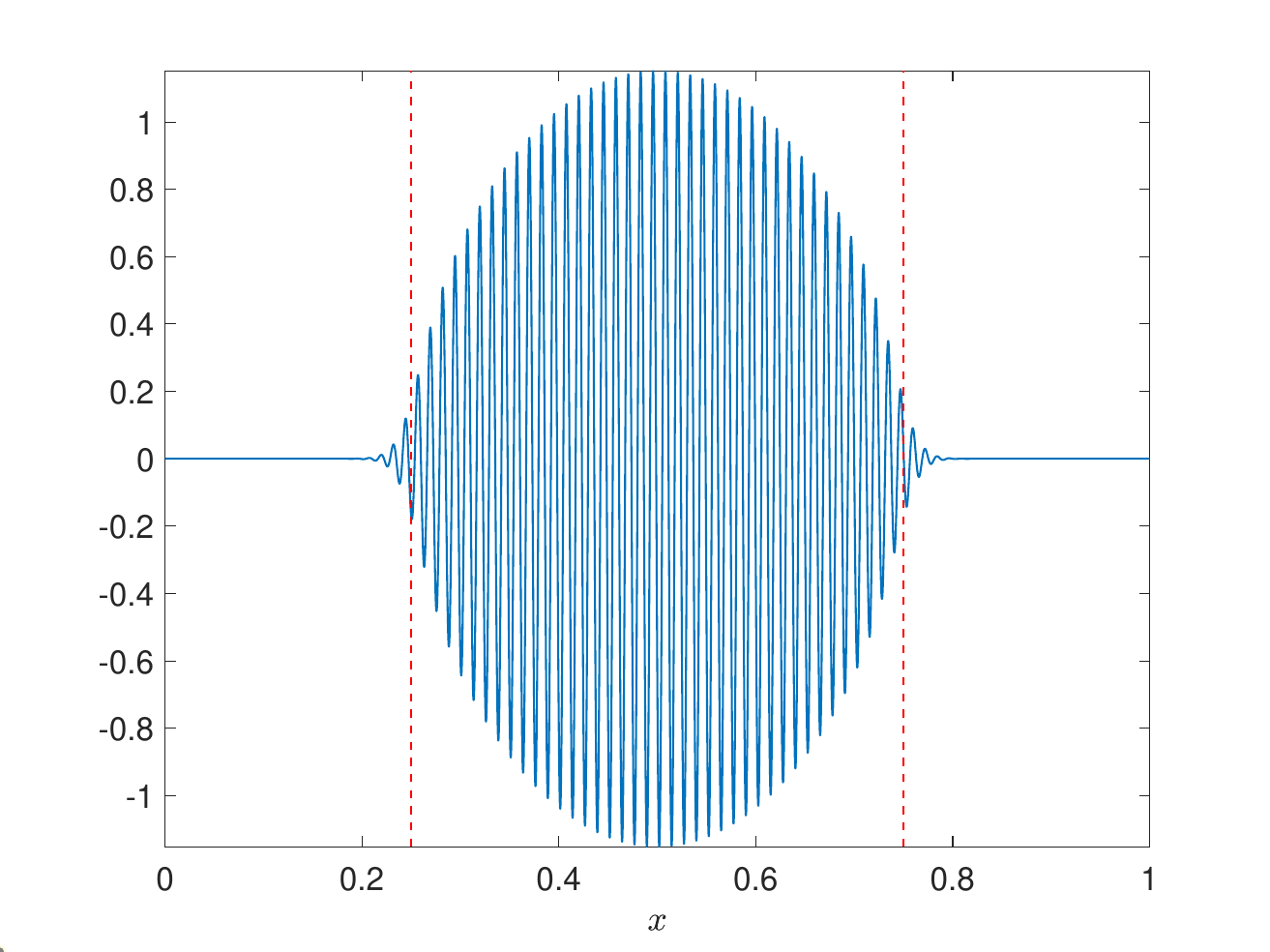}
    \caption{$r = -\cos(2\pi x)$, $\eps = 0.002$}
    \end{subfigure}
    \begin{subfigure}{.495\textwidth}\includegraphics[width=1\textwidth]{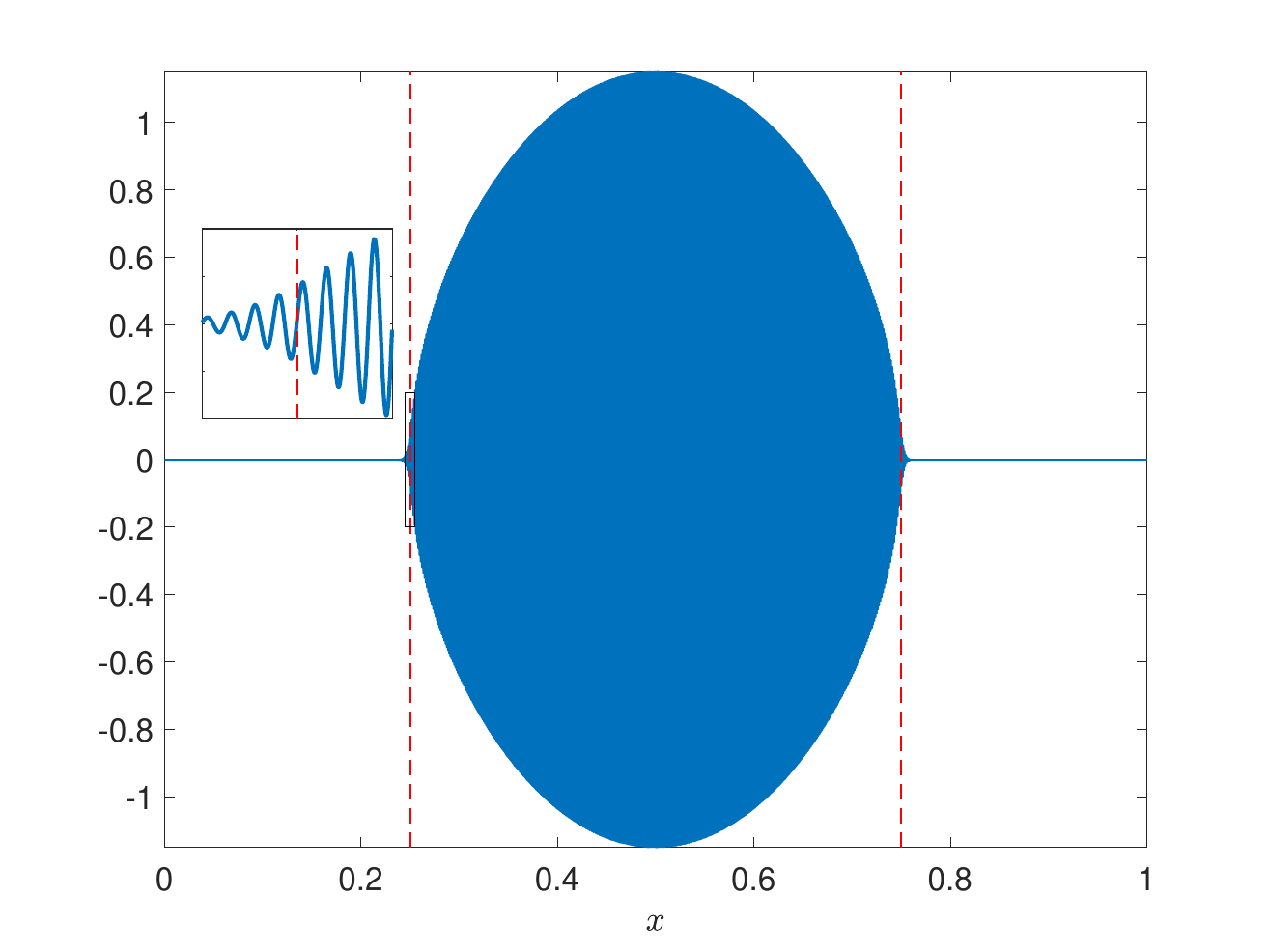}
        \caption{$r = -\cos(2\pi x)$, $\eps = 0.0002$}
        \end{subfigure}
    \caption{Solutions $u(x)$  of \cref{main_sh0} (blue solid curves) with the cubic nonlinearity $N(u) = -u^3$ for varying $r(x)$ and $\eps$. Panel (a) is the homogeneous case corresponding to \cref{homog_sh}, whereas (b)-(d) are spatially heterogeneous, with dashed red vertical lines indicating at $x=0.25,0.75$ where $r(x)=0$, and hence where a naive theory would predict patterning confinement.
    {With $r'=2\pi$ denoting the modulus $|r'(x)|$ at a root of $r(x)$, then in panel (b) 
    we have 
    that the pattern bleeds beyond the red-dash, with the bleed extent  on the scale of the prediction given in   \cref{appc}, that is $(8/r')^{1/2}\eps^{1/2} \approx 0.11$.}    
    The inset in (d) shows the pattern transition region over $x \in [0.245, 0.255]$. Simulation details can be found in \cref{appendix_numerics}.}\label{fig:homog_het_comparison}
\end{figure}

\begin{figure}
    \centering
    \begin{subfigure}{.495\textwidth}\includegraphics[width=1\textwidth]{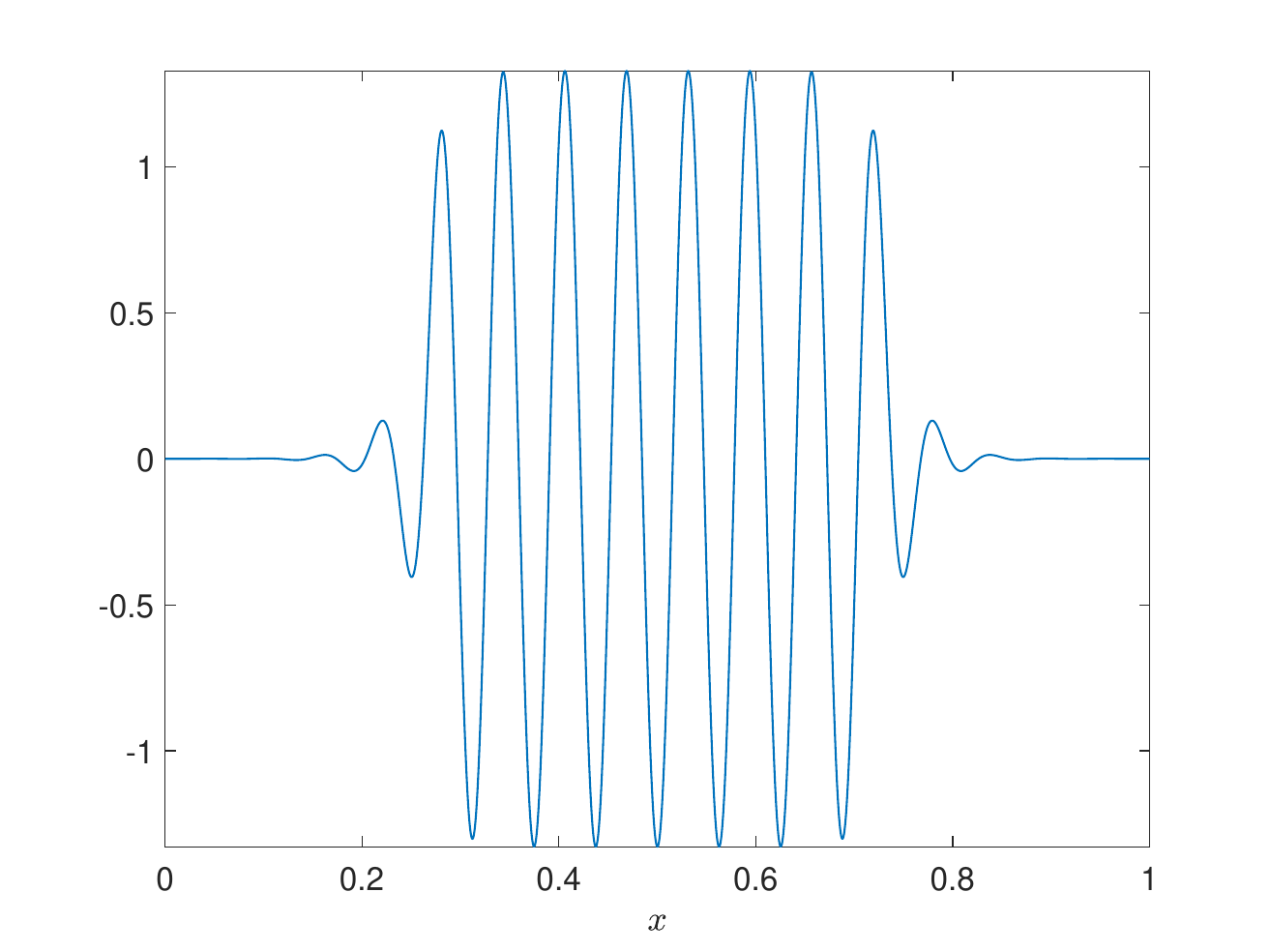}
    \caption{$\eps = 0.01$}
    \end{subfigure}
    \begin{subfigure}{.495\textwidth}\includegraphics[width=1\textwidth]{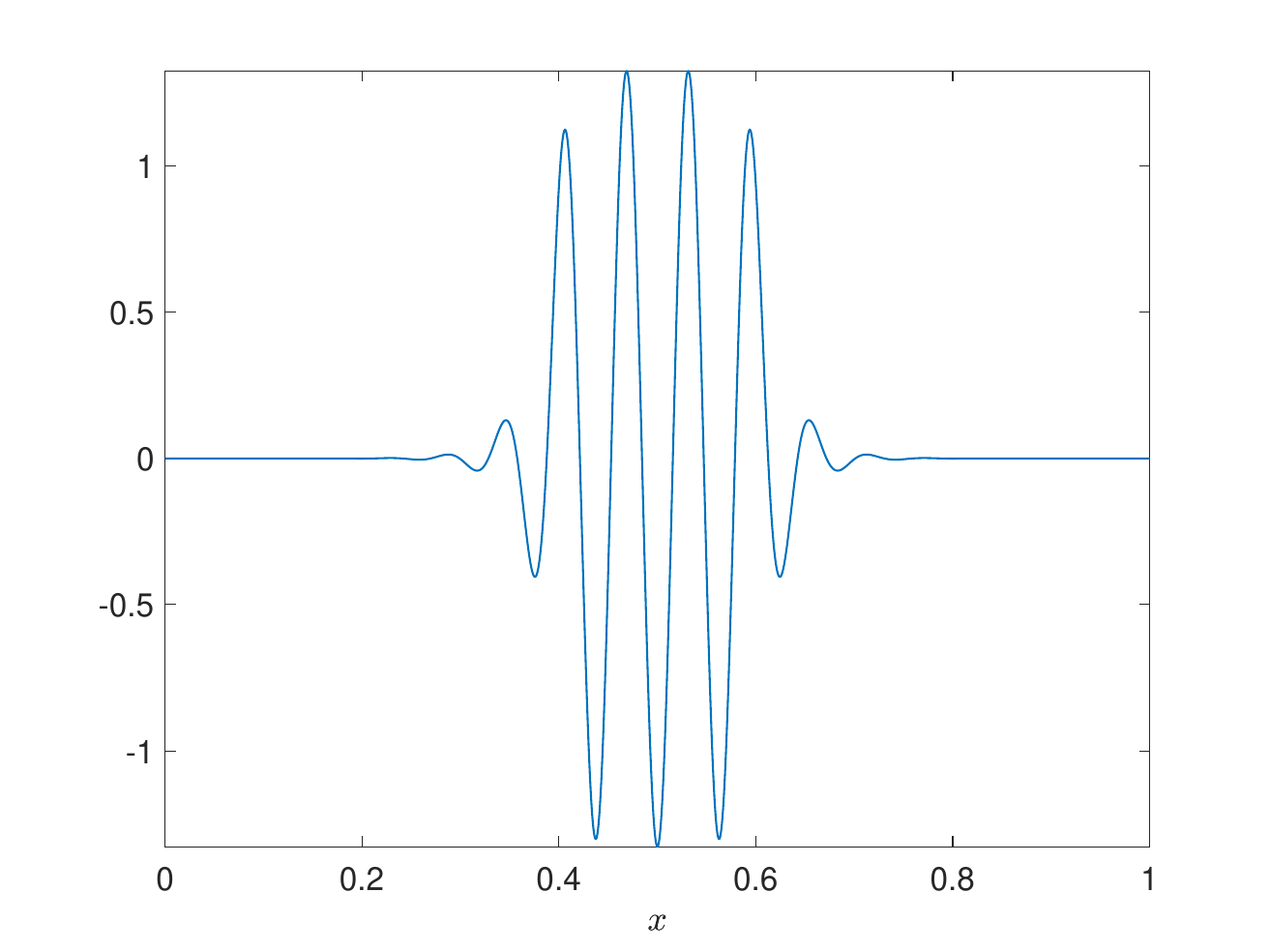}
    \caption{$\eps = 0.01$}
    \end{subfigure}

    \begin{subfigure}{.495\textwidth}\includegraphics[width=1\textwidth]{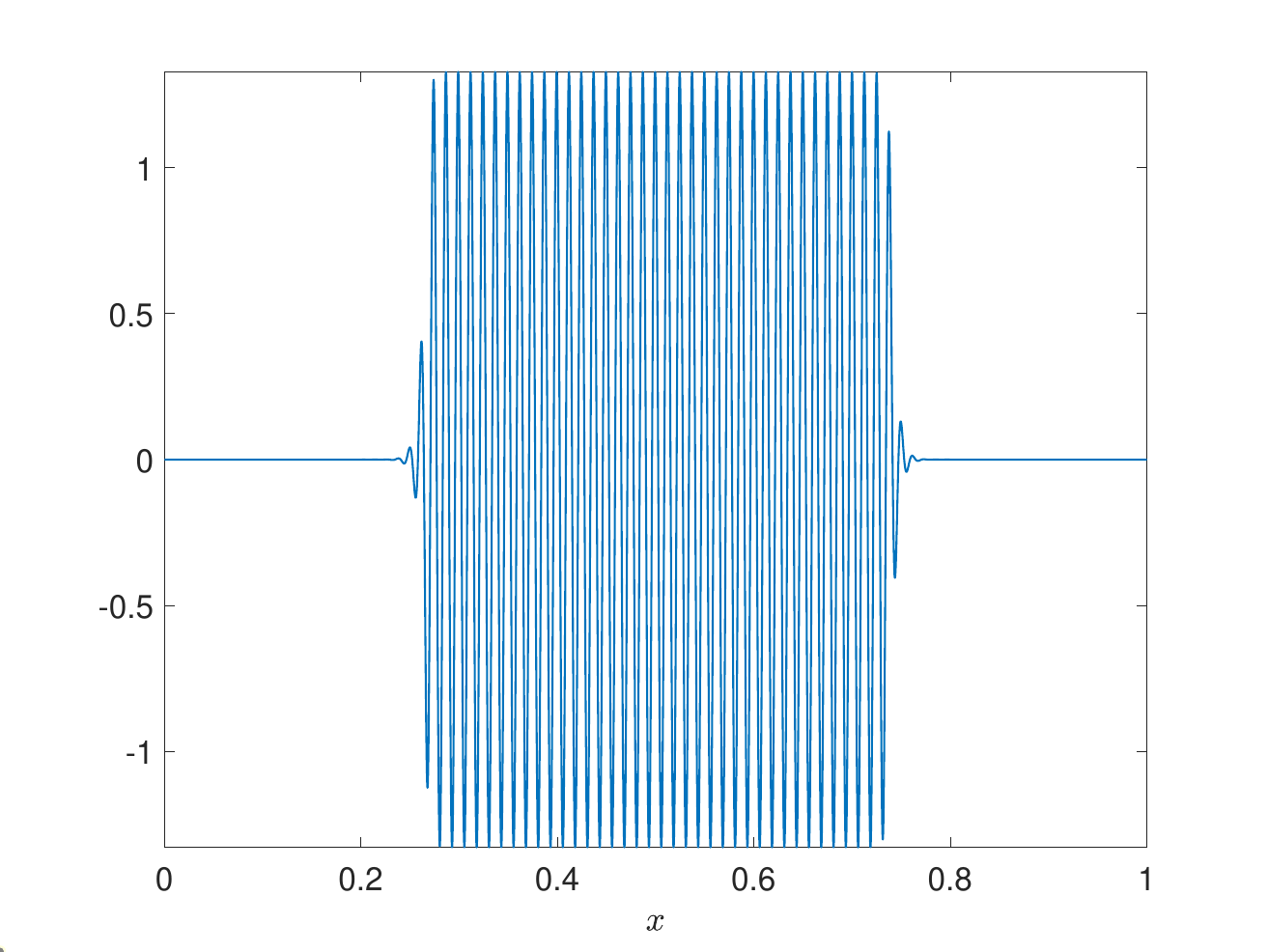}
    \caption{$\eps = 0.002$}
    \end{subfigure}
    \begin{subfigure}{.495\textwidth}\includegraphics[width=1\textwidth]{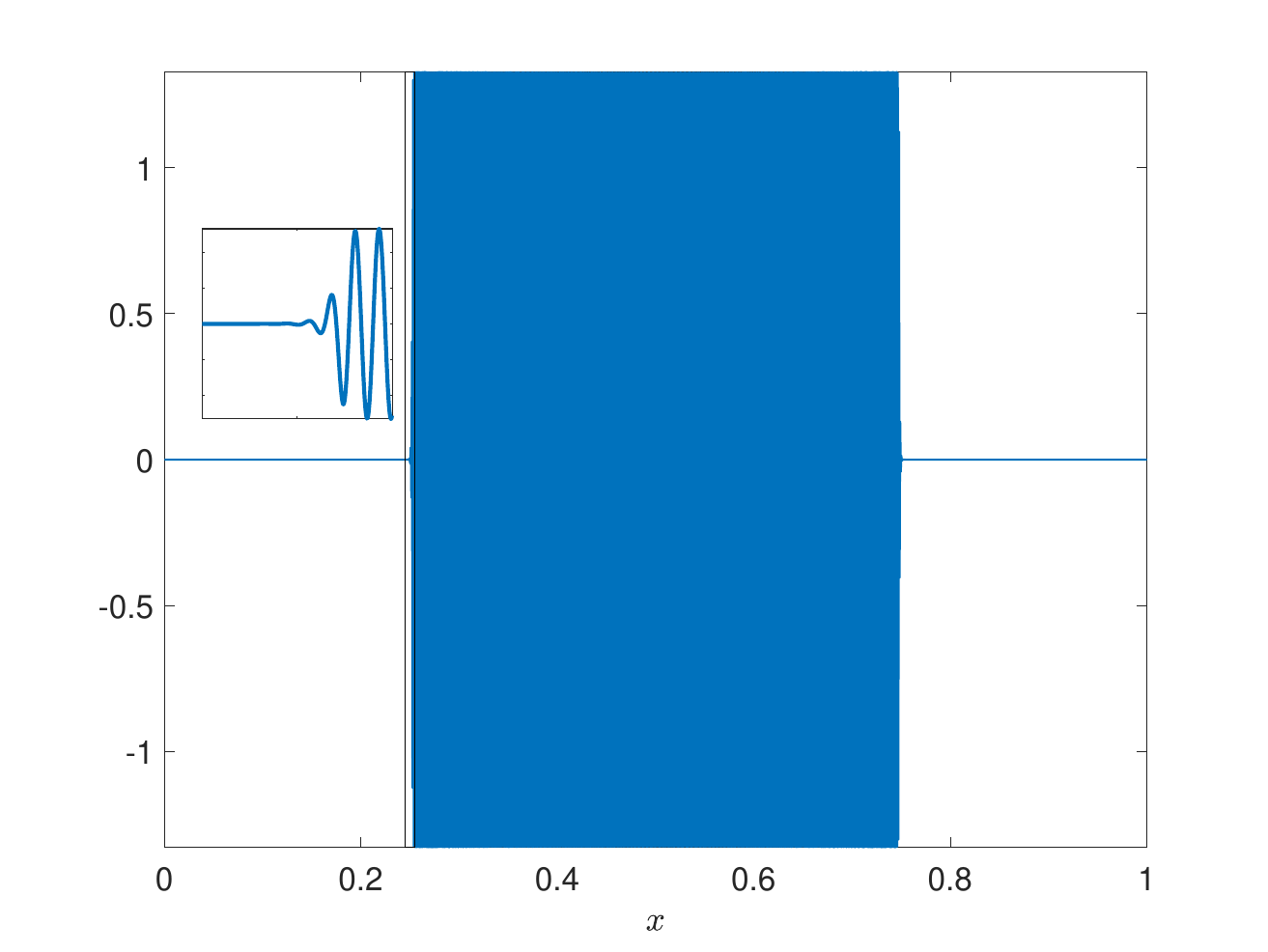}
    \caption{$\eps = 0.0002$}
\end{subfigure}
    \caption{Solutions $u(x)$  of \cref{homog_sh} (blue solid curves) with the quadratic-quintic nonlinearity $N(u) = 2u^3-u^5$ with $r_h=-0.7$. Different initial data were used for the simulation as described in the text. {The inset in (d) shows the pattern transition region over $x \in [0.245, 0.255]$. Note that the $x$ axis in the inset of (d) is the same as in \cref{fig:homog_het_comparison}(d), but the $y$ axis here is much larger.} Simulation details can be found in \cref{appendix_numerics}.}\label{fig:snaking_example}
\end{figure}

The rest of this paper is organized as follows. In \cref{sec_Localisation} we develop a linear stability theory in the limit of small $\eps$ using WKB theory. This gives an exact analogy between pattern-forming conditions in the heterogeneous model and such conditions in a local homogeneous variant corresponding to pattern confinement for $r(x)>0$.  In \cref{sec_Boundary_Matching}, we resolve the boundary layer around the bifurcation points where the leading-order outer WKB solutions are singular, showing how solution envelopes are predicted to decay according to the linear theory. In \cref{sec_Numerics}, we demonstrate that the theory generically fails to predict confinement regions when the bifurcation is locally subcritical, for which we propose an alternative prediction based on Maxwell points of a local analog system, which is only successful for some choices of parameters and nonlinearities.   Finally we discuss these results in \cref{sec_Discussion}, explaining further directions emerging from this work, and highlighting important barriers to classical bifurcation-theoretic paradigms. 

\section{Localisation of Turing instabilities}\label{sec_Localisation}

We now study the canonical Turing instability for this system about the homogeneous steady state $u=0$, leading to a linear problem of the form \cref{main_sh0} with $N(u)=0$. A Turing instability then requires linear stability for homogeneous perturbations, but an instability for a spatially varying perturbation. While such bifurcations are readily studied for constant coefficient systems, the fact that we have $r(x)$ as a function of $x$ entails that the linear stability theory is more involved, as naive expansions in terms of trigonometric eigenfunctions would not diagonalize the linear operator, and hence one cannot study a single mode's stability to deduce how perturbation growth rates depend on wavenumbers. We make use of the small parameter $\eps$ to employ a WKB approximation of the linearised system to arrive at analogous results from previous studies of second-order systems \cite{krause_WKB,gaffney2023}. 

Our first requirement is stability with respect to homogeneous perturbations. Thus we consider a perturbation of the form 
$ u(x,t)= p_0(t;x)$, essentially treating the $x$ dependence as a parameter, whereupon 
 \beq\label{homogeneous_mode}
        \pd{p_0}{t} = ( r(x)  - 1)p_0  . 
    \eeq
    Thus, to ensure stability with respect to homogeneous perturbations, we require }
    \beq \label{hcnsrt}
    1-r(x) > 0, ~\quad  ~\quad x\in[0,1].
    \eeq

    We also note that the solution to \cref{homogeneous_mode} will only satisfy the boundary conditions \cref{bcs} if $r(x)$ does, and henceforth also assume that this is the case. As mentioned in \cite{krause_WKB}, if this assumption is violated, one may expect spatially inhomogeneous boundary layers to form.

\subsection{The WKB solution}
We now consider inhomogeneous perturbations and analyze these asymptotically using WKB approximations. Linearity together with the homogeneous boundary conditions entails we can consider a weighted sum of separable solutions and thus we focus on a single separable solution\footnote{We will proceed formally and neglect details of orthogonality/completeness of the solutions we find. In principle this can be shown using variational methods as the spatial functions will be good approximations to solutions of a self-adjoint eigenvalue problem \cite{lindblom1991improving, kovavc2021wkbj}.}, which is invariably exponential in time. Hence, we seek a solution of the form $u(x,t)=\mathrm{e}^{\lambda t}p(x),$ whereupon 
 \begin{align}
      \quad 0 = (r(x)-\lambda) \, p(x)-\left(1+\eps^2 \pdd{}{x}\right)^2 p(x)  =: -{\cal L}p(x) -\lambda p(x) , \label{linearequation}
    \end{align}
    noting that non-linear terms in the expansion of $N(u)$ have been dropped as only linear terms of $p(x)$ are retained in a linear stability analysis about $u=0$. We note that $\lambda$ is real, as the linear operator ${\cal L}$, with the generalised Neumann boundary conditions, is fully self-adjoint. In addition,   we are focused on whether an unstable solution exists for a spatially heterogeneous perturbation and thus we only consider $\lambda> 0$ below. 
 
 To proceed with the WKB analysis, we consider an  expansion for $p(x) $ of the form 
 \begin{align}\label{WKBexp} 
     \quad p(x) = \exp\left(\frac{i\varphi(x)}{\eps}\right) s(x) = \exp\left(\frac{i\varphi(x)}{\eps}\right) (s_0(x) +\eps s_1(x) + O(\eps^2)),
    \end{align}
with $s_i(x)$ remaining ord(1) as $\eps \to 0$. For simplicity of notation, we will drop the $x$ dependence of $r$, $\varphi$ and the $s_i$ functions. Through direct manipulation,  with the only constant denoted by $\lambda$, we find
\beq\label{orders}
O(\eps^0) \quad\quad  &  0 =  s_0
\left( 
\left[1-\varphi '^2\right]^2+\lambda -r \right)  \\ O(\eps^1) \quad\quad    &  0 =  s_1
\left( 
\left[1-\varphi '^2\right]^2+\lambda -r \right)
+ 4i s'_0\varphi '\left( 
1-\varphi '^2
\right)     +    2is_0\left(\varphi''-3\varphi '^2\varphi''
\right)
\eeq
Thus, from the $O(\eps^0) $  constraint, we find that the possible solutions for $\varphi(x)$ satisfy   
    \begin{align}
        \pm \varphi_\pm(x) = \int_y^x \sqrt{1 \pm \sqrt{r(\alpha) - \lambda}} \, \mathrm{d}  \alpha, \label{varphi}
    \end{align}
    where     $y$ is, currently, an arbitrary constant.  From the $O(\eps^1) $  constraint we have 
       \begin{align}\label{ssl}
       \frac{s_0'}{s_0} = -\frac 1 2 \frac{\left(\varphi''-3\varphi '^2\varphi''
\right)}{\varphi '\left( 
1-\varphi '^2
\right)} = -\frac 1 2  \frac{\left(
\varphi '\left( 
1-\varphi '^2
\right)
\right)'}{\varphi '\left( 
1-\varphi '^2
\right)}. 
    \end{align}
    Hence, restoring the explicit $x$ dependencies,  we have that the general solution for $s_0(x)$ is given by 
    \begin{align}
     \quad s_{0\pm}(x) = \frac{s_{00\pm}}{\left|\varphi'_\pm(x)\left((\varphi'_\pm(x))^2 - 1\right)\right|^{1/2}} = 
     \frac{s_{00\pm}}{\left|\varphi'_\pm(x)|^{1/2} |r(x)-\lambda\right|^{1/4}}, \quad  
    \end{align}
    where $s_{00\pm}$ is a constant of integration that, at this stage, may be complex.

   { We first of all note that whenever $\varphi_\pm$ is not real, the WKB solution would have weighted sums that contain one of the four terms: 
\begin{equation}\label{etas0} \mathrm{e}^{\eta_+/\eps}, ~~\mathrm{e}^{-\eta_+/\eps}, ~~\mathrm{e}^{\eta_-/\eps}, ~~\mathrm{e}^{-\eta_-/\eps}, 
\end{equation} 
    where $\eta_\pm = \mbox{Im} (\varphi_\pm(x)).$ However, the extremely rapid variation in the oscillation envelope amplitude  associated with each of the above four terms suggests that we  do not consider such solutions below, as we  discuss and motivate  in \cref{appc}. Thus, hereafter, we primarily focus on  
 real  $\varphi_{\pm}$. Then} collecting the most general real linear combination of the WKB separable solution generates
    \begin{equation}
        \boxed{ \quad \begin{aligned}
            u_+  &=  \frac{s_{00,+} e^{\lambda t} }{\left|\varphi_+'(x)\right|^{1/2}\left|r(x) - \lambda \right|^{1/4}} \left(A \cos\left(\frac{\varphi_+(x)}{\eps} \right) + B \sin\left(\frac{\varphi_+(x)}{\eps}\right)\right)
            \\
          u_-  & = \frac{s_{00,-}e^{\lambda t}}{\left|\varphi_-'(x)\right|^{1/2}\left|r(x) - \lambda \right|^{1/4}} \left(C \cos\left(\frac{\varphi_-(x)}{\eps}\right) + D \sin\left(\frac{\varphi_-(x)}{\eps}\right)\right),
        \end{aligned}\quad }\label{ueqn}
    \end{equation}
   whenever $x$ is in a region with $\Im(\varphi_{\pm}(x))=0$ for the solutions $u_+(t,x),~u_-(t,x)$ respectively, with 
 $A,B,C,D,{ s_{00\pm}}\in \mathbb R$. {We also note that as these are solutions of the linearised Swift-Hohenberg equation in the context of a linear Turing instability, we only consider solutions that possess 
 $O(1)$ amplitudes, noting further discussion on this point for solutions with complex $\varphi_\pm$ is presented in \cref{appc}. 
 More generally, the requirement that $\varphi_\pm (x)$ is real may entail that these  WKB solutions are    only be defined in subsets of the domain (a.k.a.~their support) and, as in \cite{krause_WKB,gaffney2023}, we will deduce conditions determining their support.}

\subsubsection{The constraint \tmath{\varphi_\pm(x)} is real}

{We are interested in  real WKB solutions that grow in time and thus solutions with $\lambda>0$. We have also imposed  $$\Im(\varphi_\pm(x))=0,$$ possibly restricting the definition of the above solutions to a subset $\Omega=[a,b]\subseteq [0,1]$. In addition to $\lambda>0$, note that we already have $r(x) <1$ for stability with respect to homogeneous perturbations.}

 {Then, without loss,} defining 
the lower limit of the integral in \cref{varphi} for $\varphi_\pm(x)$ to be given by $y=a$,  we will show that 
$$ \mbox{Im}(\varphi_\pm(x)) =0 , \textrm{ for all } \,  x\in \Omega=[a,b] \subseteq [0,1]$$
if and only if
\begin{equation}\label{rcons} 
r(x) \geq  \lambda  , \textrm{ for all } \,  x\in \Omega{.}
\end{equation}
Starting with $r(x)\geq \lambda$ we have that $ (1 \pm \sqrt{r(x) - \lambda})^{1/2} $ is real, 
immediately yielding that $\varphi_\pm(x)$ is real. For the converse, by contradiction suppose the existence of an $x^*\in \Omega$ for which $r(x^*)<\lambda$ and that $\Im(\varphi_\pm(x)) = 0$ for all $x\in \Omega$. If $x^*=a$ then $\varphi_\pm(a+\delta)$ is not real for sufficiently small $\delta>0$ by continuity and we are done.  For $x^*>a$, either 
$\varphi_\pm(x)$ is not  real for at least one point on $[a,x^*)$  and we are done or 
$\varphi_\pm(x)$ is   real on $[a,x^*)$. With the latter, we have that  $\varphi_\pm(x^*-\delta)$ is real for any sufficiently small $\delta>0$, but $\varphi_\pm(x^*)-\varphi_\pm(x^*-\delta)$ is not real as the integrand in the definition of  $\varphi_\pm(x)$  is not real on $[x^*-\delta,x^*]$ by continuity. Hence $\varphi_\pm(x^*)$ cannot be real, completing the demonstration that the converse holds. 

However, while points $x\in\Omega$ with $r(x)=\lambda$ have $\varphi_\pm(x)$ real, it is not clear whether the solution of the form \cref{ueqn} can exist due to a blow-up in the denominator. We only consider cases  where $r(x)-\lambda$ has either no roots or only simple roots, in which case the roots are at the boundary of $\Omega$. {As further motivated in \cref{appc}, in the analytical development below we additionally restrict $r(x), ~\lambda$ such that roots of $r(x)-\lambda$ do not occur within $O(\varepsilon^{1/2})$ of each other or within $O(\varepsilon^{1/2})$ of the boundary. Nonetheless, such constraints are weak and  when satisfied generate numerical observations that are  in accord   with  the behaviour of WKB solutions satisyfying Im$(\varphi_\pm(x))=0$  on testing in the context of {locally} supercritical bifurcations, as we will explicitly explore  below.} As such, solutions of the form of \cref{ueqn}  extensively inform how heterogeneity can control the location of localised patterns, which we proceed to consider.

\subsubsection{Boundary conditions and preventing blow up}
 ~~
In constructing WKB solutions, we also need to address constraints  from the boundary conditions and the possibility of a breakdown of the WKB solution due to a blowup from the contributions
$$ \frac{1}{\left|\varphi_\pm'(x)\right|^{1/2}\left|r(x) - \lambda \right|^{1/4}} $$
to $u_+(t,x)$ and $u_-(t,x)$ of \cref{ueqn}. 

First of all we consider the possibility of a blowup induced by $|\varphi_\pm'(x)|=0.$ Noting 
$$ |\varphi_\pm'(x) |^2 = 1 \pm \sqrt{r(x) - \lambda} $$
we have $ |\varphi_\pm'(x)|$ will only attain zero for an unstable solution if $1 -r(x) = - \lambda <0$, which violates the constraint that the homogeneous steady state is stable everywhere in $x\in[0,1]$, \cref{hcnsrt}. Thus, a blowup due to $|\varphi_\pm'(x)|=0$ cannot occur anywhere in the domain for the solutions that can generate a Turing instability. 

Thus we have the constraints of the boundary conditions and the possibility of a breakdown of the WKB solution due to a root  of $|r(x) - \lambda |$. 
We consider two possibilities in the first instance. With $\lambda>0$ fixed, the first case is given by $r(x)> \lambda $ for all $x\in[0,1]$, while  the second case is given by  $r(x)>\lambda$ only within a simply-connected region $x\in \Omega=(a,b)\subset[0,1]$ where $0< a<b< 1$ and $r(x)-\lambda $ possesses only simple zeros at $x=a,b$ {that are not within $O(\varepsilon^{1/2})$ of each other or the boundary.}

{\bf Case 1} $r(x)>\lambda $ for all $x\in[0,1]$. There is no prospect of a blowup in the WKB solution, so the remaining constraint is that of the generalised homogeneous Neumann boundary conditions. Satisfying these at $x=0$ for $u_+(t,x), ~u_-(t,x)$ of \cref{ueqn} we have that taking $y=0$ without loss of generality subsequently requires that $B=D=0$, while  $A,~C$ can be taken to be unity again without loss of generality, so that the solutions become 
    \beq
            u_\pm  &=  \frac{s_{00,\pm}e^{\lambda t}}{\left|\varphi_\pm'(x)\right|^{1/2}\left|r(x) - \lambda \right|^{1/4}}  \cos\left(\frac{1}{\eps}  \int_0^x \sqrt{1 \pm\sqrt{r(\alpha) - \lambda}} \, \mathrm{d}  \alpha \right).
                   \quad \label{ueqn1}
    \eeq
Satisfying the  boundary conditions at $x=1$ to leading order then  additionally requires the wavenumber constraint 
\begin{equation}\label{wvc} 
  \int_0^1 \sqrt{1 \pm \sqrt{r(\alpha) - \lambda}} \, \mathrm{d}  \alpha = n_\pm\pi \eps, 
\end{equation}
where $n_\pm$ is a positive integer.

{\bf Case 2} We have $r(x)>\lambda $  only for $x\in \Omega=(a,b)\subset[0,1]$ where $\Omega$ is simply connected and $0<a<b<1$, with  $r(x)-\lambda $ possessing  simple zeros at $x=a,b$  {that are not within $O(\varepsilon^{1/2})$ of each other or the boundary. We}  have a potential blowup, and thus a loss of validity of the WKB solution, on approaching $x=a,b$. {Also, outside $\Omega$ and sufficiently away from the points where $r(x)=\lambda$ for a WKB solution to be valid, we have $\varphi_\pm(x)$ is not real and thus any 
WKB solution would have weighted sums that contain one of the four terms: 
  \begin{equation}\mathrm{e}^{\eta_+/\eps}, ~~\mathrm{e}^{-\eta_+/\eps}, ~~\mathrm{e}^{\eta_-/\eps}, ~~\mathrm{e}^{-\eta_-/\eps}, \label{etas}
  \end{equation}
    where $\eta_\pm = \mbox{Im} (\varphi_\pm(x)).$
It is worth noting that analogy to the use of the Airy function in the standard  WKB approximation for second order systems \cite{bender2013advanced} would suggest  that solutions of the form of \cref{ueqn} can be matched via connection formulae and an inner region solution to 
an evanescent solution, that is a solution containing  terms 
from \cref{etas} that decay asymptotically fast away from the point where $r(x)=\lambda$, with the solution zero to all asymptotic orders. The standard WKB approximation also suggests  an analogous solution that  exponentially grows with an exponent scaling with $1/\varepsilon$ in this region. 
However, such behaviour is not well-documented for fourth order systems and this level of detail is typically not required at present, as explicitly discussed and motivated in \cref{appc}. 
Hence, for now,  we only consider solutions of \cref{ueqn} that tend to zero as the location of the root $r(x)=\lambda$ is approached to prevent blow-up. 
We also neglect solutions including terms from \cref{etas} that rapidly grow in space  away from points with $r(x)=\lambda$, as also detailed in \cref{appc}, where we also motivate use of the crude, but asymptotically correct,  approximation that    evanescent solutions are zero.

 We will also indicate where this approximation breaks down in the numerical observations later due to the impact of $\eps$ not  being sufficiently  small for the approximation to hold but, for now, we take it that the WKB solutions of \cref{ueqn} tend to zero when $x\rightarrow a, ~r(x)\rightarrow \lambda$. This  is achieved without loss by setting  $y=a$, $A=C=0$, $B=D=1$, whereupon we have the solutions,
\beq
u_\pm =      \dfrac{s_{00,\pm}e^{\lambda t}}{\left|\varphi_\pm'(x)\right|^{1/2}\left|r(x) - \lambda \right|^{1/4}}  \sin\left(\dfrac{1}{\eps} \psi_\pm(a, x)  \right),     ~~~~~ x \in \Omega = (a,b), 
                  \label{ueqn2} 
\eeq
where
\beqs
   \psi_\pm(a,x) &=&  {\displaystyle \int_a^x} \sqrt{1 \pm\sqrt{r(\alpha) - \lambda}}   \,  \mathrm{d}  \alpha.
    \eeqs
   Noting the above discussion, we simply take 
$$ u_\pm = 0 , ~~~~~  x\not  \in \Omega = (a,b).$$ 
Analogously requiring the absence of blow-up at the righthand boundary 
gives the wavenumber constraint 
\begin{equation}\label{cn2}
   \psi_\pm(a,b)=\int_a^b  \sqrt{1 \pm \sqrt{r(\alpha) - \lambda}} \, \mathrm{d}  \alpha = n_\pm\pi \eps, 
\end{equation}
with $n_\pm$ a positive integer. Thus taking  $(a,b)$ as the region for localised patterns of temporal growth rate $\lambda>0$ induces an asymptotically small error  in the support of the pattern as evanescent solutions are anticipated to cause a bleed of asymptotically  small extent for the pattern across the point $r(x)=\lambda$. This will also induce an analogous asymptotically small correction to the wavenumber condition due to the correction in the patterning extent, with brief further discussion in \cref{appc}.}

{Nonetheless even with these crude approximations,  which are still  accurate as $\varepsilon$ tends to zero, we have  simple conditions on  $r(x)$ that give the region of localisation for Turing-type instabilities.
Such constructions can be readily generalised as required to generate solutions when $r(x)>\lambda $ on non-simply connected domains or domains that partially or fully} include the boundaries. More generally, on the regions where $r(x)>0$ there is a growth rate for which solutions such as \cref{ueqn2} exist provided the wavenumber constraint \cref{cn2} is satisfied, which will generally be true for a suitable,  sufficiently small, choice of $\eps$. 
Further, in the construction of the WKB solutions we note that localisation for the Turing instability once $r(x)$ is spatially varying is determined by the roots of $r(x)$ according to leading order WKB solutions. We will numerically show in \cref{sec_Numerics} that for  classes of nonlinearities associated with {locally} supercritical bifurcations, this instability criterion correctly predicts pattern confinement approximately within regions where $r(x)>0$.

{Note that} these observations are consistent with the fact that in the absence of spatially heterogeneous coefficients, that is for constant $r$, the conditions for a Turing instability are  $ 1 - r >0, $ for the homogeneous steady state to be stable and $r>0$ for a spatially varying perturbation to be unstable (together with a wavenumber constraint). Thus, as seen previously for {locally supercritical Turing bifurcations of} reaction-diffusion and cross reaction-diffusion systems \cite{gaffney2023,krause_WKB}, the condition for the spatially varying perturbation to be unstable is inherited pointwise once coefficients become spatially varying whilst in the parameter regime where WKB solutions are legitimate.

Finally, we note that the localisation of instabilities in this asymptotic linear theory is delimited by where the WKB solution breaks down due to a prospective blowup. As already mentioned, such behaviours of the WKB system are not well-documented for fourth-order systems.  {Hence, we return to consider a detailed analysis of how the solution of \cref{linearequation} behaves in the vicinity of $r(x) \sim \lambda$. We still assume that  we do not have growth with an asymptotically large exponent on emerging from the inner region as further discussed in \cref{appc}.
Thus, in particular, we proceed to examine the structure of the localised inner-region solution as it transitions from its oscillatory form of \cref{ueqn} to an evanescent solution.}

\section{Solution behaviour of unstable solutions near a WKB turning point}\label{sec_Boundary_Matching}
    The solution $p(x)$ to \cref{linearequation} given in terms of the functions $\varphi(x)$ and $s_0(x)$ above can be thought of as an outer solution away from turning points. In the argument above, we have forced this outer solution to be zero to prevent blowup, but we can resolve the actual behaviour across the turning point through {an inner solution scaling} which we now do in this section.
    
    From {the } outer WKB solution we have potentially two {types of} turning points, {one} when $r(x^*) = \lambda$ and  {one} when $1 - \sqrt{r(x^*) - \lambda} = 0$. Note that the singular point $r = 0$ corresponds to the bifurcation point in the classical theory which is not exactly equal to the turning point in WKB, that is $r(x^*) - \lambda = 0$, though we expect to achieve {an} arbitrarily good approximation to the classical bifurcation point as $\eps \to 0$ (see \cite[Theorem 4.4 and Proposition 11]{gaffney2023} for a more careful discussion of this argument). Further, the second singular point is \underline{not} of concern here, as we require the homogeneous steady state (HSS) to be stable to homogeneous perturbations (i.e.~$r(x) - 1 < 0$ {for all $x$}). 
    
We note that when $\eps$ is sufficiently small {and the heterogeneity function $r(x)$ is fixed, the turning points are sufficiently isolated to consider them in isolation, with decaying behaviour on one side and matching into the WKB solutions of the previous section on the other. See \cref{appc} for more details, for instance Eqs.~(\ref{val1},\ref{c4}). }

\subsection{Inner solution using contour integration}

We consider the following expansion of the coefficient $r(x)$ near $x=a$:
\beq
  r(x)=\lambda+(x-a) \rho + \Ord{(x-a)^2}.
\eeq
Let $y=(x-a)/\eps$ so that the inner problem reads
\begin{equation}
  \label{eq:1}
  0=\eps \rho y p - ( p + 2 p_{yy}+p_{yyyy}).
\end{equation}

Motivated by the fact that we are looking for the generalisation of an Airy function, which may be written  in terms of a contour integral, we consider the generalisation to $p(z)$ denoting a holomorphic function of $z\in\mathbb{C}$, with $\Re(z)=y$, that satisfies 
\begin{equation}
  \label{eq:10}
  0=\eps \rho z p - ( p + 2 p_{zz}+p_{zzzz}).
\end{equation}
Hence  
$ \Re(p(\Re(z)))$ 
and 
$ \Im(p(\Re(z)))$ are two independent solutions to  the original (real) problem \eqref{eq:1} and thus 
 we look for a solution of \cref{eq:10} in the form of the contour integral
\beq
  p(z)=\int_C f(t) e^{-zt} \dd t,
\eeq
where $t\in\mathbb{C}$, and the contour $C$ is to be identified. Then,
\beq
  zp(z) = \int_C f(t) \left(-\frac{d}{dt} e^{-zt}\right) \dd t = \int_C \frac{d f}{dt} e^{-zt} \dd t +\left[-f(t) e^{-zt}\right]^{C_{\rm end}}_{C_{\rm start}}
\eeq
and hence 
\beq
  0=\eps \rho  \left[-f(t) e^{-zt}\right]^{C_{\rm end}}_{C_{\rm start}}+\int_C  e^{-zt} \left(\eps \rho \frac{d f}{dt} - (1+2 t^2 + t^4)f \right)\dd t.
\eeq
{Setting the integrand to $0$ and solving for $f(t)$, we find that
\beq \label{eq:contourInt}
  p(z) = \int_C  \exp\left[\frac{1}{\eps\rho}\left(t+\frac{2}{3}  t^3 + \frac{1}{5} t^5\right) -zt\right]\dd t
\eeq
is a solution of \eqref{eq:10} if
\beq
  \exp\left[\frac{1}{\eps\rho}\left(t+\frac{2}{3} t^3 + \frac{1}{5} t^5\right) -zt\right]^{C_{\rm end}}_{C_{\rm start}} =0,
\eeq
and the integral converges. The contour $C$ cannot be closed (otherwise we have $p=0$ due to Cauchy's theorem) and to force the boundary term to vanish,  the contour  $C$ is chosen such that it both starts and ends at  limits of $t$ with the real part of 
$$ \frac{1}{\eps\rho}\left(t+\frac{2}{3} t^3 + \frac{1}{5} t^5\right) -zt   $$
tending to {\b minus infinity} in the same limit  for all real $z$ {so that the exponential decays}. 

For convenience, we now rescale with 
$$ z = \frac 1 {\eps} \xi, ~~~~ t = i s $$
to define $p_*(\xi)$ with 
\beq
  p(z) =: i  p_*(\xi) = i  p_*({ \eps} z) =   i \int_{\tilde{C}} \exp\left(\frac{1}{\eps^*} \psi(s;\xi) \right) \dd s,
\eeq
where $\tilde{C}$ is the mapped contour and
\begin{equation} \label{eq:scaling}
\psi(s;\xi) = i \left[(1-\Lambda)s-\frac{2}{3} s^3+\frac{1}{5} s^5\right], \quad \eps^* = \rho \eps, \quad \Lambda = \rho \xi.
\end{equation}
Note that $\psi$ is an odd function of  $s$ and $\Lambda$ is just a rescaling of $\xi$ by the local gradient of the heterogeneity.
Further, with an overloading of the symbol $a$, which also previously  denoted the location of the inner region,   we define  $s=a+ib$ to determine  
    \begin{align}
      \Im(\psi) &= a\left( (1-\Lambda) -\frac{2}{3} a^2 + 2b^2 + \frac{1}{5} a^4 - 2 a^2 b^2 + b^4\right), \label{eq.ImPsi}\\
      \Re(\psi) &= -b\left( (1-\Lambda) -2 a^2 + \frac{2}{3} b^2 +  a^4 - 2 a^2 b^2 + \frac{1}{5} b^4\right).\label{eq.RePsi}
    \end{align}

In principle, we now have solutions to the original (real) problem \eqref{eq:1} in terms of the contour integral, which can be used to construct  the inner solution of the leading-order WKB asymptotics via the linearly independent solutions  
$$ \Re(p(\Re(z)))= -\Im(p_*(\Re(\xi))),
~~~~~~  \Im(p(\Re(z)))= \Re(  p_*(\Re(\xi))). $$
 However, we would like to have an explicit form of these solutions, at least as an approximation of the contour integral representations, to construct the inner solution and match it with the outer solutions given by \eqref{ueqn}. We now proceed via a steepest descent argument 
 {for small values of $|\eps^{*}|$} in order to simplify this representation to a form suitable for interpretation and matching to the outer solution. We split this into $\Lambda>0$ and $\Lambda<0$, as these will correspond to being on different sides of the turning point.

\subsection{Asymptotics for \tmath{\Lambda>0}}
{ First, as we are assuming the stability of the homogeneous steady state, we always have $\Lambda<1$. Next,} we identify the saddle points of the integrand. If $\Lambda>0$, then we have two  real  roots in $s^2$ for  $\psi'(s)=(1-\Lambda)-2 s^2 + s^4=0$. 
Thus the  saddles are at 
\begin{align}
  s_{++}&:= + (1+\Lambda^{1/2})^{1/2},\label{s++Lposit}\\
  s_{-+}&:= - (1+\Lambda^{1/2})^{1/2}=-s_{++},\\
  s_{+-}&:= + (1-\Lambda^{1/2})^{1/2},\label{s+-Lposit}\\
  s_{--}&:= - (1-\Lambda^{1/2})^{1/2}=-s_{+-},
\end{align}
and, due to symmetry, we may drop $s_{-+}$ and $s_{--}$.

The steepest descent contour (SDC) $C_s$ is given by  $C_s\equiv \Im(\psi)=\alpha$ where $\alpha$ is the value of $\Im(\psi)$ at the chosen saddle point {(note} that the symbol $\alpha$ has been overloaded and redefined here, having previously denoted a dummy integration variable in \cref{varphi}). \bk In particular, with a contour parameterized by $\tau$, we have 
\begin{align*}
  C_s\equiv \Bigg\{(a,b)\, \Bigg| \, a&=\tau,~~
  b=\pm \left(-1+\tau^2\pm\frac{1}{\tau}\left(\frac{4}{5} \tau^6 - \frac{4}{3} \tau^4 + \Lambda \tau^2 +\alpha \tau\right)^{1/2}\right)^{1/2}  \Bigg\}.
\end{align*}
As both $s=s_{++}$ and $s=s_{+-}$ are real, we have for these saddle points that
\begin{align*}
  \alpha_{++}:=\alpha(s=s_{++})=-\frac{4}{15} (1+\Lambda^{1/2})^{1/2} (-2 + \Lambda^{1/2}+3 \Lambda),\\
   \alpha_{+-}:=\alpha( s=s_{+-})=-\frac{4}{15} (1-\Lambda^{1/2})^{1/2} (-2 - \Lambda^{1/2}+3 \Lambda).
  \end{align*}

For analytical estimates, we need to determine the direction of the steepest descent and the asymptotes. The direction  is given by the angles $\pi/4$, $3\pi/4$ at $s_{++}$, $s_{+-}$ respectively, {which we obtain from the reciprocal value of the square root of the {negative of the} second derivative at the saddle point, $1/\sqrt{-\psi''(s)}$}. 
The asymptotes follow from the fact that the contour away from the saddles is given by $\psi(s) \sim i(-2 s^3/3+s^5/5+s(1-\Lambda))\sim i \frac{1}{5} s^5$.
Hence, to have the integrand $e^{\psi(s)/\eps^*}$ asymptotically small (and a convergent integral) for $s\to+\infty$, i.e. $\psi(s)\to -\infty$, we need to have
\beq
  e^{\frac{1}{\eps^*}\psi(s)} \sim \exp\left(\frac{1}{\eps^*} i \frac{1}{5} |s|^5 e^{i 5 \theta}\right) \to 0 \mbox{ as }|s|\to+\infty \mbox{ for } s=|s| e^{i\theta}.
\eeq
Hence, $e^{i 5 \theta}=i$ and thus $\theta=\frac{\pi}{10}+\frac{4k}{10} \pi$ for $k\in\{0,1,\ldots,4\}$, where $\theta$ is the angle of the asymptotes.

These qualitative results are compared against numerical results for $\Lambda=0.6\in(0,1)$ in   \cref{SDCs++} (see also \cref{SDCs+-,SDCcontours} in \cref{appendix_lambda>0}). One can observe that the estimates of the steepest descent contour at the saddles and at infinity do match and that the behaviour at $\xi=+\infty$ is as desired (thus, ensuring  the boundary term vanishes).


\begin{figure}
\centering
\begin{subfigure}{.485\textwidth}
   \centering
   \includegraphics[width=.6\linewidth]{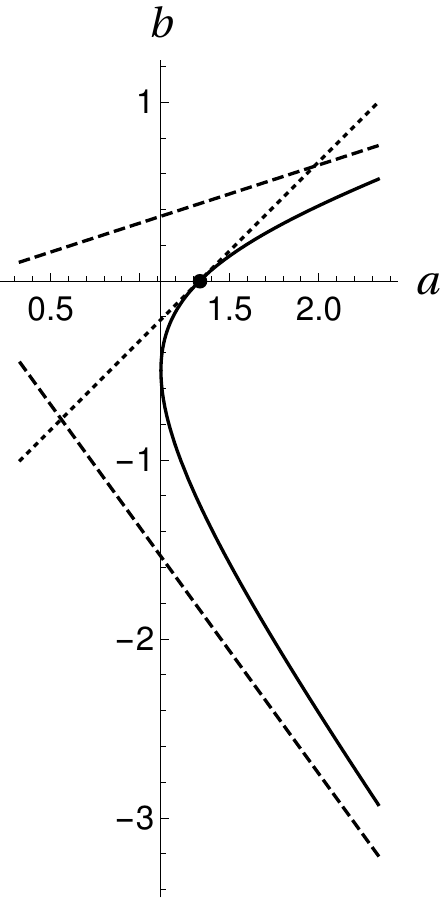}
   \caption{Plot of the steepest descent curve (with a constant phase), in the complex plane.}
\end{subfigure}~~~~
\begin{subfigure}{.485\textwidth}
   \centering
   \includegraphics[width=.95\linewidth]{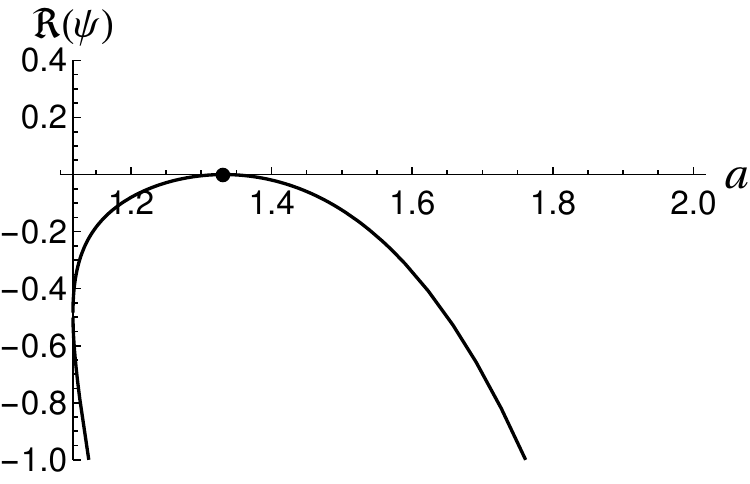}  
\caption{Real part of $\psi$ along the contour with the horizontal axis corresponds to $\tau=a$, which parameterises the curve confirming that it is the (steepest) descent curve.
}
 \end{subfigure}
    \caption{\label{SDCs++}Steepest descent contour for $\Lambda=0.6$ passing through $s_{++}$. The dashed lines indicate the revealed asymptotes and tangent at the saddle. In particular, we have $s_{++}= 1.33$, $\alpha_{++} = -0.2$, the tangent {angle} of the SDC at $s_{++}$ is $\pi/4$ and the asymptotes are {at the angles} $\pi/10$ and $17\pi/10$.}
\end{figure}

One can show from \cref{eq.ImPsi}, \cref{s++Lposit} and \cref{s+-Lposit} that the saddles \emph{do not} lie on the same contour and hence all four contours are admissible (though two of them are simple reflections as follows from their symmetry). These asymptotes then generate four independent solutions which satisfy \cref{eq:1}, namely $s_{++}$ and $s_{+-}$ each with real and imaginary parts generating a distinct solution as noted above.

Finally, one can approximate the contour integral near the saddle, following the Laplace method  once sufficiently away from $\Lambda=0$; {this approximation for  
the solution associated with} the contour passing through the saddle $s_{++}$ is
\begin{equation}
  \label{eq:p++4LambdaPosit}
p_{++}(\xi)\sim \sqrt{\frac{\pi \eps^*}{2}}  \Lambda^{-1/4} (1+\Lambda^{1/2})^{-1/4} \cos\left[\frac{1}{\eps^*}\frac{4}{15}(-2+\Lambda^{1/2}+3 \Lambda) (1+\Lambda^{1/2})^{1/2} \right],
\end{equation}
while there is another {approximate} solution when replacing $\cos$ with $\sin$ in the expression (corresponding to the real and imaginary part of the complex solution).

Similarly, the {Laplace method approximation to the} solution corresponding to the $s_{+-}$ saddle is
\begin{equation}
  \label{eq:p+-4LambdaPosit}
p_{+-}(\xi)\sim \sqrt{\frac{\pi \eps^*}{2}}  \Lambda^{-1/4} (1-\Lambda^{1/2})^{-1/4} \cos\left[\frac{1}{\eps^*}\frac{4}{15}(-2-\Lambda^{1/2}+3 \Lambda) (1-\Lambda^{1/2})^{1/2} \right],
\end{equation}
where again we have another {approximate} solution when replacing $\cos$ with $\sin$.

Note that this is exactly the outer WKB solution given by \cref{ueqn} provided $r(x)$ is itself a linear function, that is when $r(x)=\lambda+\rho (x-a)$. Therefore, this contour integral representation of the inner solution should match the WKB outer solution. However, the above approximation of the contour integral is still only valid if it is sufficiently far away from the turning point (so that the Laplace method works or, intuitively, the Gaussian is not spread out too far from the saddle point which would mean that the approximation of the contour by a straight line in the steepest descent direction is insufficient).

\subsection{Asymptotics for \tmath{\Lambda<0}}

In the situation when $\Lambda<0$, one can repeat the analysis above for the positive case with a few key but technical differences: the saddle points are complex and the two of interest ($s_{++},~s_{+-}$) are complex conjugates; the SDCs have the same asymptotes but different tangents at the saddles, as they are no longer constant in $\Lambda$; the two saddles $s_{++},~s_{+-}$ lie on the same contour while $\Re(\psi)$ is larger at $s_{+-}$. 

Finally, the {method for constructing the} contour integral approximation is the same as in the positive case once sufficiently away from $\Lambda=0$.
We again have four independent solutions corresponding to the real and imaginary parts of the saddle points $s_{++},~s_{+-}$. {Note that  the SDC corresponding to the saddle $s_{+-}$ passes through two saddles ($s_{++}$ and $s_{+-}$) and hence has two contributions. However, due to $\Re(\psi)|_{s_{+-}}>\Re(\psi)|_{s_{++}}$, the contribution of the neighbourhood of $s_{++}$ to the contour integral along the $s_{+-}$ contour (passing through $s_{++}$) does not contribute to the leading-order asymptotics. See Appendix \ref{appendix_lambda<0} for more details.}

We finally obtain the asymptotic solution
\begin{multline} 
  p_{++}(y) \sim \sqrt{2\eps^* \pi} \left(-\psi_{++}'' e^{2i{\phi_{++}}}\right)^{-1/2} \exp(H) \\
             \times \cos\left(\frac{1}{\eps^*}\frac{4}{15} \left(1+(1-\Lambda)^{1/2}-3\Lambda\right)(1-\Lambda)^{1/4} \cos(\varphi)\right),\label{eq:p++4LambdaNeg}
\end{multline}
where
\begin{multline*}
H = \frac{-\sin(\varphi)(1-\Lambda)^{1/4}}{240\eps^*}  \left[184+56(1-\Lambda)^{1/2}-132\Lambda\right. \\ \left.+\frac{120\Lambda}{(1-\Lambda)^{1/2}\sin^2(\varphi)} + \frac{\Lambda^2}{1-\Lambda}\frac{15}{\sin^4\varphi}\right] 
\end{multline*}
and $\phi_{++}$ is the angle of the SDC at $s_{++}$, {and thus given by 
\begin{align}\nonumber 
 ~~~ \phi_{++} &= \mathrm{Arg}\left[\left((-i+\sqrt{-\Lambda})\left[(-1+\sqrt{1-\Lambda})\cos(\varphi)+i(1+\sqrt{1-\Lambda})\right]\right)^{-1/2}\right],\\~~~
    \phi_{+-} &= \mathrm{Arg}\left[\left((-i-\sqrt{-\Lambda})\left[(-1+\sqrt{1-\Lambda})\cos(\varphi)-i(1+\sqrt{1-\Lambda})\right]\right)^{-1/2}\right], \label{phipp0}
\end{align}
with an analogous expression for  $\phi_{+-}$; 
(see \cref{{appendix_lambda<0}} for details, in particular \cref{phipp}).} There is again another solution with $\sin$ instead of $\cos$ and $-\psi_{++}'' e^{2 i \phi_{++}}>0$ by the construction of the steepest descent curve.


The other saddle, $s_{+-}$, with larger $\Re(\psi)$ yields an approximate solution
\begin{multline} 
  p_{+-}(y) \sim \sqrt{2\eps^* \pi} \left(-\psi_{+-}'' e^{2i{\phi_{+-}}}\right)^{-1/2} \exp(H)\\
             \times \cos\left(\frac{1}{\eps^*}\frac{4}{15} \left(1+(1-\Lambda)^{1/2}-3\Lambda\right)(1-\Lambda)^{1/4} \cos(\varphi)\right), \label{eq:p+-4LambdaNeg}
\end{multline}
{with  $\phi_{+-}$ defined as above.
Note that} there is an additional solution when replacing $\cos$ with $\sin$, as before.


In summary, the contour integral representation contains four independent solutions for both cases of $\Lambda>0$, and $\Lambda<0$, with these approximations available away from the turning point $\Lambda=0$, that is   $|x-a|\gg \eps^{2/3} \rho^{-1/3}$. These solutions show similar characteristics as Airy functions, as one might expect from a WKB approximation. Two solutions are oscillatory with an exponentially decaying envelope while the other pair of solutions are oscillatory with an exponentially growing envelope.

\subsection{Approximation of the contour integral near the turning point}

In this key situation, the above approximations invoking  Laplace's method {are no longer valid as this method  relies}
on the admissibility of the SDC replacement by a tangent line. We shall take advantage of the fact that each of the two pairs of the real saddles 
coalesce as {$\Lambda\to 0^+$ and then separate out into two complex conjugate pairs once $\Lambda$ has become negative, so for $0<|\Lambda|\ll 1$ the saddles are  close to coalescence.} 

{ This invites the use of the  method} of coalescing saddles (see \cite[Chap 23]{temme2014asymptotic}, \cite[Chap 9]{olver1997asymptotics}, or the original paper developing the technique \cite{chester1957extension}), where the main idea is to find a suitable change of variables into a cubic function in the exponent so that one can use the known integral representation and asymptotics of Airy functions, which we denote as $\Ai(z)$. Namely, we have \cite{temme2014asymptotic}
\begin{multline} \label{eq:2}
  \frac{1}{2\pi i} \int_{C_{\Ai}}e^{\frac{1}{\eps}(\frac{1}{3} t^3 - \eta t)} f(t) \dd t \sim \eps^{1/3}\left[\frac{1}{2}\left(f(\sqrt{\eta})+f(-\sqrt{\eta})\right)+\Ord{\eps}\right] \Ai\left(\eta \eps^{-2/3}\right)\\-\eps^{2/3}\left[\frac{1}{2}\frac{1}{\sqrt{\eta}}\left(f(\sqrt{\eta})-f(-\sqrt{\eta})\right)+\Ord{\eps}\right] \Ai'\left(\eta \eps^{-2/3}\right),
\end{multline}
as $\eps\to 0^+$   where the contour $C_{\Ai}$ is one of the three Airy contours with the asymptotes of $(-1)^{1/3}$. In our case, we have that
\beq
  \int_C e^{\frac{1}{\eps^*}\psi(s)} \dd s \sim \int_{\tilde{C}} e^{\frac{1}{\eps^*} \Xi(t)} \frac{\dd s}{\dd t} \dd t,
\eeq
with  $\Xi(t)=\frac{1}{3} t^3-\eta t+A$ the cubic corresponding to the Airy functions.  Note that $\eta, ~A$ are functions of $\Lambda $, and thus $\xi$, that are determined in \cref{appendix_approx_contours}  and $\tilde{C}$ is one of the Airy contours.

The largest contributions to the transformed contour integral arise from the neighbourhood of the coalescing saddles, where $f(t)=\frac{\dd s}{\dd t}$ can be explicitly identified. One can show (see \cref{appendix_approx_contours} for details) that the contour integral representation of the solution near the turning point is the real or imaginary part of 
\begin{equation}
  \label{eq:PacrossTurningPoint}
  p(\xi) \sim 2 \pi i e^{\frac{1}{\eps^*} A} \left[(\eps^*)^{1/3} \Ai\left(\eta  /(\eps^*)^{2/3}\right) M_- - (\eps^*)^{2/3} \Ai'\left(\eta  /(\eps^*)^{2/3}\right) N_-\right],
\end{equation}
with $A, \eta, M_-, N_-$ being functions of $\Lambda(\xi)$ given in \cref{appendix_approx_contours}.

To explicitly see the behaviour across the turning point, we Taylor expand and obtain a continuous function of the form
\begin{multline}
  \label{eq:finalPnearTP}
  p\sim \frac{\pi \rho^{1/3}}{6^{2/3} \Gamma(2/3)} \left[2\cos\left(\frac{8 }{15 \rho}\right) + \rho^{1/2} \left(\Theta(y) \sin\left(\frac{8 }{15 \rho}\right) \right.\right. \\ \left.\left. -\Theta(-y) \cos\left(\frac{8 }{15 \rho}\right)\right) \sqrt{|y|} + 2  \sin\left(\frac{8 }{15 \rho}\right) y\right],
\end{multline}
where $\Theta$ stands for the Heaviside step function. We verified this choice of the root on several random parameter sets and it always led to a visually correct approximation of the behaviour near the turning point (see  \cref{fig:VerifyingVia2cases} for two examples). Note that the solution is continuous but has a  discontinuity in the first derivative (due to the Heaviside step function). Thus, the WKB solution, and \cref{eq:p++4LambdaPosit}-\cref{eq:p+-4LambdaNeg}, show excellent agreement with numerical integration up to a neighborhood of the turning point, where we have a linear approximation \cref{eq:finalPnearTP}.

\begin{figure}
\centering
\begin{subfigure}{.495\textwidth}
   \centering
   \includegraphics[width=.95\linewidth]{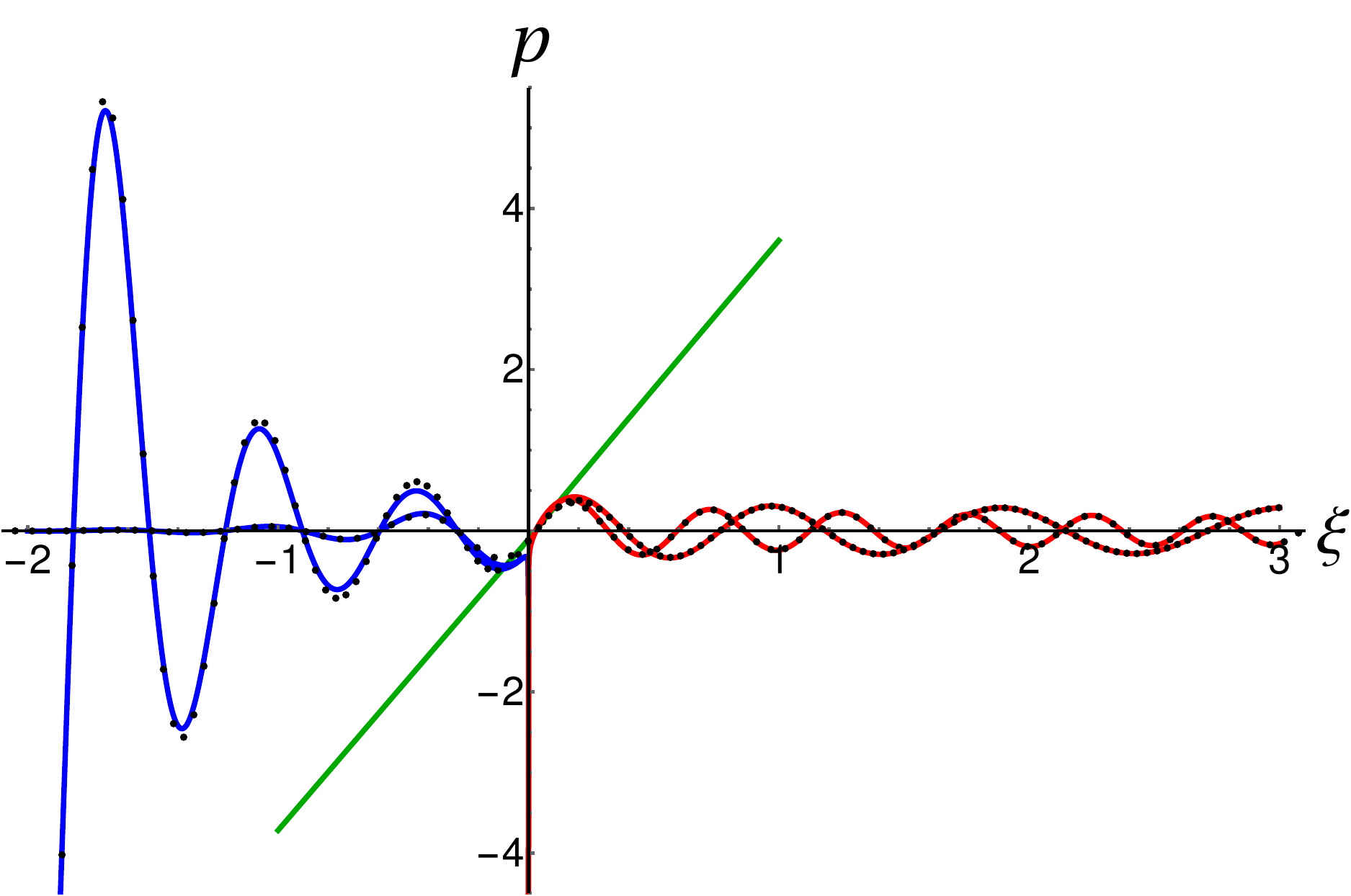}
\caption{$\rho =1.5^4$,  $\eps=0.1$}
\end{subfigure}
\begin{subfigure}{.495\textwidth}
   \centering
   \includegraphics[width=.95\linewidth]{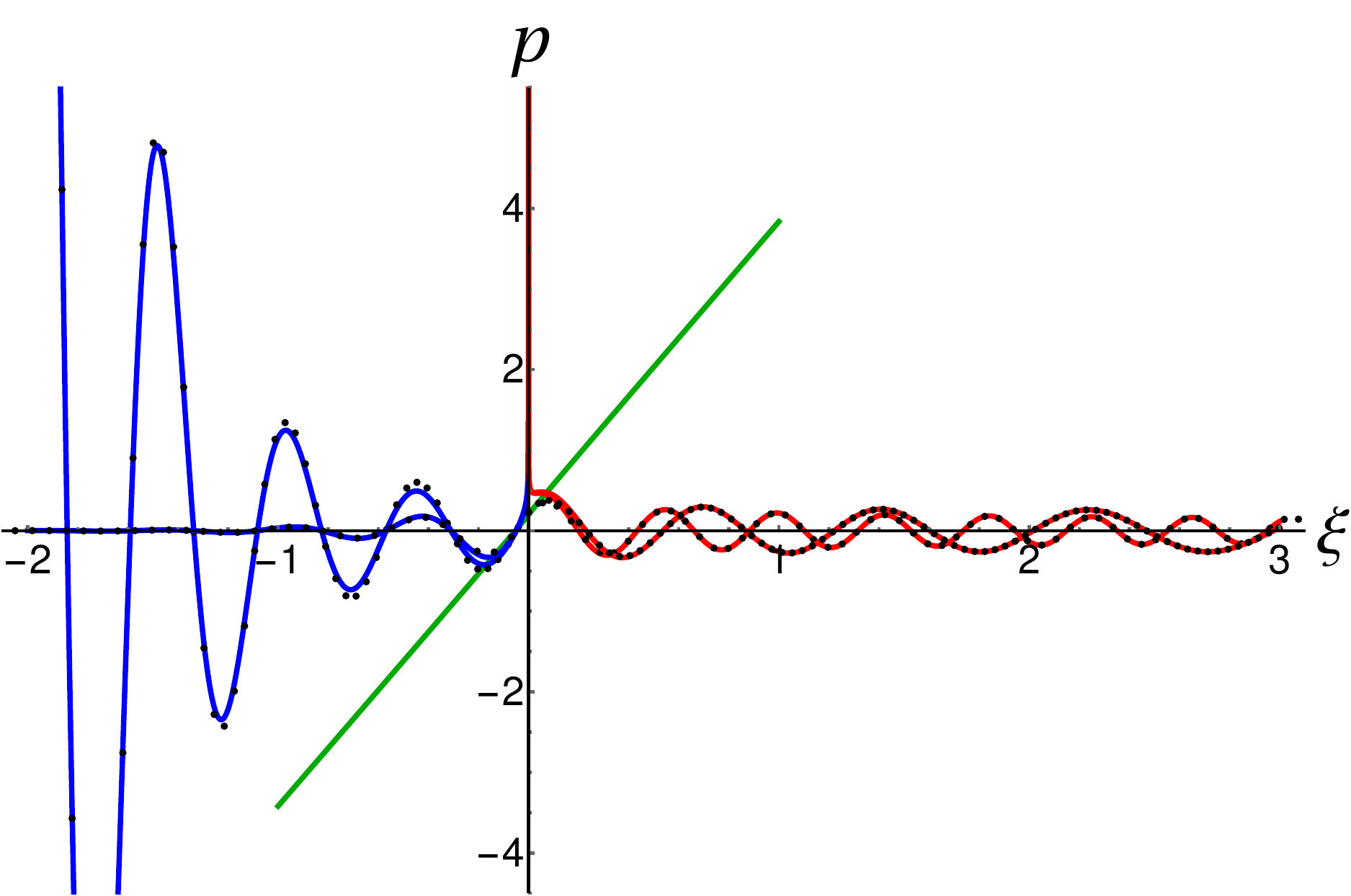}
\caption{$\rho =1.5^4$,  $\eps=0.12$}
\end{subfigure}

\begin{subfigure}{.495\textwidth}
   \centering
   \includegraphics[width=.95\linewidth]{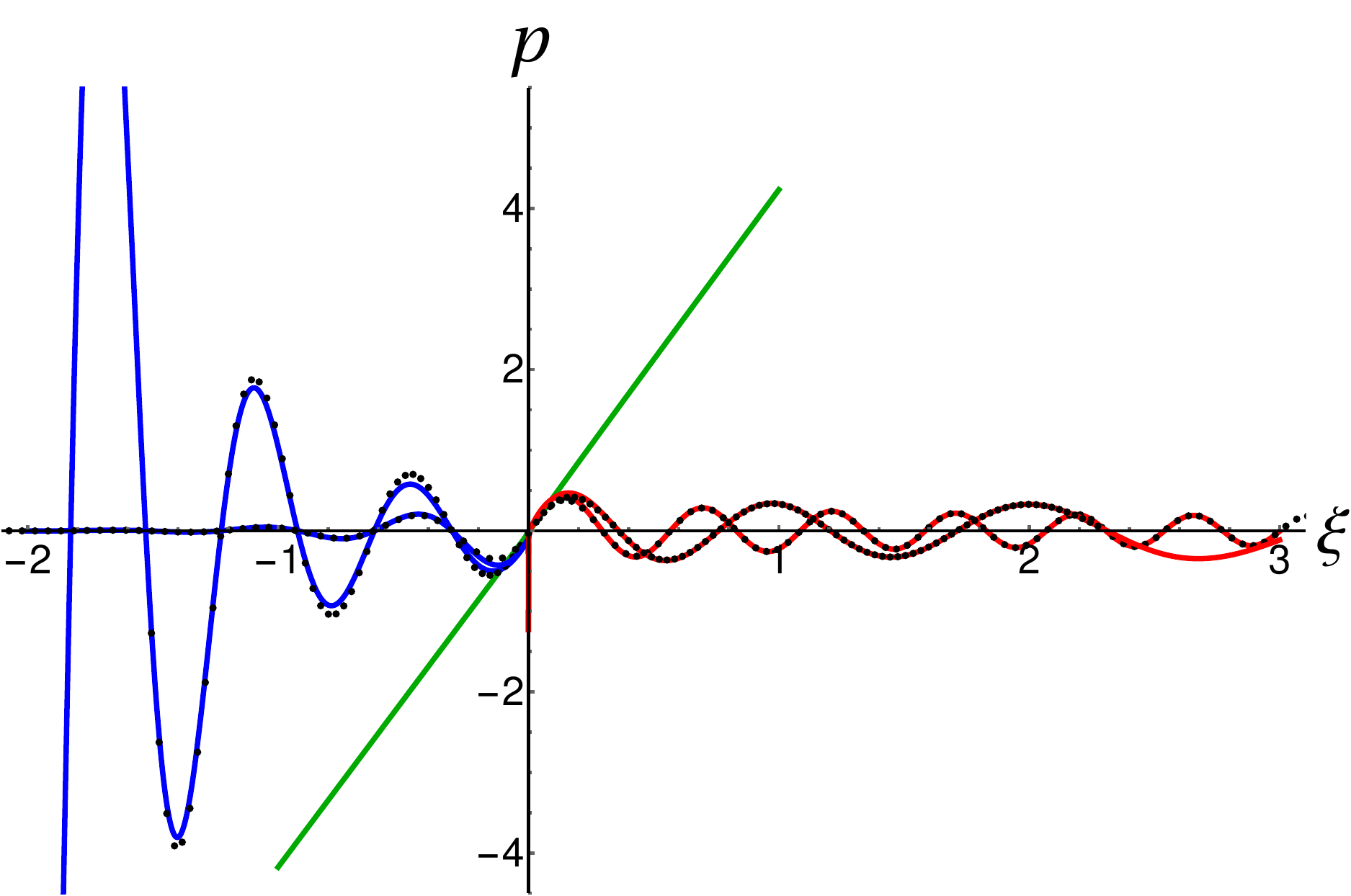}
\caption{$\rho =1.4^4$,  $\eps=0.1$}
\end{subfigure}
\begin{subfigure}{.495\textwidth}
   \centering
   \includegraphics[width=.95\linewidth]{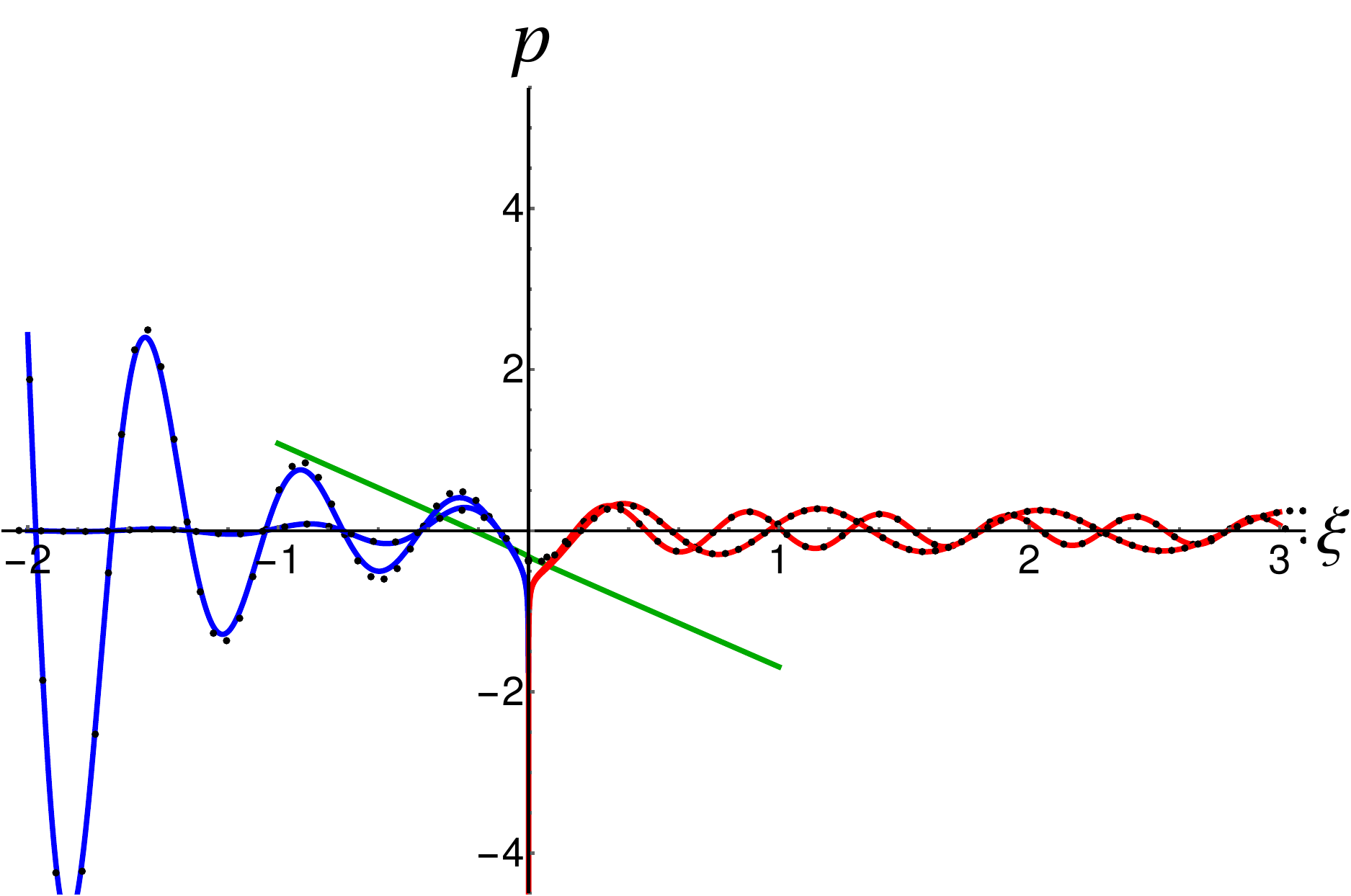}
\caption{$\rho =1.6^4$,  $\eps=0.1$}
\end{subfigure}
    \caption{\label{fig:VerifyingVia2cases} {Solid lines represent the approximate solutions: red for $\Lambda>0$ in\cref{eq:p++4LambdaPosit},\eqref{eq:p+-4LambdaPosit}, blue for $\Lambda<0$ in\cref{eq:p++4LambdaNeg},\eqref{eq:p+-4LambdaNeg}, green for the approximation across the turning point in \cref{eq:finalPnearTP}. Each dot is the numerically calculated contour integral, \cref{eq:contourInt} along the SDC {$C_s$} for a given $\xi$. Note that the approximation away from the turning point estimates the solution {well. However, it diverges at the turning point but in this region }the final approximation captures the transition from an exponentially growing to an exponentially decreasing solution. These plots also serve as a verification of the final form of expression, \cref{eq:finalPnearTP}, that is the chosen complex root of $(-1)^6$, in various situations.}}
\end{figure}

Note that this knowledge of the solution behaviour reveals that the envelope is $\Ord{1}$ near the turning point, and the leading order behaviour is actually a rescaled Airy function as follows from  \cref{eq:PacrossTurningPoint} {and \cref{Eq.A16}},  
\beq\label{eq:final_form_boundary_layer}
  p \sim \underbrace{2 \pi i e^{i\frac{1}{\eps^*} \frac{8}{5}} (\eps^*)^{1/3} {M_-(0)}}_{\mbox{constant}} \Ai\left((-i)^{3/2}  (\eps^*)^{2/3} \Lambda(\xi) \right), {\mbox{ as } \xi\to 0}.
\eeq
Hence, the decay rates of the pattern tails correspond to the envelope behaviour of the Airy function matching those identified above in the outer WKB solution, as in Eqs.~\cref{eq:p++4LambdaPosit}-\cref{eq:p+-4LambdaNeg}.

\section{Simulations of Heterogeneous Pattern Localisation}\label{sec_Numerics}

\begin{figure}
    \centering
    \begin{subfigure}{.495\textwidth}\includegraphics[width=1\textwidth]{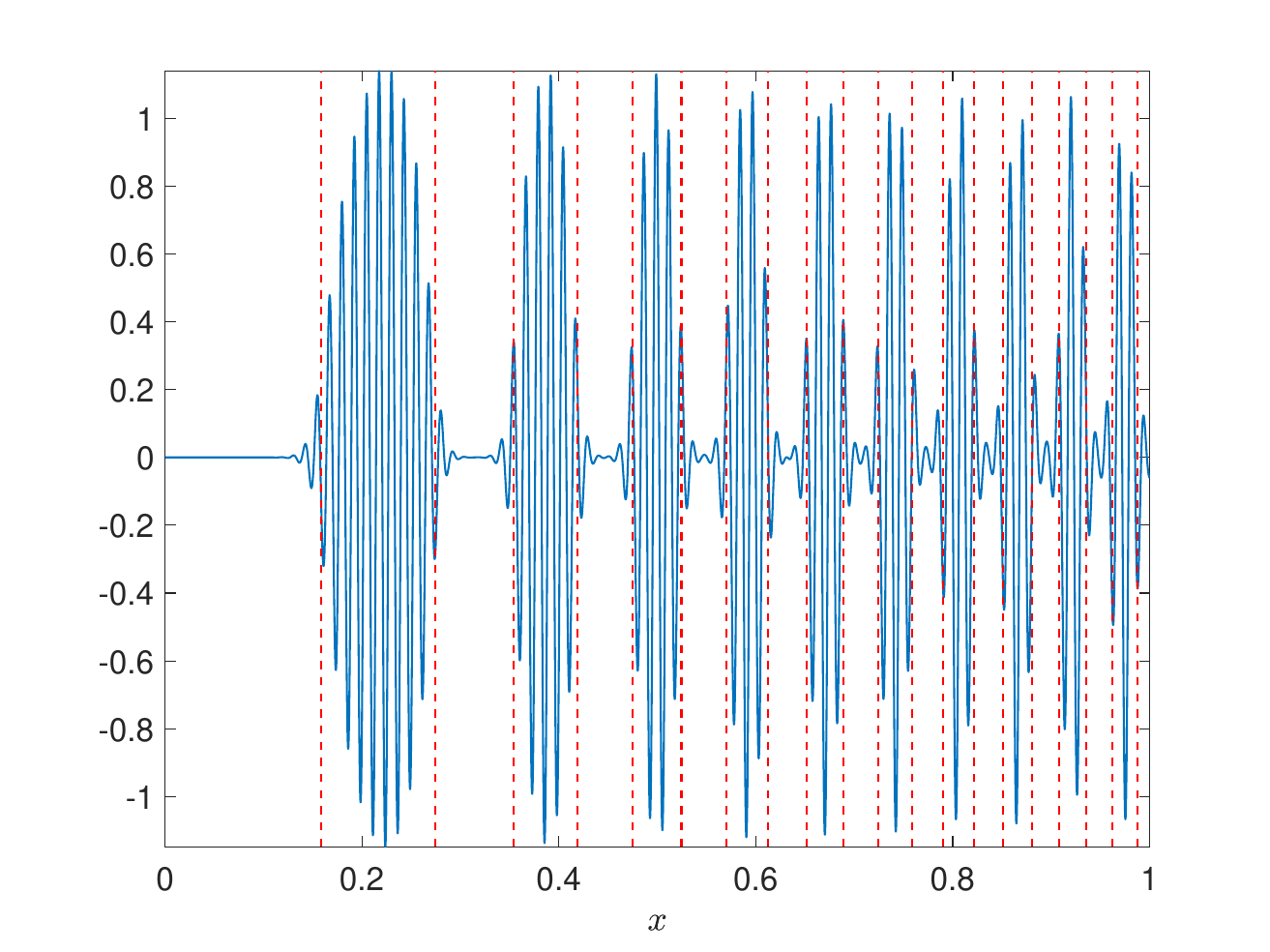}
    \caption{$N(u) = -u^3$, $\eps = 0.002$}
    \end{subfigure}
    \begin{subfigure}{.495\textwidth}\includegraphics[width=1\textwidth]{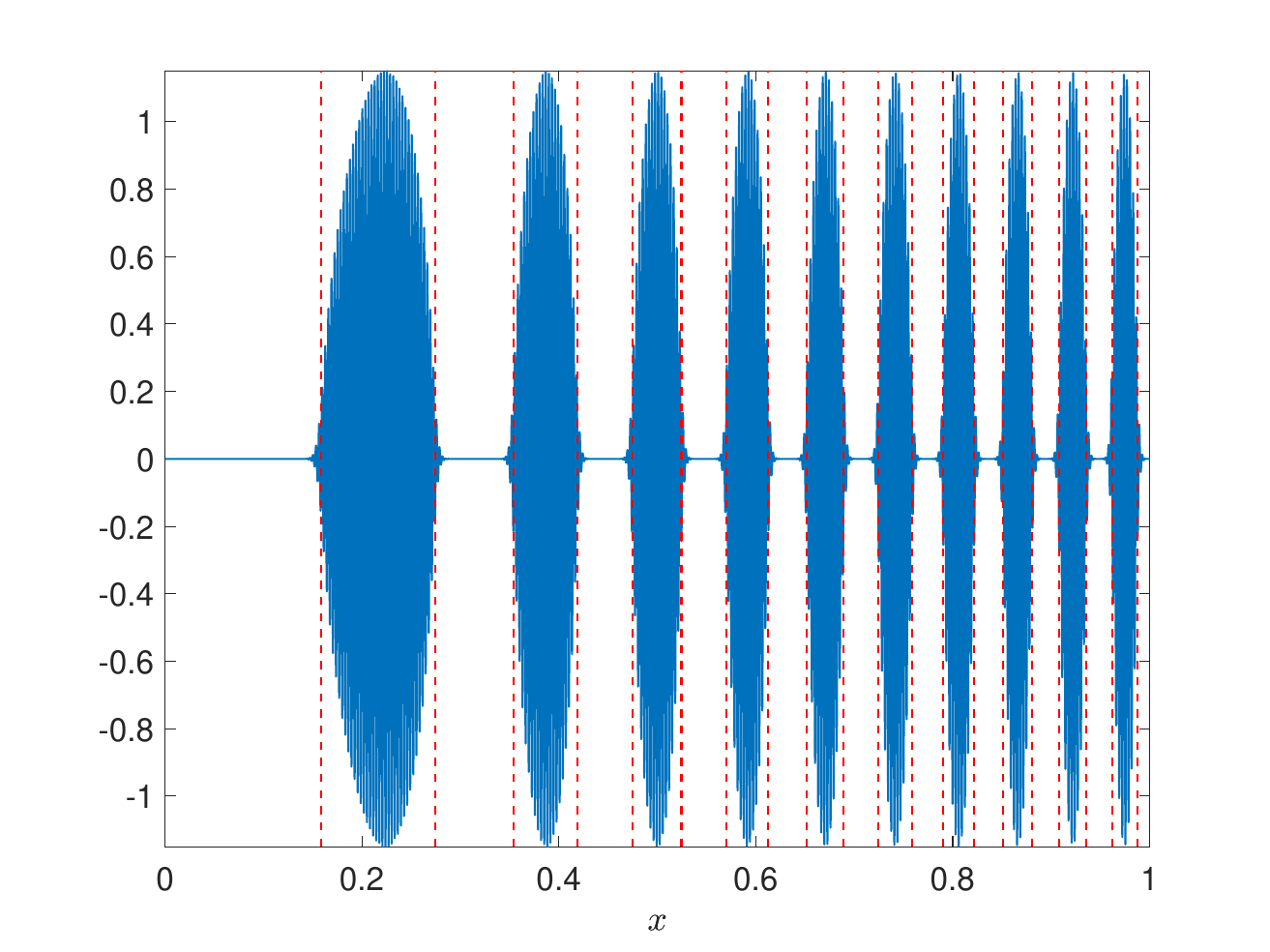}
    \caption{$N(u) = -u^3$, $\eps = 0.0005$}
\end{subfigure}
\begin{subfigure}{.495\textwidth}\includegraphics[width=1\textwidth]{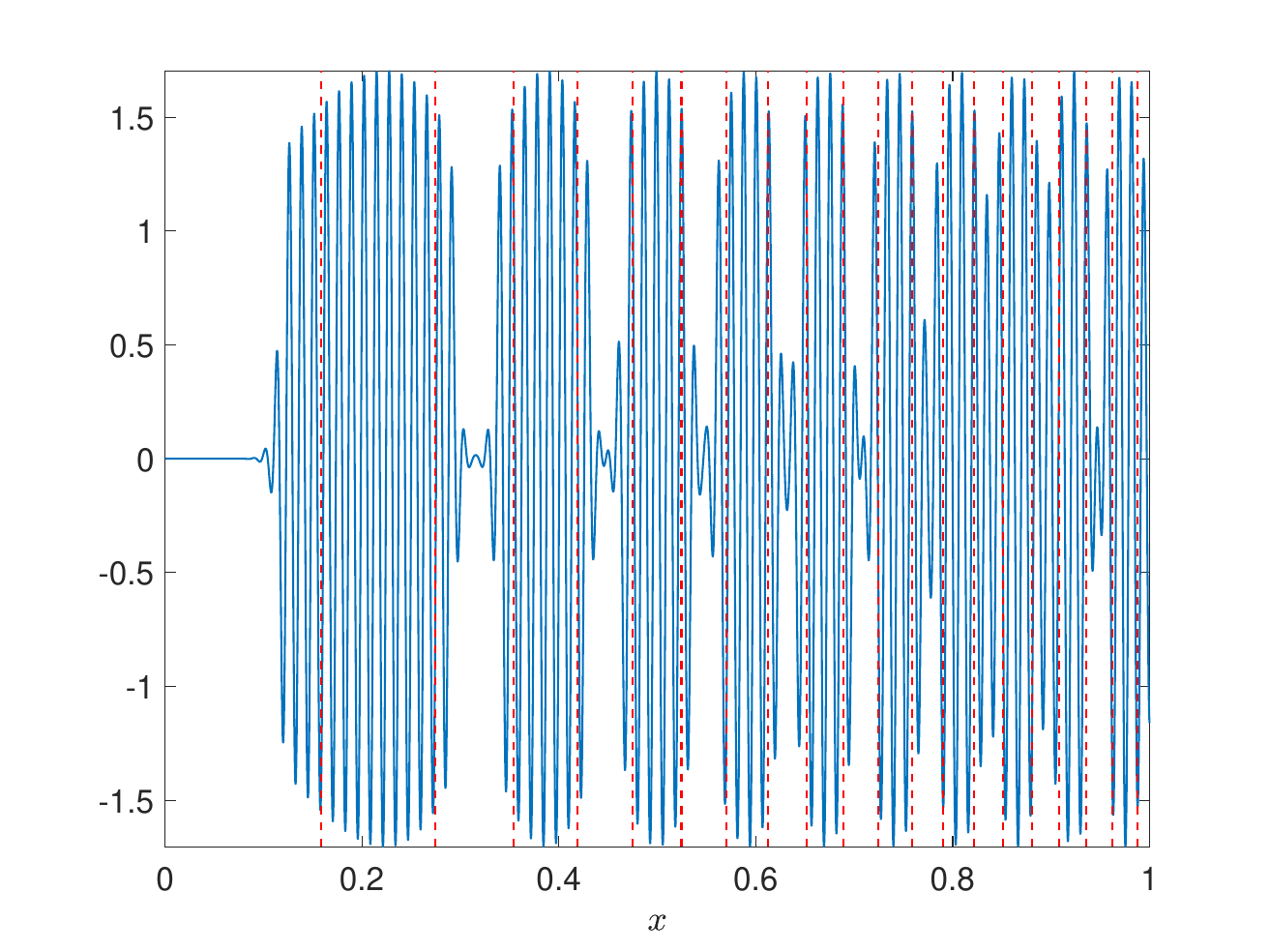}
    \caption{$N(u) = 2u^3-u^5$, $\eps = 0.002$}
    \end{subfigure}
    \begin{subfigure}{.495\textwidth}\includegraphics[width=1\textwidth]{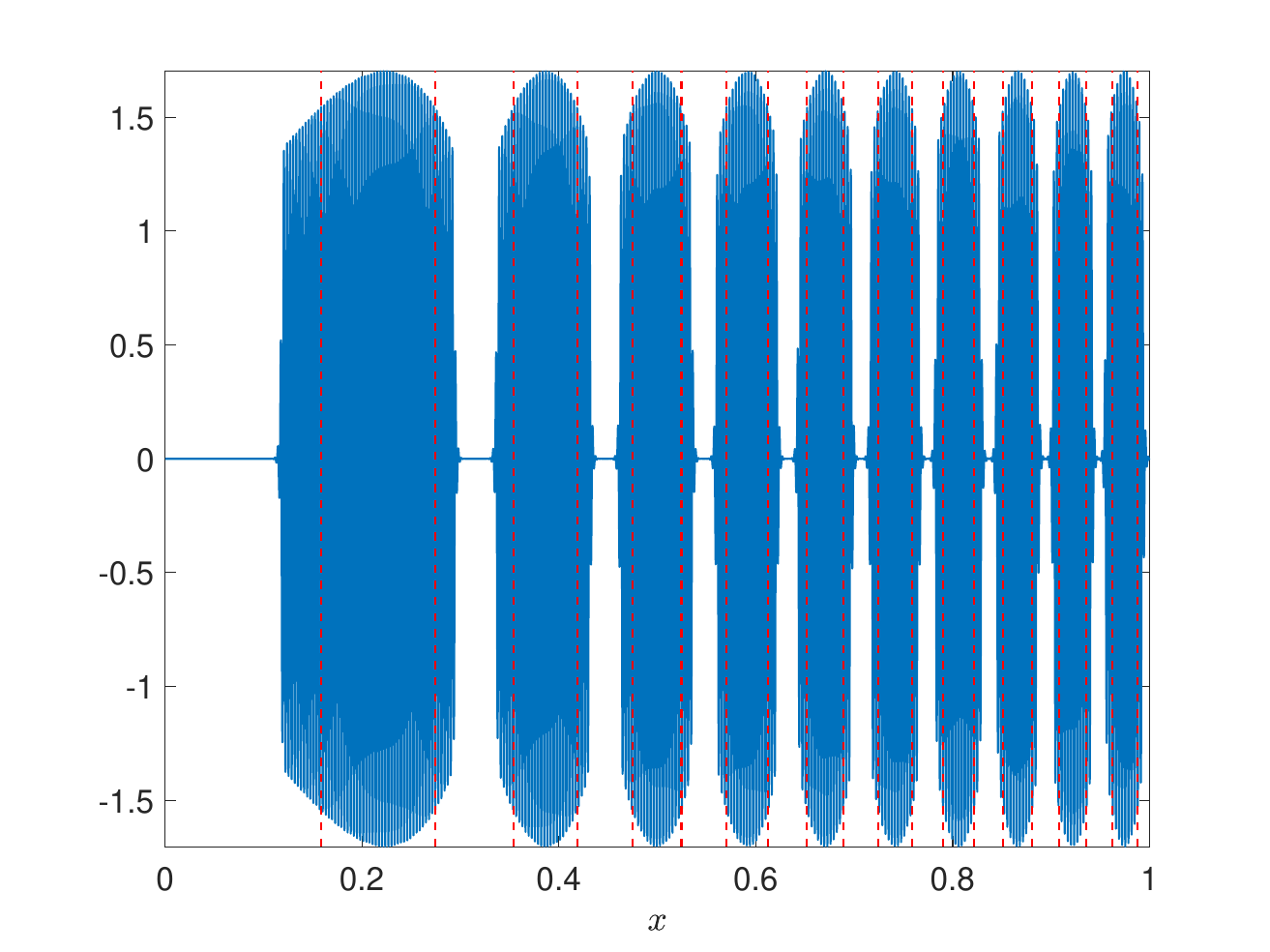}
    \caption{$N(u) = 2u^3-u^5$, $\eps = 0.0005$}
\end{subfigure}
    \caption{Solutions $u(x)$  of \cref{main_sh0} (blue solid curves) with $r(x) = -\cos(20\pi x^2)$ and varying nonlinearity $N(u)$ and $\eps$. Panels (a) and (b) are the cubic case corresponding to a locally supercritical nonlinearity. {In panel (a)  additional, evanescent, WKB solutions cause  a bleed of the patterning beyond the vertical red dash lines, where $r(x)=0$ can clearly be seen for large $x$, with the effective removal  of such bleeding for the reduced  $\varepsilon$ of panel (b).  Panels (c) and (d) in contrast, present numerical observations  for  a} cubic-quintic nonlinearity, corresponding to a locally subcritical instability. The dashed red vertical lines indicate where $r(x)=0$, and hence where a naive theory would predict patterning confinement. Simulation details can be found in \cref{appendix_numerics}.}\label{fig:super_sub_comparisons}
\end{figure}

Here we show how a notion of `local criticality' can impact the extent of patterning, and influence the tails of regions exhibiting confined patterns. We numerically simulated a large set of choices of the nonlinearity $N(u)$, focusing on polynomials up to seventh degree, a variety of trigonometric and more complex kinds of heterogeneity $r(x)$, as well as how the resulting solutions behave as $\eps$ is varied. We refer to \cref{appendix_numerics} for details of our numerical methods, as well as for details of an implementation of the model using a rapid interactive web simulator \cite{walker2023visualpde} that can be found at \href{https://visualpde.com/sim/?preset=Heterogeneous-Swift-Hohenberg}{this simulation link}\footnote{\url{https://visualpde.com/sim/?preset=Heterogeneous-Swift-Hohenberg}}. Below we present a small subset of these simulations to illustrate what we have learned, organized by the qualitative types of behaviour observed.

\begin{figure}
    \centering
    \begin{subfigure}{.495\textwidth}\includegraphics[width=1\textwidth]{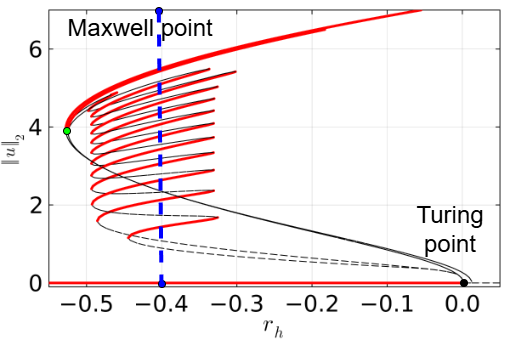}\caption{}
    \end{subfigure}
    \begin{subfigure}{.495\textwidth}\includegraphics[width=1\textwidth]{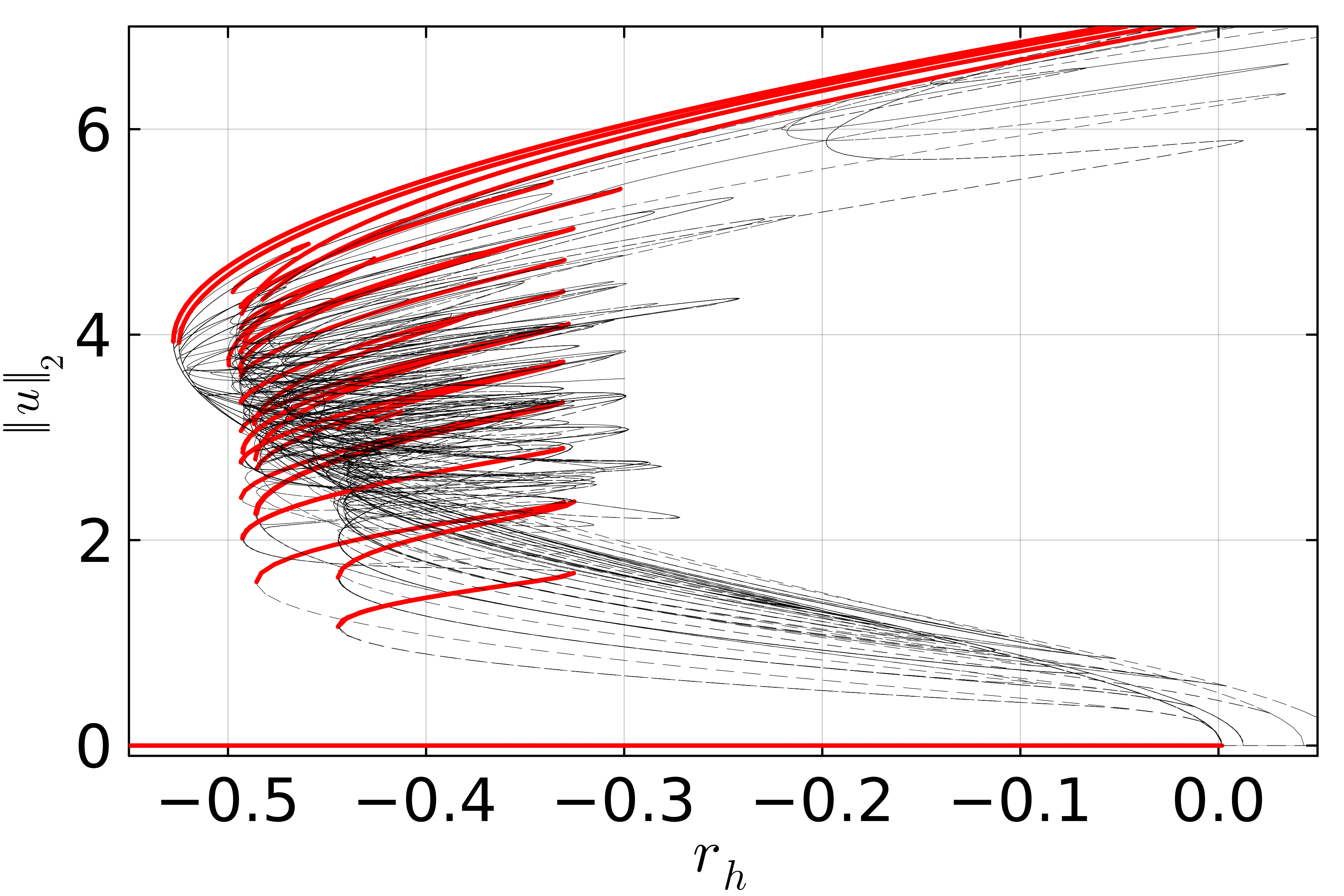}\caption{}
\end{subfigure}
    \caption{ {Bifurcation diagrams corresponding to steady state solutions of \cref{homog_sh} as $r$ is varied for $N(u)=2u^2-u^3$ and $\eps = 0.025$. Panel (a) shows the main branches emanating from the subcritical instability, leading to a stable branch after a secondary fold bifurcation as well as the two main snaking branches which emerge from this unstable state, whereas panel (b) shows all connected primary, secondary, and ternary solution branches connected to this region that originate up to $r=0.1$. Stability is indicated with red curves as stable and black dashed curves as unstable. We also indicate the approximate Maxwell point using a blue dashed vertical line,  with localised solution branches `snaking' around this point. Details about the continuation algorithm using BifurcationKit \cite{veltz:hal-02902346} can be found in \cref{appendix_numerics}}.}\label{fig:bifurcation_diagrams}
\end{figure}

We first show in panels (a) and (b) of \cref{fig:super_sub_comparisons} that the linear theory developed predicts patterning regions even for complicated spatial heterogeneity $r(x)$ (in contrast to the simple heterogeneity of \cref{fig:homog_het_comparison}), as long as the nonlinearity leads to a locally supercritical bifurcation. 

{
As expected, for sufficiently small $\eps$, the pattern formation is confined approximately to regions where $r(x)>0$. However, when $\eps$ is not sufficiently small, as in panel (a) for larger $x$ in particular, there are decaying tails that cause a bleed of patterning from these regions. 
A condition for the effective absence of such bleeding
for the heterogeneity of \cref{fig:super_sub_comparisons} 
is determined in \cref{appc} via \cref{c4}, giving   
\begin{equation} \label{fineqn} \eps \ll \frac{\pi^2} {32r'} \approx 0.0025 
\end{equation}
where $r'$ denotes $|r'(x)|$ at a root of $r(x)=0$ and the numerical estimate is for  $x\approx 1$. 
This is violated  for panel (a), but holds  for panel (b), consistent with the numerical observation that there is effectively pattern separation for panel (b) but pattern bleeding   for larger $x$ within panel (a). While the derivation of \cref{fineqn} takes advantage of the form of $r(x)$ in \cref{fig:super_sub_comparisons}, it nonetheless demonstrates that once the rate of change in the heterogeneity becomes sufficiently large for fixed $\eps$, so as to induce poorly separated roots of $r(x)=0$,  the pattern separation predicted by the simple treatment of WKB functions above fails due to pattern bleed.}


\bk


In panels (c) and (d) of \cref{fig:super_sub_comparisons}, we change the nonlinearity such that the corresponding spatially homogeneous model exhibits a subcritical instability for $r=0$. In this case, the confinement is no longer predicted well by the linear theory, even for small $\eps$. We also observe that the tails confining the patterning regions appear more rapid than in the {locally} supercritical case, as one might expect from a larger amplitude solution rapidly losing stability at $x$ changes.

To understand this, we make use of the bifurcation structure of the homogeneous problem \cref{homog_sh} in the vicinity of a subcritical Turing bifurcation. We numerically continue a solution, using the same nonlinearity $N(u) = 2u^3-u^5$ as in \cref{fig:super_sub_comparisons}(c)-(d), and plot the resulting branched structure in \cref{fig:bifurcation_diagrams}. {Panel (b)} shows that there are an enormous number of branches even in the homogeneous case, though we will be most interested in the main/topmost (red) stable branch depicted in panel both panels, as this branch will correspond to domain-filling Turing patterns. Denoting this equilibrium patterned solution with $u^*$, we can compute its energy $E\left(u^*|_r\right)$ using \cref{energy} as a function of $r$ along the branch. Note that $E\left(u^*|_{r=0}\right)>0$ in all of the subcritical bifurcations we explored, and that the energy of this branch decreases as $r$ decreases. We define the Maxwell point of this domain-filling pattern branch (denoted as $\tilde{r}$)  to be the point where the patterned solution and the trivial solution $u=0$ are equally energetically favorable (i.e.~$E\left(u^*|_{r=\tilde{r}}\right) = E(0)$). This point is shown in \cref{fig:bifurcation_diagrams}(a) as the dashed {blue} line, as it tends to organize branches of localised solutions \cite{beck2009snakes}. As we simulate solutions to the heterogeneous system \cref{main_sh0}, we locally compute the values of $r$ where a corresponding homogeneous problem undergoes the fold bifurcation of the patterned state (the {solid} green circle in \cref{fig:bifurcation_diagrams}(a)), as well as the corresponding Maxwell point. {In other words, we find the value of $x$ for which $r(x)=r_h$ is at the Maxwell point or the fold of the homogeneous system \eqref{homog_sh}, which is a `local' analogue of these structures in the heterogeneous context. While we include the snaking structure in panel (a), we remark that, in contrast to \cite{kao2014spatial}, our focus is on how the top periodic branch's properties (i.e.~the Maxwell and fold points) in the homogeneous model influence the system in the heterogeneous case, rather than on how the snaking structure is changed due to heterogeneity.}

\begin{figure}
    \centering
    \begin{subfigure}{.495\textwidth}\includegraphics[width=1\textwidth]{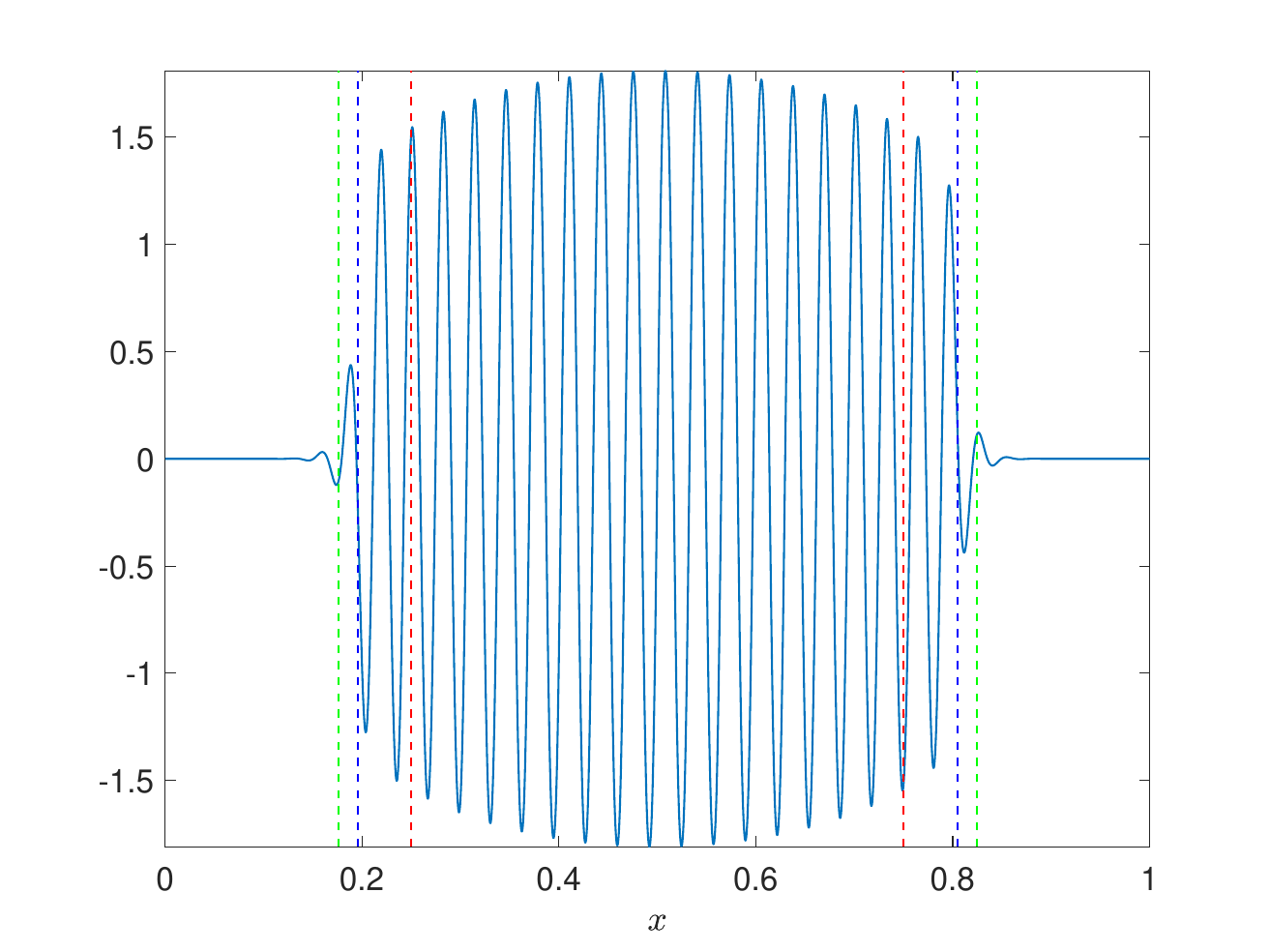}
    \caption{$r(x) = -2\cos(2\pi x)$, $\eps = 0.005$}
    \end{subfigure}
    \begin{subfigure}{.495\textwidth}\includegraphics[width=1\textwidth]{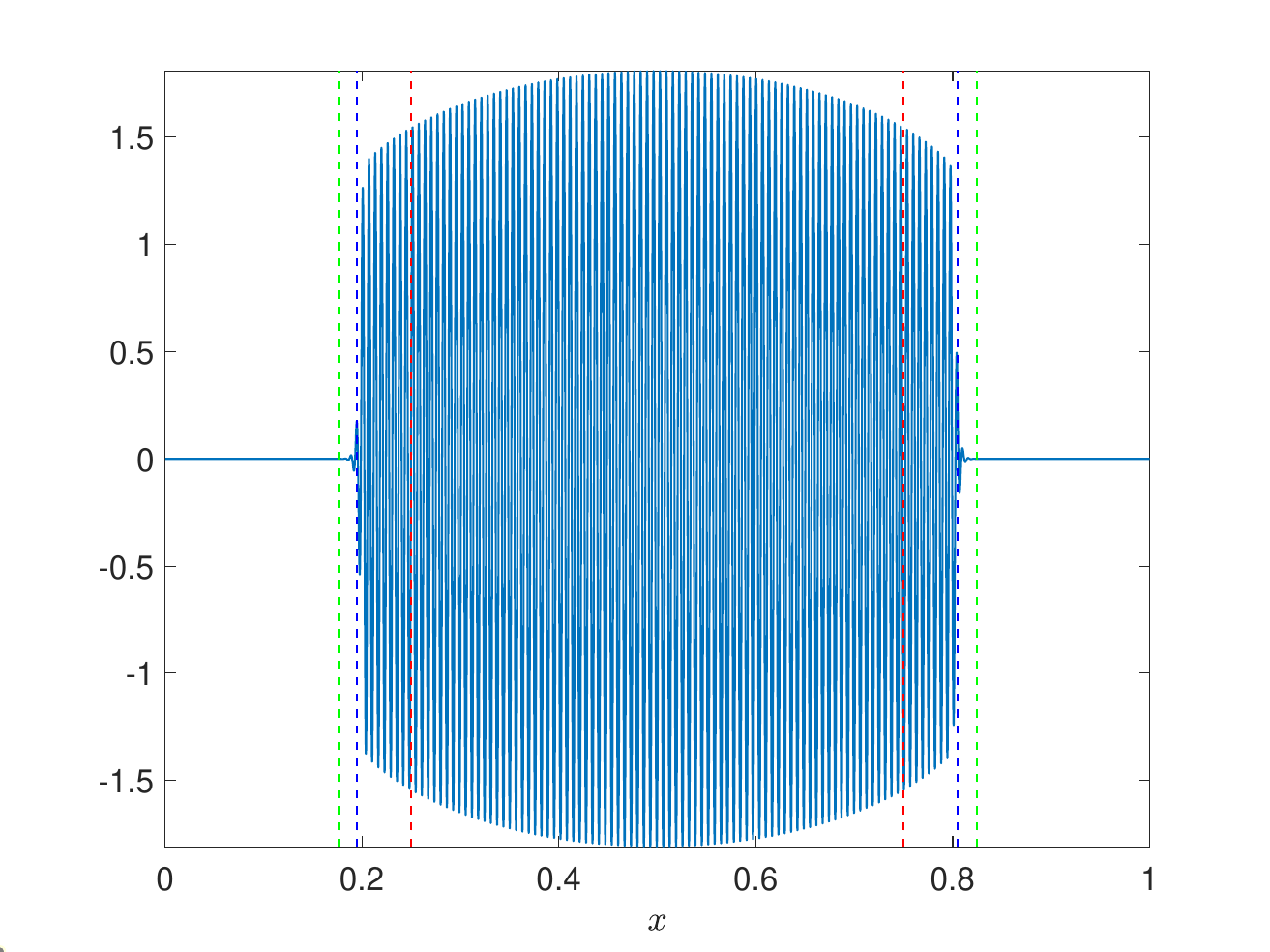}
    \caption{$r(x) = -2\cos(2\pi x)$, $\eps = 0.001$}
\end{subfigure}

\begin{subfigure}{.495\textwidth}\includegraphics[width=1\textwidth]{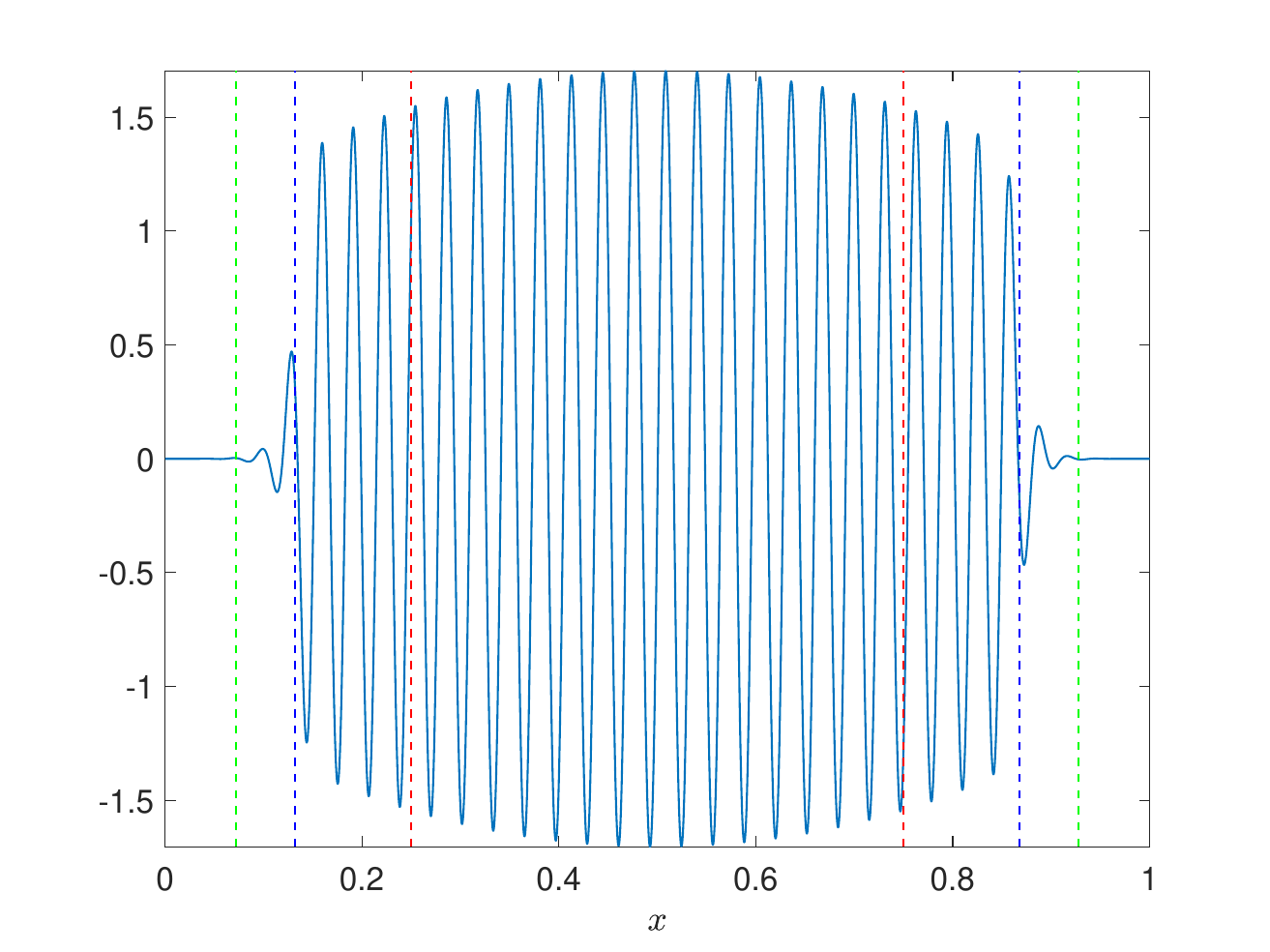}
    \caption{$r(x) = -\cos(2\pi x)$, $\eps = 0.005$}
    \end{subfigure}
    \begin{subfigure}{.495\textwidth}\includegraphics[width=1\textwidth]{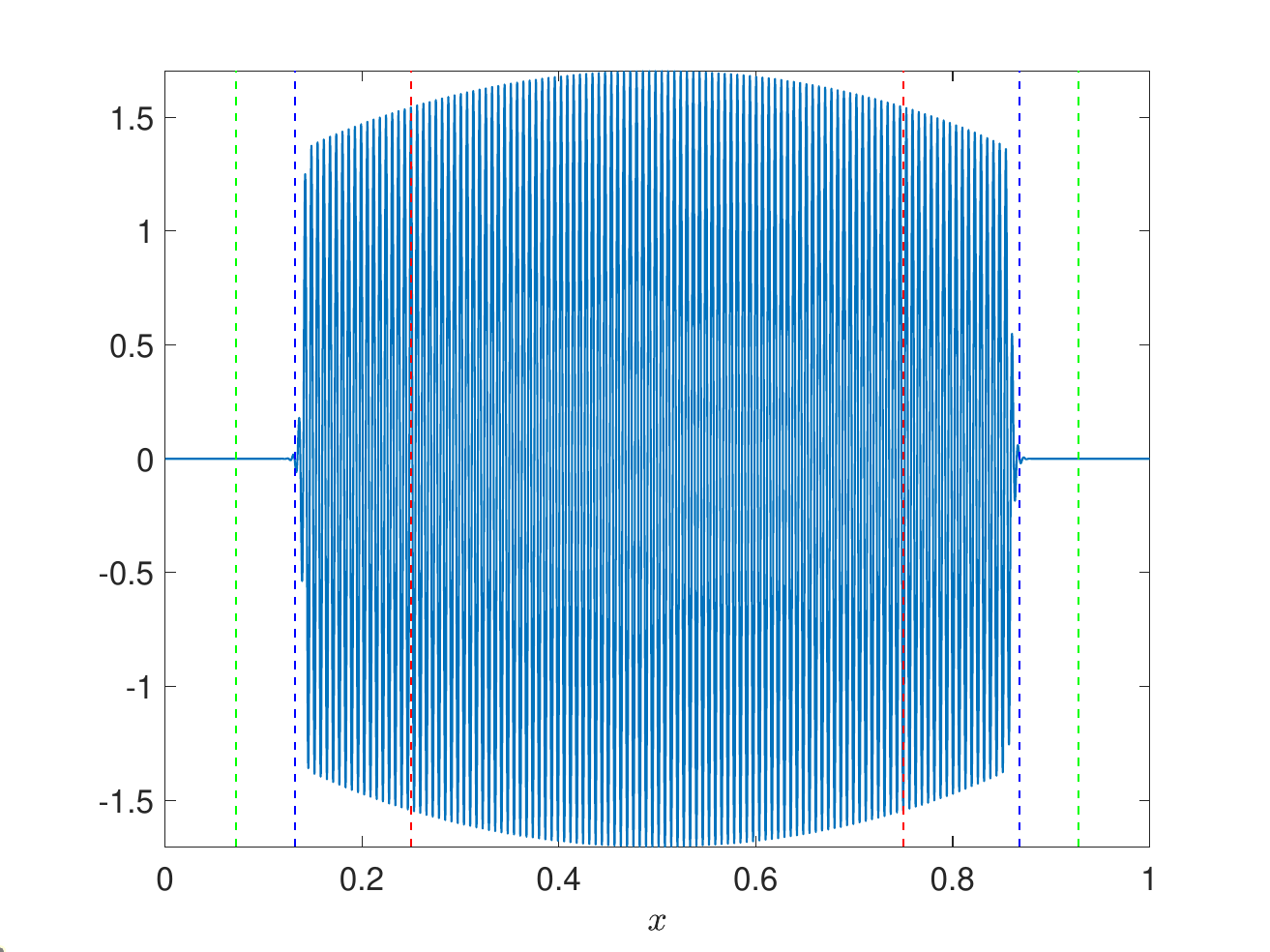}
    \caption{$r(x) = -\cos(2\pi x)$, $\eps = 0.001$}
\end{subfigure}

\begin{subfigure}{.495\textwidth}\includegraphics[width=1\textwidth]{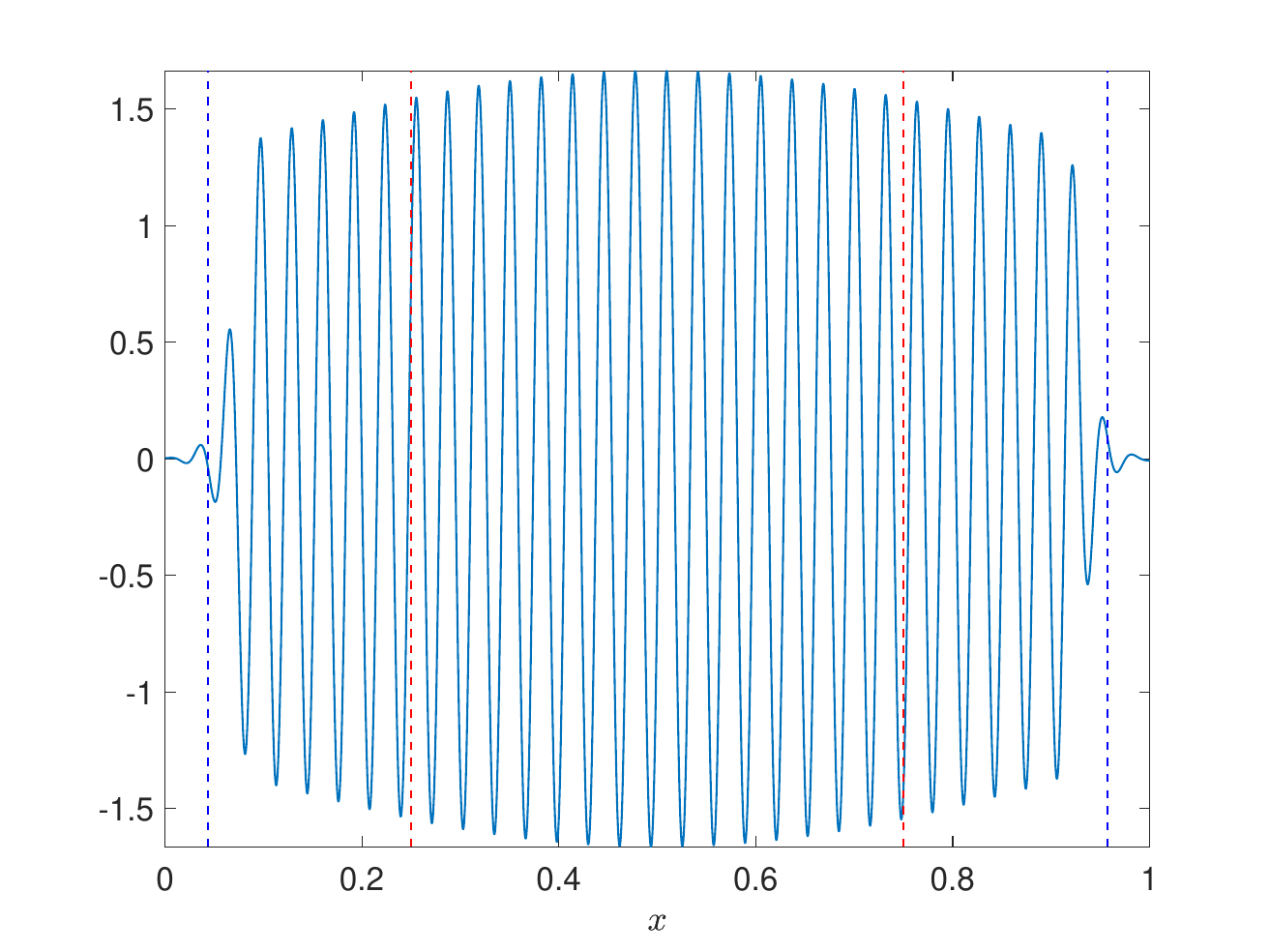}
    \caption{$r(x) = -0.7\cos(2\pi x)$, $\eps = 0.005$}
    \end{subfigure}
    \begin{subfigure}{.495\textwidth}\includegraphics[width=1\textwidth]{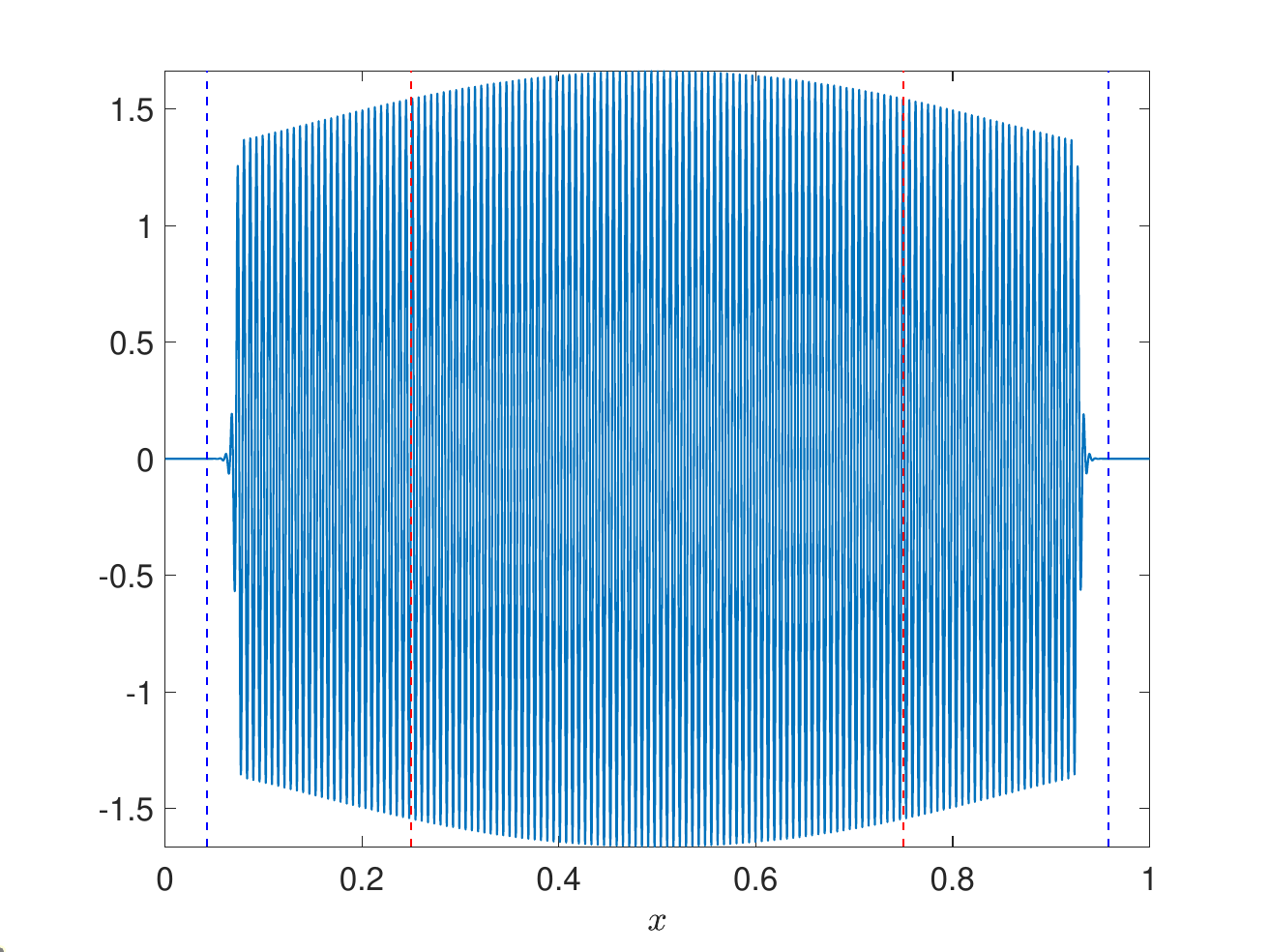}
    \caption{$r(x) = -0.7\cos(2\pi x)$, $\eps = 0.001$}
\end{subfigure}
    \caption{Solutions $u(x)$  of \cref{main_sh0} (blue solid curves) with $N(u) = 2u^3-u^5$ and varying heterogeneity $r(x)$ and $\eps$. The dashed red vertical lines indicate where $r(x)=0$, the dashed blue lines correspond to where $E(u^*)=0$, and the dashed green lines correspond to a local fold bifurcation as described in the text (which is now outside of the domain in panels (e) and (f)).  Simulation details can be found in \cref{appendix_numerics}. }\label{fig:sub_35}
\end{figure}

We first consider varying the amplitude of a simple cosine heterogeneity in the cubic-quintic case. We plot solutions in \cref{fig:sub_35}, using vertical red lines to denote local Turing conditions, vertical blue lines to denote local Maxwell points, and vertical green lines to denote local fold points. As before, we observe a sharp drop in pattern amplitude, particularly for smaller values of $\eps$ (cf \cref{fig:homog_het_comparison}). Roughly speaking, the Turing and fold points seem to fail to locate the region of confinement, whereas the Maxwell point approximates it well for most choices of $r(x)$ and $\eps$. We do observe noticeable disagreement in panel (f), and for many other nonlinearities we see different kinds of disagreements with the local theory proposed here, but nevertheless find some evidence that local Maxwell points can predict pattern confinement, particularly for odd-ordered nonlinearities with moderate amplitude heterogeneities. As one might expect, in the subcritical case the local linear stability theory presented in \cref{sec_Localisation} fails in all cases to account for the confinement of the pattern, though it is usually a subset of the patterning regions found in most cases (with some exceptions, as noted below).

\begin{figure}
    \centering
    \begin{subfigure}{.495\textwidth}\includegraphics[width=1\textwidth]{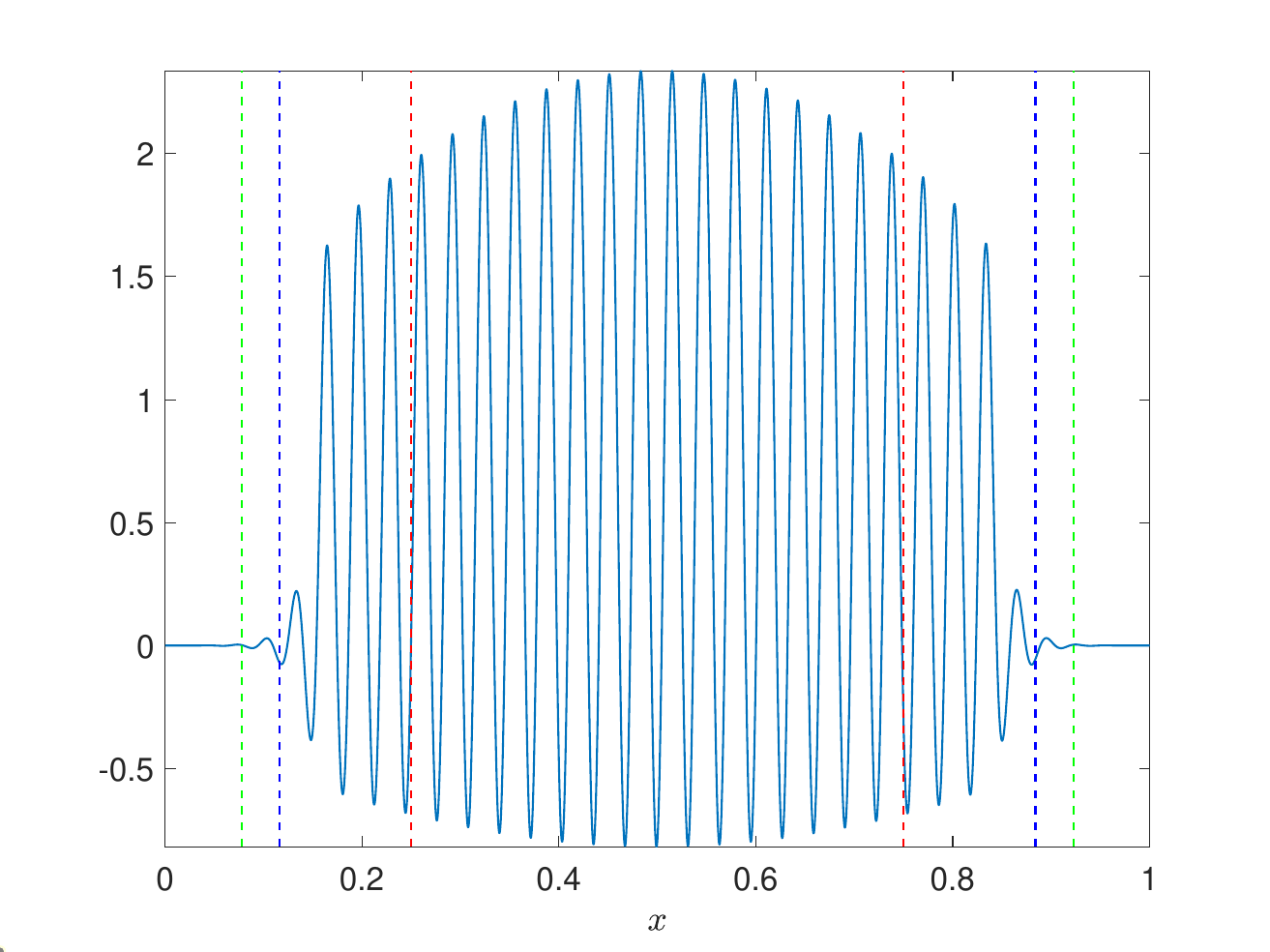}
    \caption{$r(x) = -0.6\cos(2\pi x)$, $\eps = 0.005$}
    \end{subfigure}
    \begin{subfigure}{.495\textwidth}\includegraphics[width=1\textwidth]{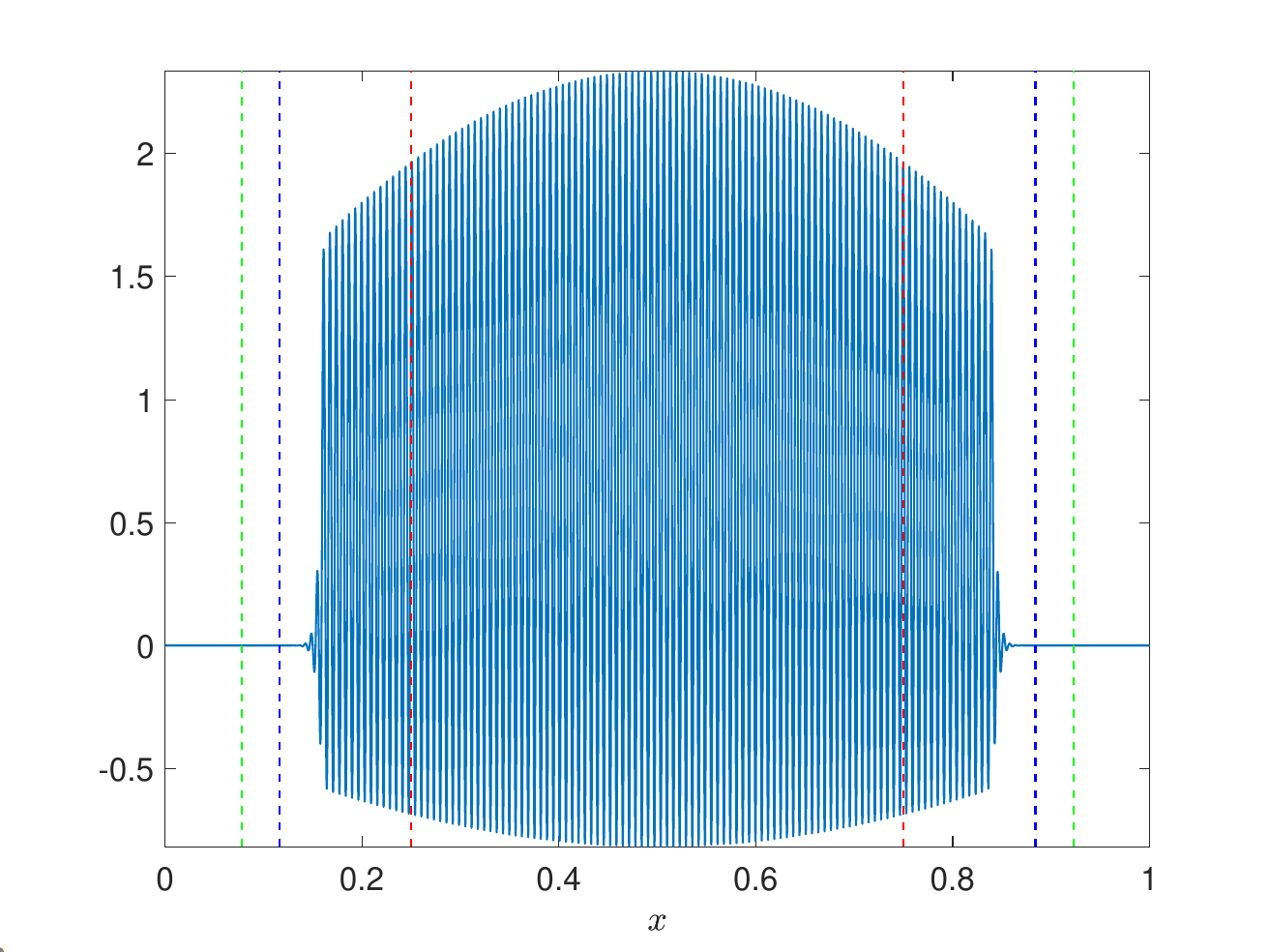}
    \caption{$r(x) = -0.6\cos(2\pi x)$, $\eps = 0.0005$}
\end{subfigure}

\begin{subfigure}{.495\textwidth}\includegraphics[width=1\textwidth]{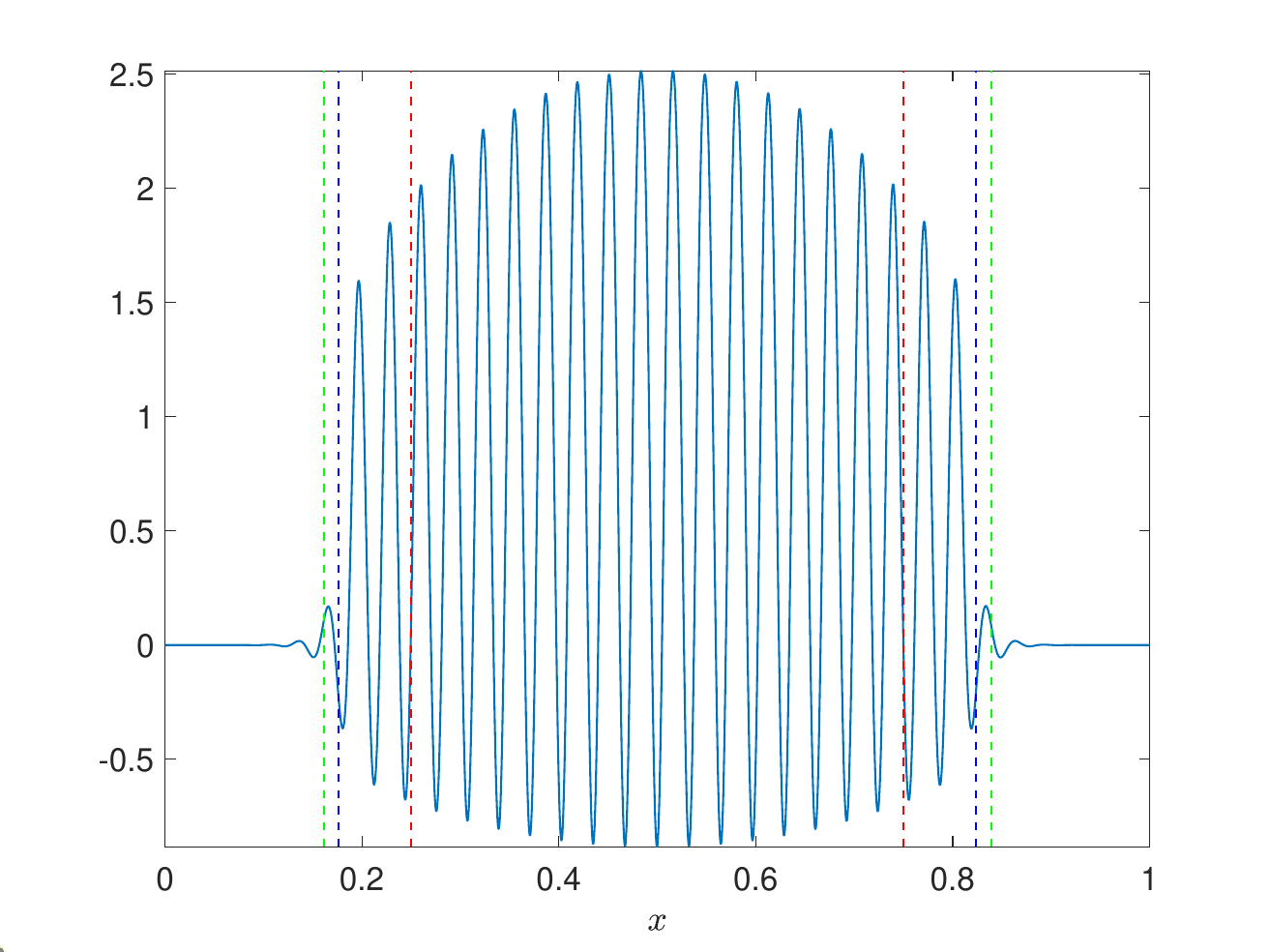}
    \caption{$r(x) = -\cos(2\pi x)$, $\eps = 0.005$}
    \end{subfigure}
    \begin{subfigure}{.495\textwidth}\includegraphics[width=1\textwidth]{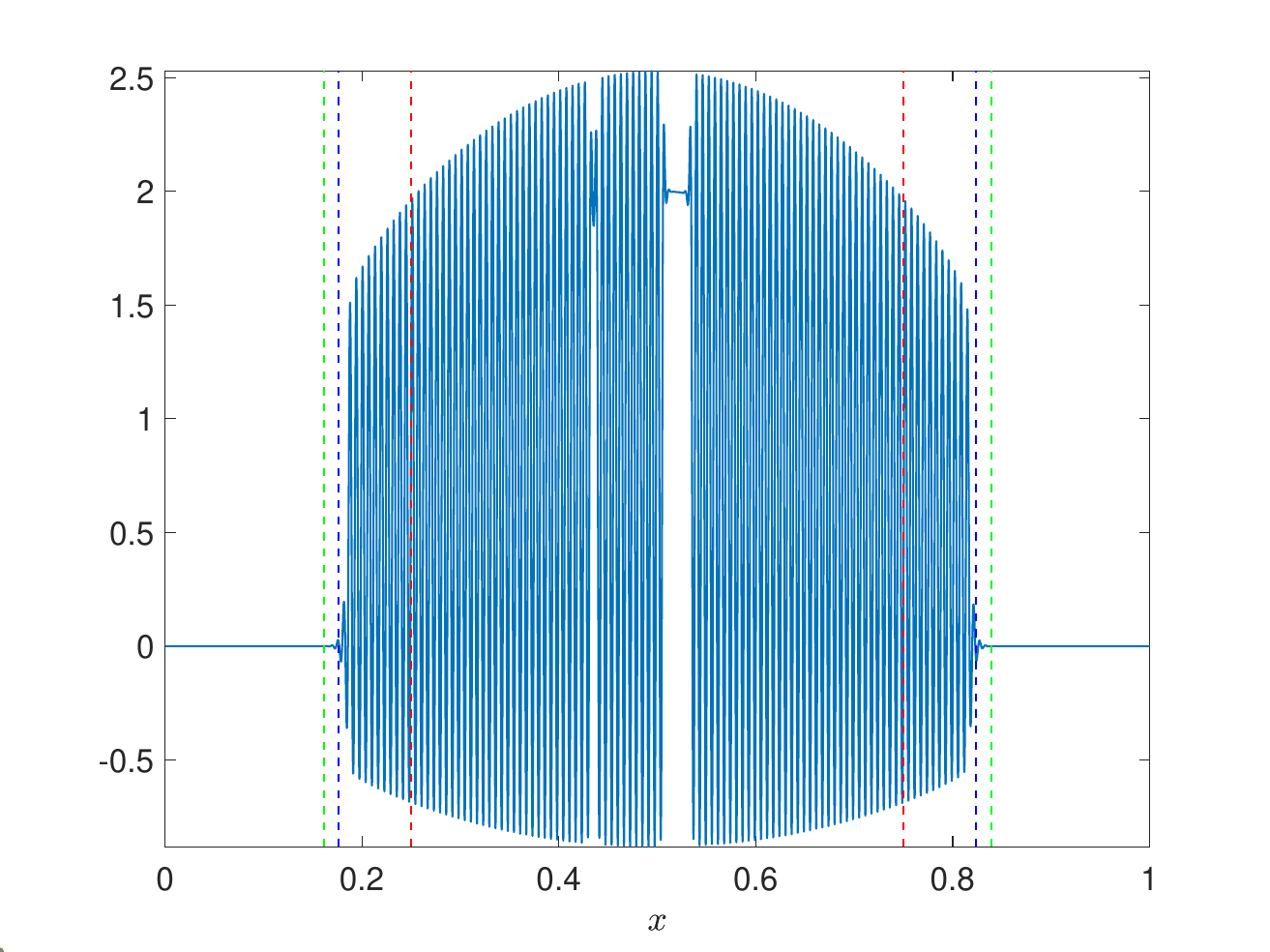}
    \caption{$r(x) = -\cos(2\pi x)$, $\eps = 0.0005$}
\end{subfigure}

\begin{subfigure}{.495\textwidth}\includegraphics[width=1\textwidth]{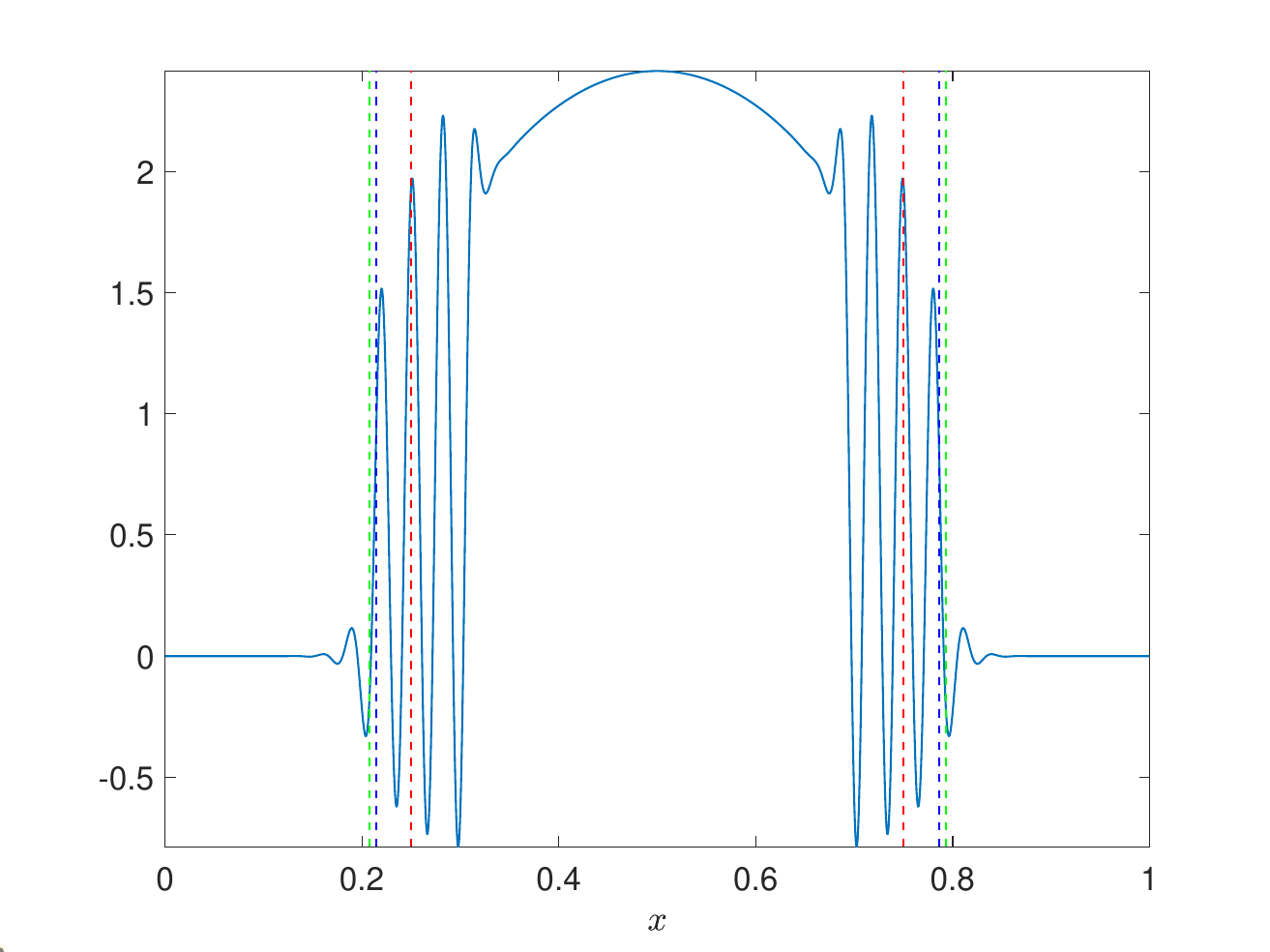}
    \caption{$r(x) = -2\cos(2\pi x)$, $\eps = 0.005$}
    \end{subfigure}
    \begin{subfigure}{.495\textwidth}\includegraphics[width=1\textwidth]{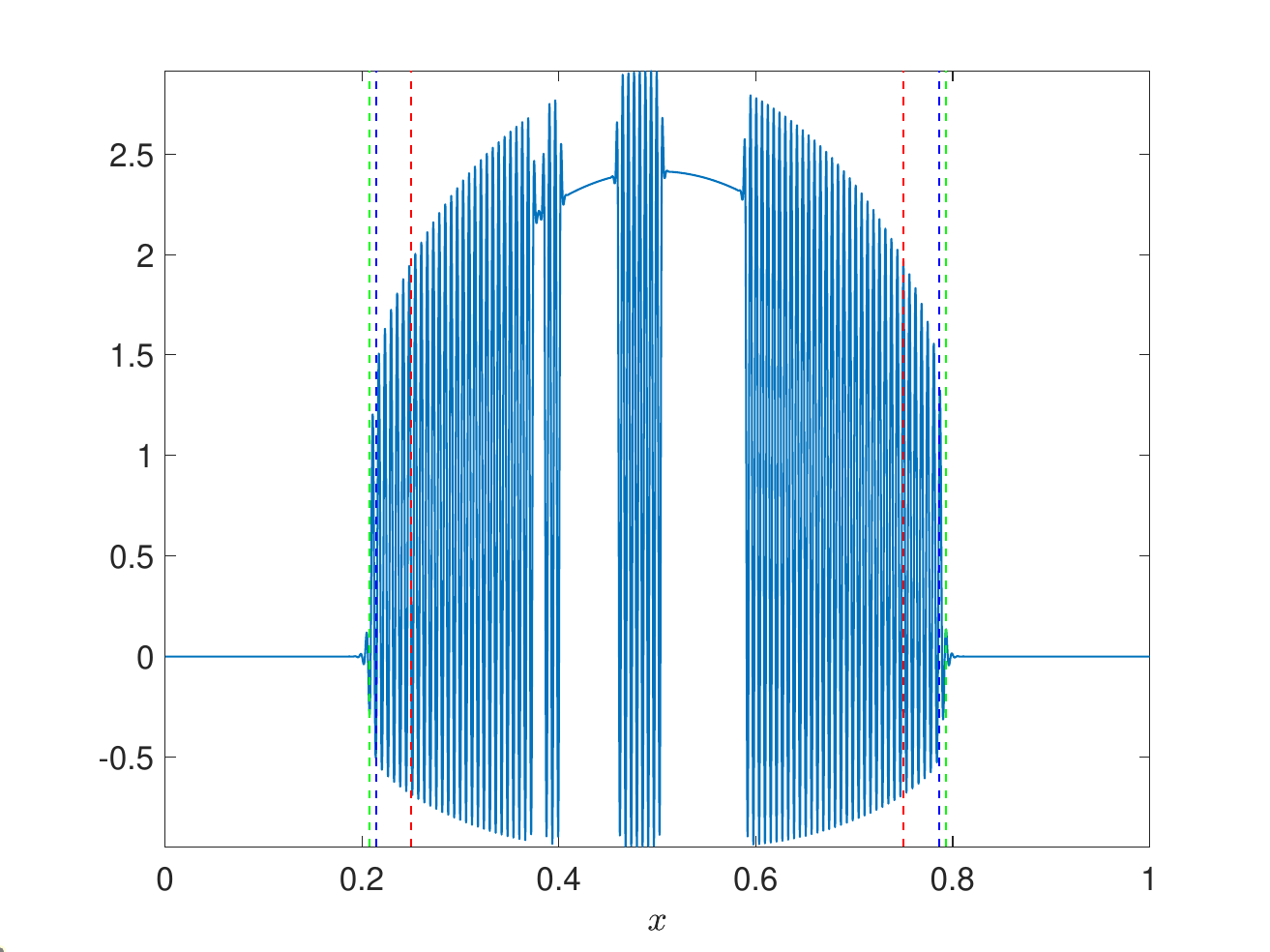}
    \caption{$r(x) = -2\cos(2\pi x)$, $\eps = 0.0005$}
\end{subfigure}
    \caption{Solutions $u(x)$  of \cref{main_sh0} (blue solid curves) with $N(u) = 2u^2-u^3$, varying $r(x)$ and $\eps$. The dashed red vertical lines indicate where $r(x)=0$, the dashed blue lines correspond to where $E(u^*)=0$, and the green lines correspond to a local fold bifurcation as described in the text.  Simulation details can be found in \cref{appendix_numerics}.}\label{fig:sub_23_nonlinearities}
\end{figure}

In contrast, nonlinearities {including} even-ordered terms typically led to solutions where the local Maxwell point only occasionally predicted the confined pattern region. Some examples are shown in \cref{fig:sub_23_nonlinearities}. Perhaps more strikingly, this nonlinearity also led to local regions of the solution approximately following a different local equilibrium (that is, {a nonzero solution of the $x$}-dependent polynomial one gets from \cref{main_sh0} on setting $\eps=0$ and considering time-independent solutions). Such `local equilibria' solutions vary in $x$ in a predictable way, but are beyond the scope of the analysis done in preceding sections on the stability of $u=0$.} Different simulations with varied initial conditions (e.g.~using the same random perturbations of $u=0$ but taking a different random seed) also led to solutions where different parts of the domain contained oscillatory `patterned' states interspersed with more smoothly varying `local' equilibria. While it appears the Maxwell point idea works better for larger amplitudes, this is {observed to be extensively driven by} larger amplitudes drastically increasing the speed at which the solution moves through the bifurcation. Hence, a more accurate theory of the slow passage through this structure must account not only for the value of $r(x)$ where the pattern is no longer energetically favorable, but also the nonlinearity and the speed at which the heterogeneous solution passes this point. {We note that numerical continuation results in \cite{kao2014spatial} for the quadratic-cubic nonlinearity also reinforce some of the complexity in these quadratic cases, demonstrating the existence of asymmetric branches and changes in stability of different branches due to the non-zero mean of patterned solutions.}

\begin{figure}
    \centering
    \begin{subfigure}{.495\textwidth}\includegraphics[width=1\textwidth]{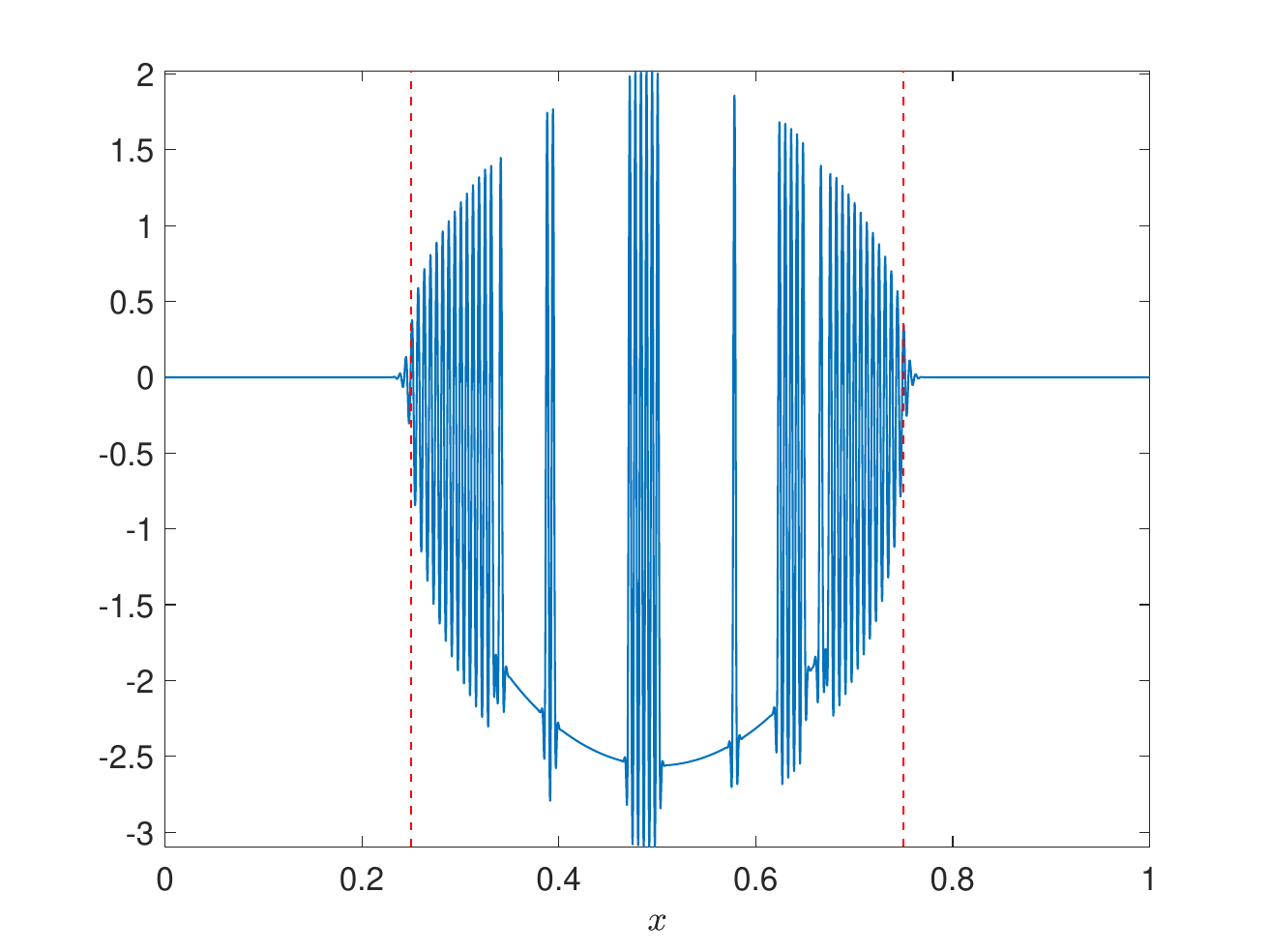}
    \caption{$N(u) = -u^2-u^3$}
    \end{subfigure}
\begin{subfigure}{.495\textwidth}\includegraphics[width=1\textwidth]{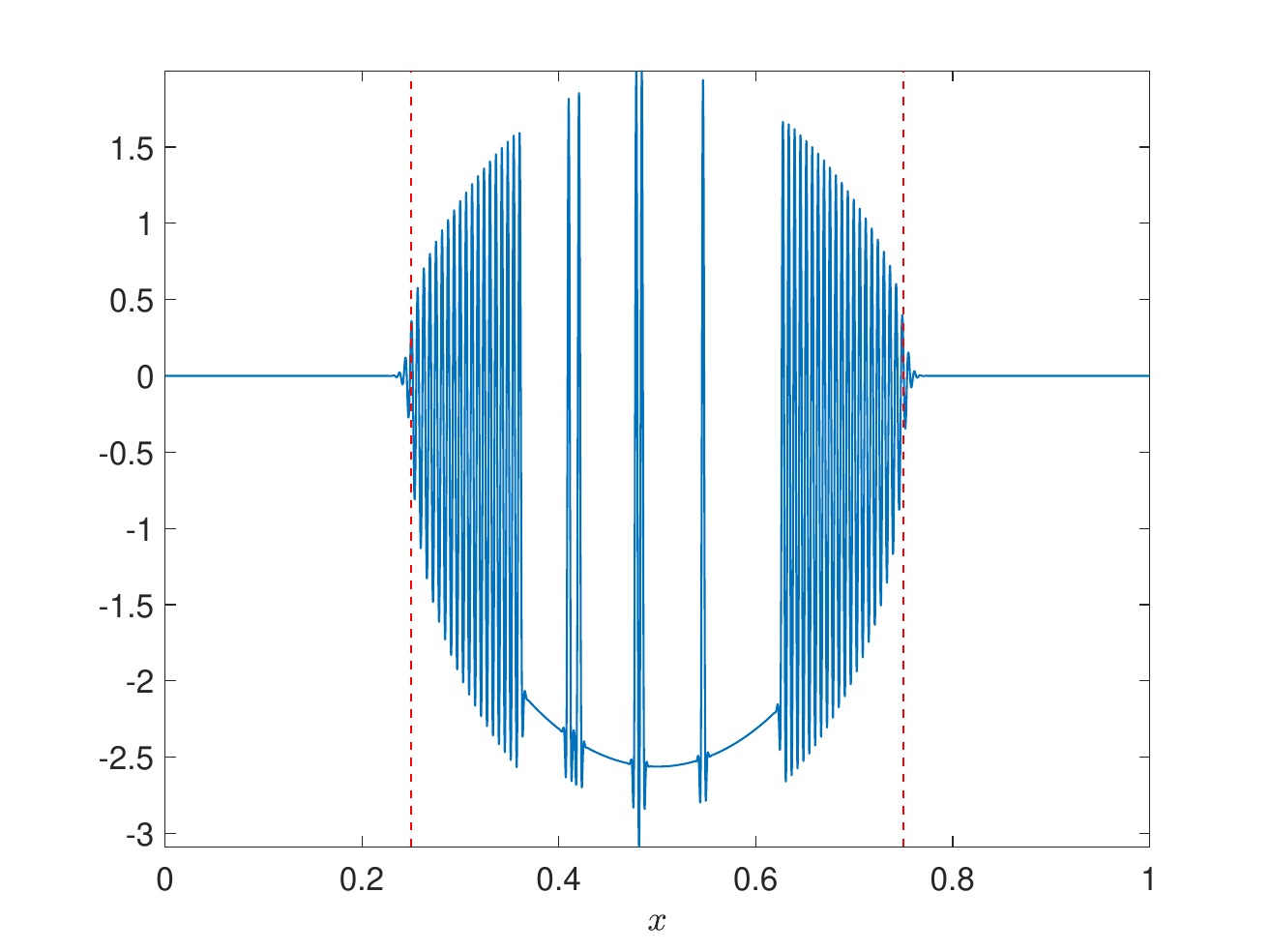}
    \caption{$N(u) = -u^2-u^3$}
    \end{subfigure}
    
    \begin{subfigure}{.495\textwidth}\includegraphics[width=1\textwidth]{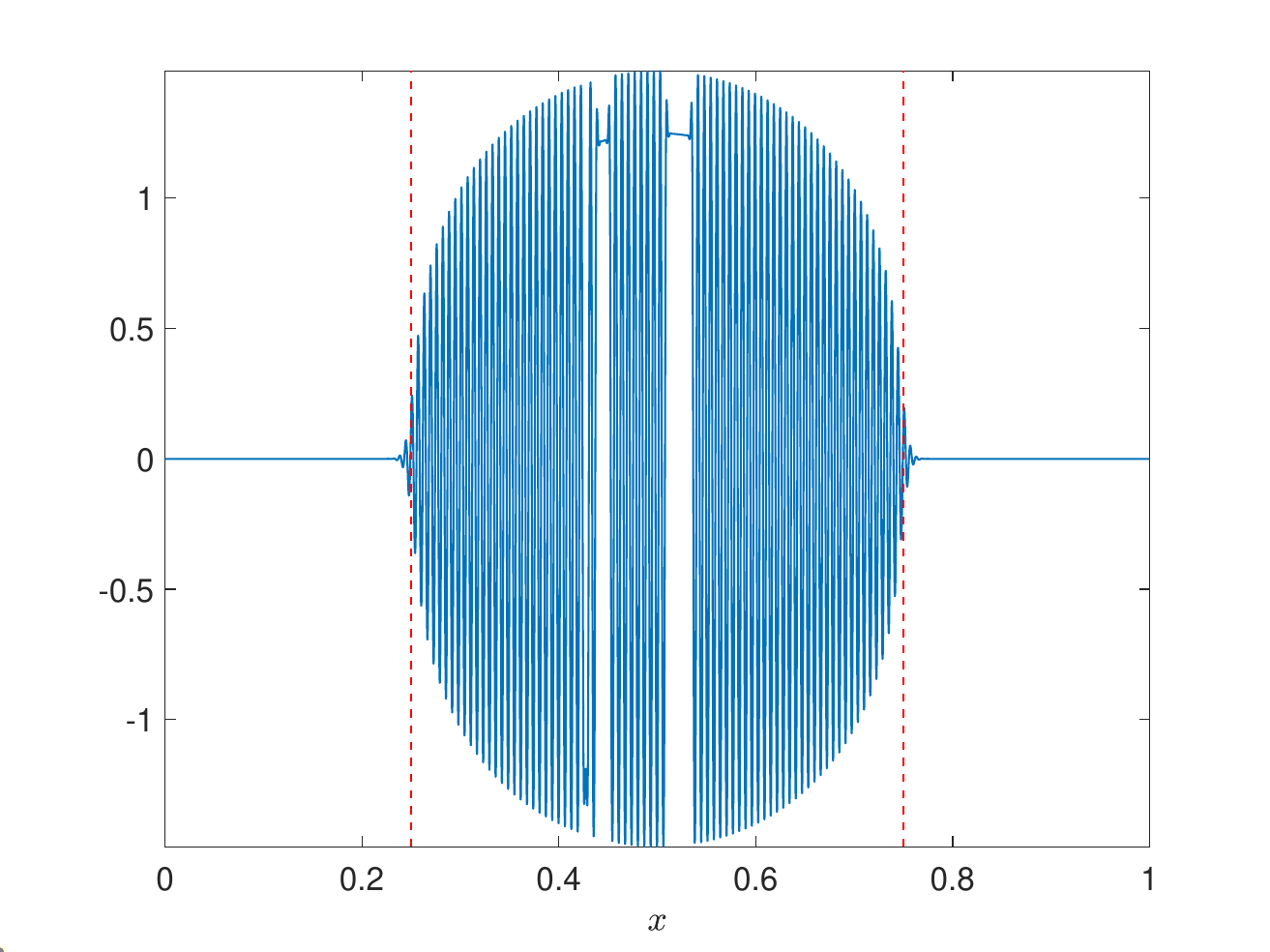}
    \caption{$N(u) = -u^3-u^5$}
\end{subfigure}
\begin{subfigure}{.495\textwidth}\includegraphics[width=1\textwidth]{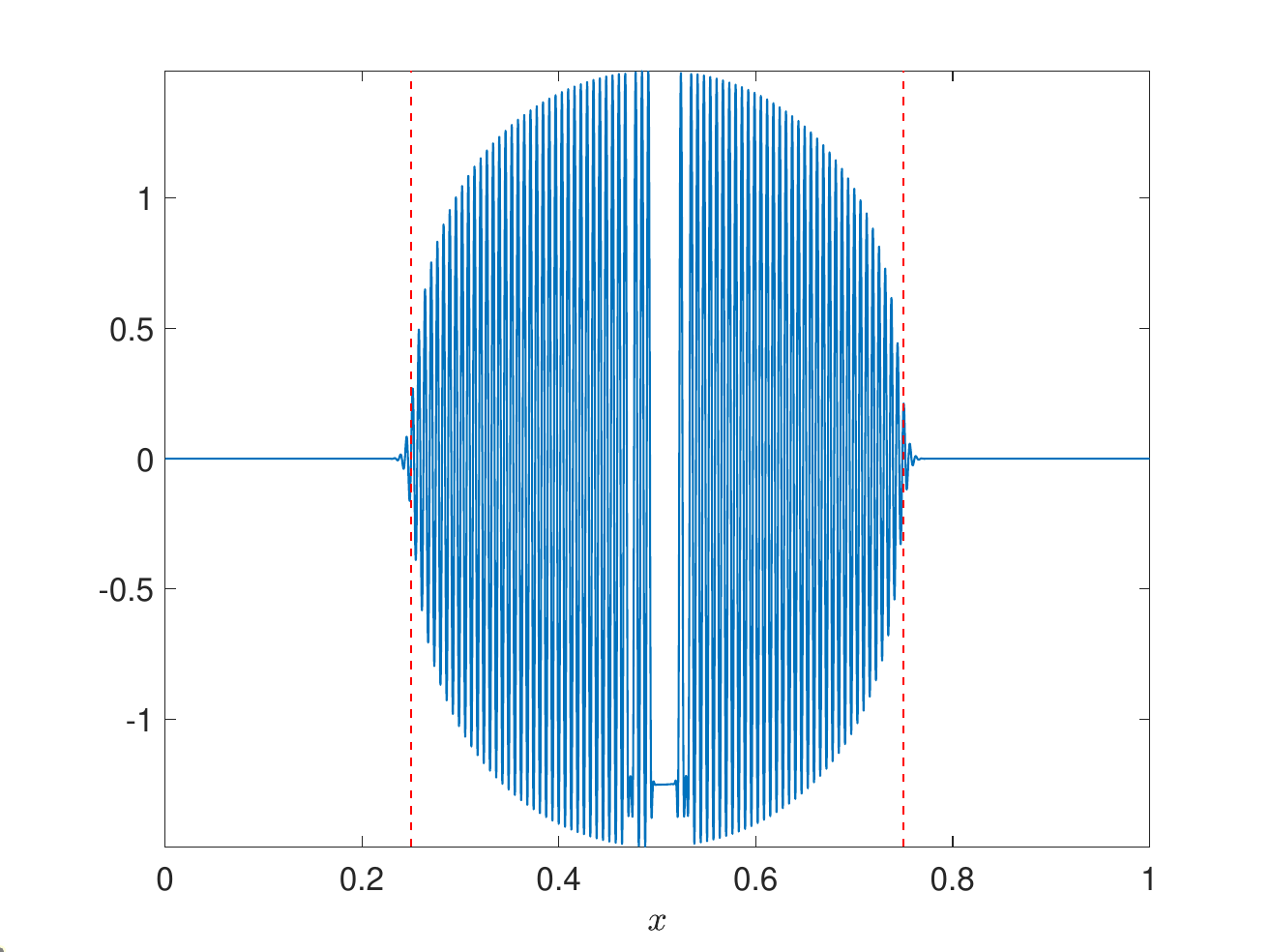}
    \caption{$N(u) = -u^3-u^5$}
    \end{subfigure}

    \caption{Solutions $u(x)$  of \cref{main_sh0} (blue solid curves) with $r(x)=-5\cos(2\pi x)$, $\eps = 0.001$, varying $N(u)$ and the random initial conditions. The dashed red vertical lines indicate where $r(x)=0$.  Simulation details, {including a description of the initial conditions,} can be found in \cref{appendix_numerics}.}\label{fig:sup_different_ICs}
\end{figure}

The existence of other local equilibria does not only plague the locally subcritical case. In \cref{fig:sup_different_ICs}, we use locally supercritical nonlinearities but a much larger heterogeneity to observe that these {interior regions of non-oscillatory} solutions can also create distorted patterns in these cases, {again due to the existence of other nonzero `locally homogeneous' equilibria}. While the patterned states are indeed approximately confined by the local Turing instability criterion predicted in \cref{sec_Localisation} (that is, $r(x)\approx 0)$), the patterned state can be made up of a variety of states involving these branches of local equilibria. Which solution emerges then is seemingly sensitively dependent on the initial condition, {with the location and structure of these interior regions varying based on the specific realization of a random initial condition.} We are unaware of any simple theory capable of explaining why some kinds of structures are observed more often than others, noting that any notion of basins of attraction for such solutions will likely be complex. {We note that the simulations shown in \cref{fig:sup_different_ICs}, as well as the final two panels of \cref{fig:sub_23_nonlinearities}, violate the condition \cref{hcnsrt}, though in the patterned region this condition is not so important as the solution is already far from the trivial equilibrium where linear stability is valid.}

We end this section by noting that we only exhibited a handful of the solutions produced in order to focus on the essence of the behaviours we observed. We also explored a more general heterogeneous model of the form,
\beq\label{generalised_sh}
    \pd{u}{t} = r(x)u-\left(q(x)^2+\eps^2 \pdd{}{x}\right)^2u + N(u,x), \quad \quad \quad x\in[0,1],
\eeq
though chose to focus our attention on the simpler model given by $q(x)=1$ and $N(u,x)=N(u)$ to present key exemplars of what we found more generally. As long as $\eps$ was sufficiently small (e.g.~we always took $\eps\leq 0.01$), locally supercritical nonlinearities always gave rise to confined regions of patterning, with tails near the boundary behaving as predicted by \cref{eq:final_form_boundary_layer}. In contrast, locally subcritical instabilities had larger regions of patterning which fell rapidly back to the trivial state $u=0$. These regions of pattern confinement in the subcritical cases could sometimes be predicted by looking for a local Maxwell point, but in other cases could not. In all cases, sufficiently large amplitude heterogeneity leading to many local equilibria could give rise to disconnected regions of pattern formation, somewhat independent of the nature of the heterogeneity and nonlinearity involved, and the final form of the observed pattern became more sensitive to initial data.

\section{Discussion}\label{sec_Discussion}

We started this paper by asking: given an observed spatial pattern, what can we say about the underlying mechanism that generated this localisation? To begin formulating an answer, we studied a simple model of slow heterogeneity in the Swift-Hohenberg equation. We extended ideas from the reaction-(cross)-diffusion literature \cite{krause_WKB, gaffney2023} to provide an asymptotic justification of a local Turing instability theory, and a resolution to the boundary-layer behaviour of this asymptotic theory.
We then explored a variety of different kinds of heterogeneity and nonlinearity, finding that this theory does well for predicting confinement of patterns in the case of locally supercritical nonlinearities. 

In particular,  we have found that predictions of localisation for pattern refinement from an already heterogeneous state are inherited from the homogeneous setting. In other words, { for  heteroegeneties $r(x)$ that are compatible with the boundary conditions, as in  \cref{bcsr}, we have at leading order  
that the linearized system will evolve pointwise away from the steady state at a specific point if the homogeneous Swift-Hohenberg system with the parameters at that point is unstable, as long as $$1\gg \epsilon\rightarrow 0.$$ Although} it is only valid at leading order for the linearized dynamics, this is a very simple, yet powerful, theory, and to the authors' knowledge has not reported before for Swift-Hohenberg (though it has been recorded for simple reaction-diffusion systems, including cross diffusion \cite{krause_WKB,gaffney2023}).  
Furthermore, in contrast to these previous studies of reaction and diffusion in the context of pattern formation from heterogeneous states, here  
there has been elucidation of the  structure of the inner solution for the WKB solutions used to investigate linear stability, and also quantification of the scale of the error in pointwise predictions of localisation, noting that the evanescent solutions decay on a scale of $\eps^{1/2}$, as deduced in detail in the appendices. This linear theory then applies for localisation in the full model for locally supercritical Turing bifurcations, although in the subcritical case it under-predicts the region of pattern formation.

As the WKB technique can be understood in the context of uniform renormalization group approximations, it opens up the possibility that asymptotic techniques can be used more generally to understand and characterise locally supercritical bifurcations from heterogeneous states. In turn, such approaches readily  find application in understanding and investigating  ``pattern upon pattern,'' that is the emergence of more refined patterns from a simpler pre-pattern (as emphasized by Turing himself \cite{turing1952chemical}).
However, and interestingly, this local framework also \bk fails to predict regions of pattern confinement in locally subcritical cases. 
Despite the simplicity of the 1D model chosen, there seem to be large gaps in our understanding of heterogeneous systems in the presence of subcriticality. 

There are numerous direct extensions of what we have studied here. While we have provided {numerical and theoretical evidence that locally supercritical instabilities lead to decaying tails of patterning in the specific cases explored, more generality remains to be shown, as does}  any analytical support that subcritical instabilities should always coincide with a more sudden decay in amplitude. {The latter} is what one may expect from the heteroclinic connection between different equilibria using ideas from spatial dynamics, (e.g.~one can imagine falling off the top branch in \cref{fig:bifurcation_diagrams}(a) onto the trivial state via such a connection). The precise unfolding of such a heteroclinic connection is likely to involve effects that are beyond all orders in $\varepsilon$. Such exponential asymptotics approaches to localised pattern formation have been carried out in the spatially homogeneous setting (see \cite{kozyreff2009physicaD,dean2011exponential} and references therein), but have not, to our knowledge, been applied to spatially heterogeneous systems. More generally, the interaction between snaking-induced localisation and heterogeneity-induced localisation is especially relevant to understand from an applied point of view. We have provided evidence that slow Airy-like envelopes may correspond to {locally} supercritical heterogeneity-induced bifurcations {for the cases considered,} but we cannot distinguish between more rapid decay due to heterogeneity or snaking in the subcritical case in general. {More generally, the interaction between snaking-induced localisation and heterogeneity-induced localisation is especially relevant to understand from an applied point of view. We note that \cite{kao2014spatial} focuses on the impact of heterogeneity on emergent snaking, but only considers heterogeneity with $r(x)<0$ across the whole domain, and hence did not observe the kind of localisation studied here.}  

{Besides these ideas, one can imagine studying these phenomena in other models, in higher dimensions (such as in the work on multidimensional localised snaking solutions \cite{hill2024role, bramburger2024localized}), or pursuing more {refined approaches than our simple formal asymptotics and numerical explorations. Examples include the formal consideration of scenarios where pattern bleed is significant, as  seen in 
Fig.~\ref{fig:super_sub_comparisons}(a). Here evanescent solutions do not decay due to the close proximity of roots of $r(x)=0$, 
 requiring a detailed characterization of  WKB solutions with 
Im$(\varphi_\pm) \neq 0$ and the associated connection formulae and Stokes phenomena \cite{trinh2015}. Further examples for future generalization include  applying more rigorous techniques found in the literature, such as those used by 
 Kovac and Klika~\cite{kovavc2022liouville}
 and Hummel et~al.~\cite{hummel2022geometric}.} 
 
 A further extension might be to consider  spatiotemporal forcing of pattern forming systems, determining parameter regimes where the system does or does not follow a naive quasi-static prediction of when and where pattern formation occurs, depending on the magnitude and frequency of the forcing {as previously pursued in \cite{dalwadi2023universal} for self-organising systems. We remark that this  study} crucially required the assumption of a locally supercritical bifurcation, as the subcritical case is, as demonstrated here, vastly more intricate.  Entirely alternative approaches, such as directly looking at how heterogeneity itself induces bifurcations \cite{vandenberg2023turing}, or considering the impact of introducing heterogeneity on localised solutions coming from homoclinic snaking mechanisms \cite{kao2014spatial}, may also prove useful.

Despite the complexity observed in our simulations, the existence of the energy functional \cref{energy} precludes the possibility of long-time spatiotemporal states, heterogeneity-induced \cite{krause2018heterogeneity, kolokolnikov2018pattern} or those arising from local Hopf instabilities \cite{patterson2023spatial}. {The existence of multiple spatially homogeneous} equilibria can likely induce a range of nontrivial behaviours such as Turing instabilities which fail to form patterned states \cite{krause2024turing}, or complex dynamics only sometimes understood via local analogues such as heteroclinic connections between homogeneous solutions \cite{patterson2023spatial}. {Versions of the Swift-Hohenberg equation with broken nonlinear symmetries \cite{houghton2011swift}, or with non-variational structure \cite{burke2009swift}, have been shown to exhibit a variety of interesting behaviours, and would be obvious models to consider to generalize the ideas presented here.} 

\cref{fig:bifurcation_diagrams}(b), despite only showing a subset of solution branches, demonstrates a variety of branches apparent even in the homogeneous form of our model as a single parameter is varied. {We note that existing continuation in the heterogeneous case has so far been confined to specific primary and secondary branches \cite{kao2014spatial}.} We anticipate that such pictures will only become more complicated with spatial heterogeneity {which not only modulates snaking structures, but also induces new solution branches via local Turing bifurcations.} While local bifurcation-theoretic approaches (such as those used in this paper and essentially all of the existing literature) are important for understanding some aspects of these systems, we also want to highlight that there is a growing need to understand more global dynamics of models with heterogeneity and multiple equilibria.

\section*{Rights Retention Statement}
For the purpose of Open Access,  a CCBY public copyright licence is applied to any Author Accepted Manuscript version arising from this submission. 

\section*{Acknowledgments}
E.~V-S. has received PhD funding from ANID, Beca Chile Doctorado en el extranjero, number 72210071.

\bibliographystyle{siamplain}
\bibliography{references}

\appendix
\section{Details of inner solution asymptotics}

\subsection{Asymptotics of the integral solution for \tmath{\Lambda>0}}\label{appendix_lambda>0}
To further illustrate the asymptotics from the main text, we add plots of the SDC, tangent at the saddle and asymptotes for the second saddle $s_{+-}$ together with contour plot of $\Im(\psi)$ in  \cref{SDCs+-} and \cref{SDCcontours}.

\begin{figure}
\centering
\begin{subfigure}{.485\textwidth}
   \centering
   \includegraphics[width=.5\linewidth]{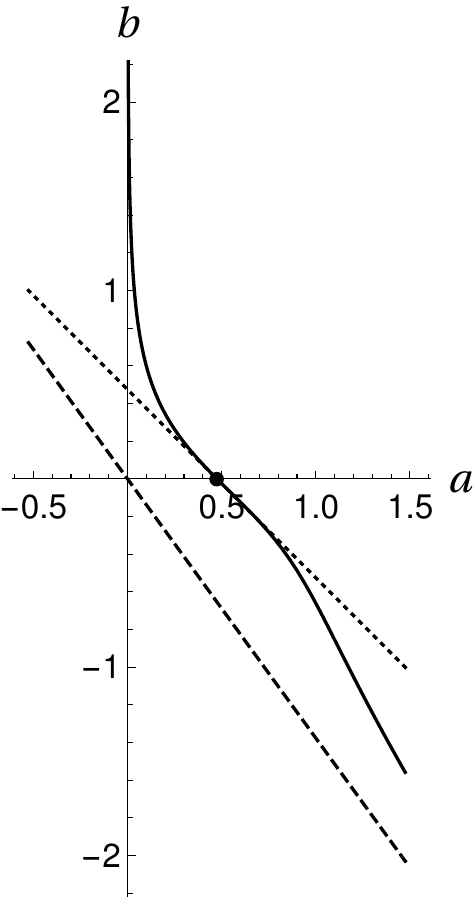}
   \caption{SDC in the complex plane.} 
\end{subfigure}~~~~
\begin{subfigure}{.485\textwidth}
   \centering
   \includegraphics[width=.95\linewidth]{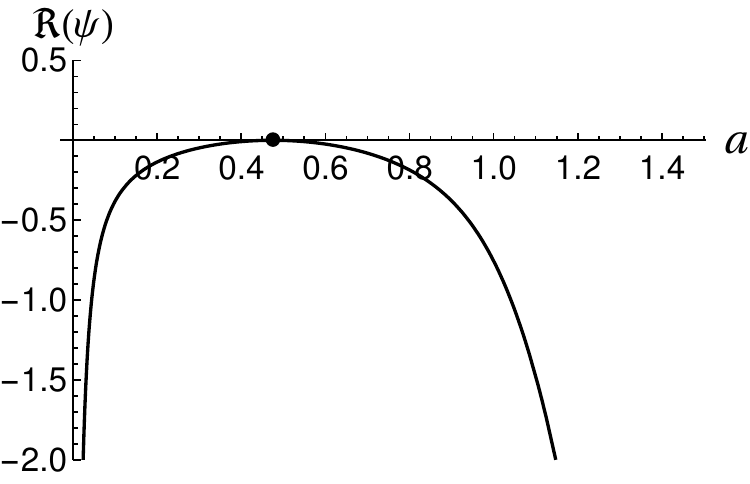}  
\caption{Real part of $\psi$ along the contour ($x$ axis being the parameter $t=a$ along the contour) confirming that it is the (steepest) descent curve.}
 \end{subfigure}
    \caption{\label{SDCs+-}Steepest descent contour for $\Lambda=0.6$ passing through $s_{+-}$, {with $s_{+-}= 0.5$, $\alpha_{+-} = -0.2$ and the tangent  angle of the} SDC at $s_{+-}$ given by  $3\pi/4$.  {The asymptotes in panel (a) are at the angles $\pi/2$ and $17\pi/10$, with one dashed line indicating the latter asymptote and a second dashed line depicting the tangent at the saddle, while the asymptote of angle  $\pi/2$ coincides with the axis.}}
\end{figure}

\begin{figure}
\centering
\begin{subfigure}{.485\textwidth}
   \centering
   \includegraphics[width=.95\linewidth]{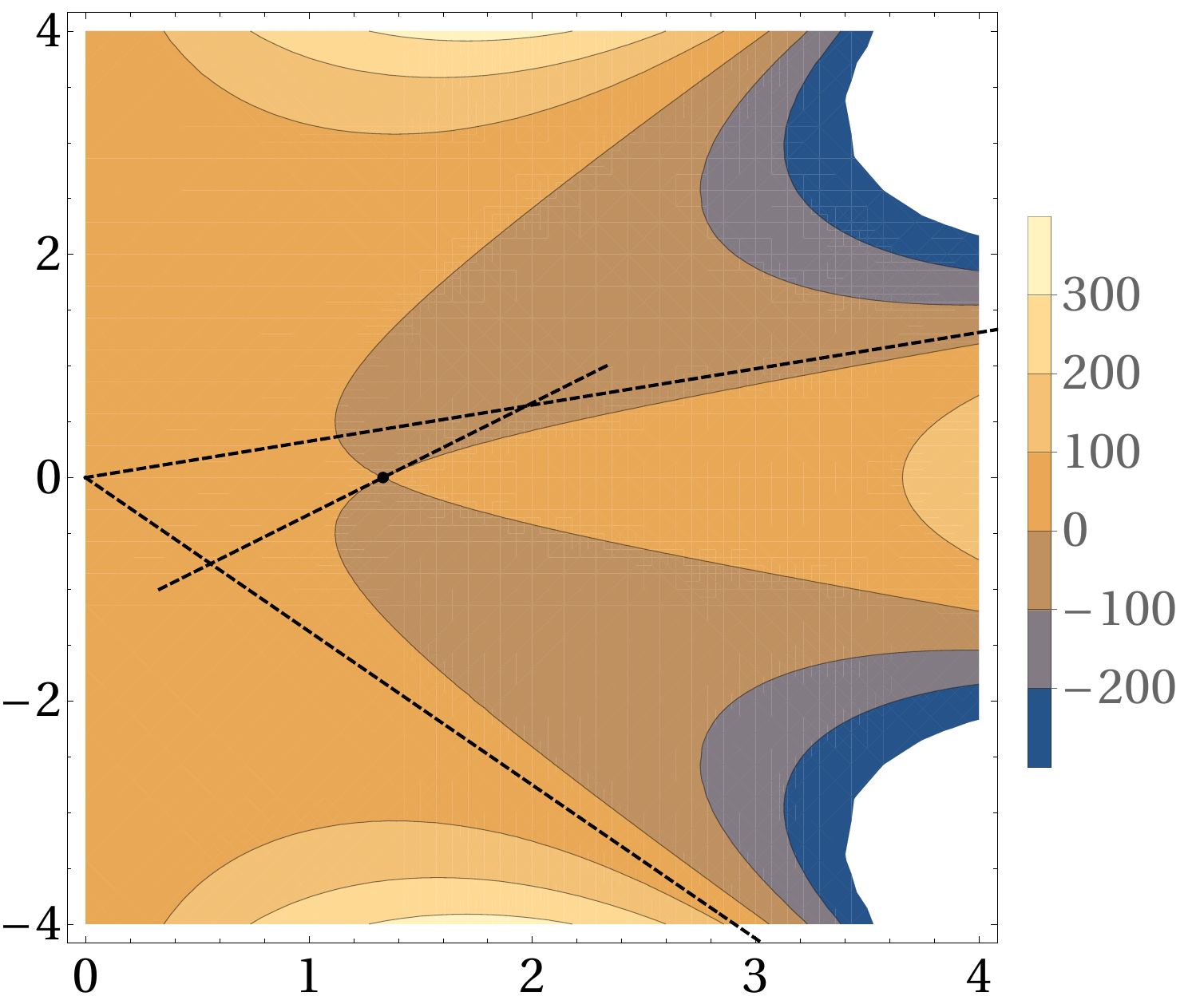}
   \caption{Contour lines of the imaginary part of $\psi$ in the complex plane.}
\end{subfigure}~~~~
\begin{subfigure}{.485\textwidth}
   \centering
   \includegraphics[width=.95\linewidth]{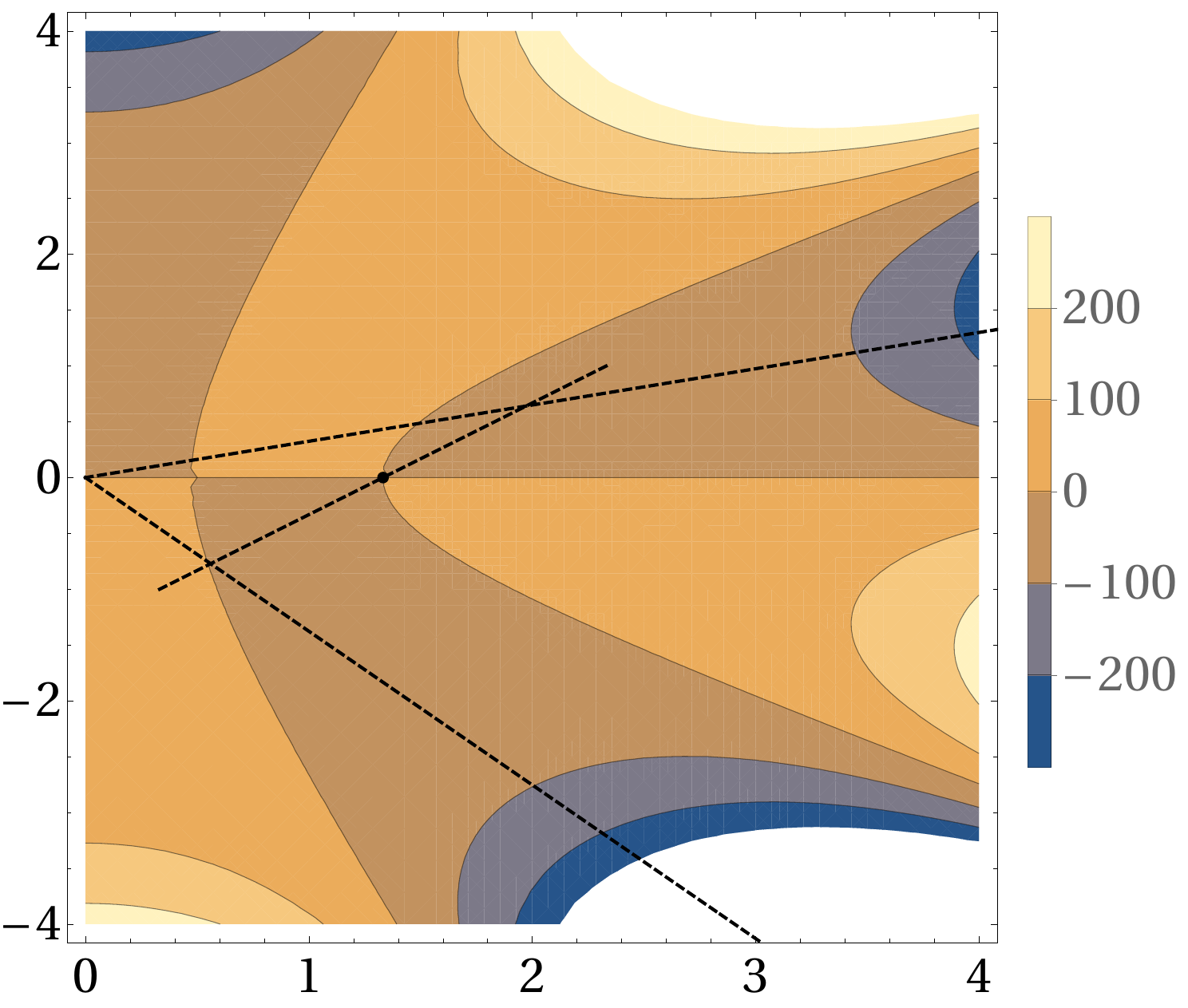}  
   \caption{Contour lines of the real part of $\psi$ in the complex plane.}
 \end{subfigure}
    \caption{\label{SDCcontours} The landscape of real and imaginary part of $\psi$ together with the highlighted location of the SDC passing through the saddle $s_{++}$ together with the calculated asymptotes and tangent at the saddle. Here, $\Lambda=0.6$.}
\end{figure}

\subsection{Asymptotics of the integral solution for \tmath{\Lambda<0}}\label{appendix_lambda<0}

In the situation when $\Lambda<0$, saddles are  {not located on the real line and their locations are given by}
\begin{align}
  s_{++}&:= + (1-\Lambda)^{1/4}(\cos(\varphi)+i\sin(\varphi)),\label{s++Lneg}\\
  s_{+-}&:= - (1-\Lambda)^{1/4}(\cos(\varphi)-i\sin(\varphi))=\bar{s}_{++},\label{s+-Lneg}\\
  s_{-+}&= -s_{+-}, s_{--}= - s_{++},\nonumber
\end{align}
where $\varphi=\frac{1}{2}\arctan(\sqrt{-\Lambda})$ and again we drop $s_{-+}$ and $s_{--}$ due to the symmetry.

The SDC can be parametrised in the same way as in the positive case, hence the asymptotes are the same. 
However, the directions of the steepest descent at the saddles $s_{++}$, $s_{+-}$ depend on $\Lambda$ (and are not constants as in the case $\Lambda>0$). They are given by angles
\begin{align}\nonumber 
 ~~~ \phi_{++} &= \mathrm{Arg}\left[\left((-i+\sqrt{-\Lambda})\left[(-1+\sqrt{1-\Lambda})\cos(\varphi)+i(1+\sqrt{1-\Lambda})\right]\right)^{-1/2}\right],\\~~~
    \phi_{+-} &= \mathrm{Arg}\left[\left((-i-\sqrt{-\Lambda})\left[(-1+\sqrt{1-\Lambda})\cos(\varphi)-i(1+\sqrt{1-\Lambda})\right]\right)^{-1/2}\right], \label{phipp}
\end{align}
where the subscript denotes the corresponding saddle. In  \cref{TangentForLambdaNeg} we illustrate the dependence by plotting these angles for (complex arguments of) all the four saddles. Note that one can expand this expression for the tangent {angle} for small negative $\Lambda$ and get that the tangent {angle of the} SDC at $s_{++}$ approaches $0$ while {taking the value} $-\pi/2$ at $s_{+-}$. Hence there is a discontinuity of the contour across the turning point $\Lambda=0$. This is not unexpected, as two pairs of saddles coalesce into one and then split.

\begin{figure}
\centering
   \includegraphics[width=.65\linewidth]{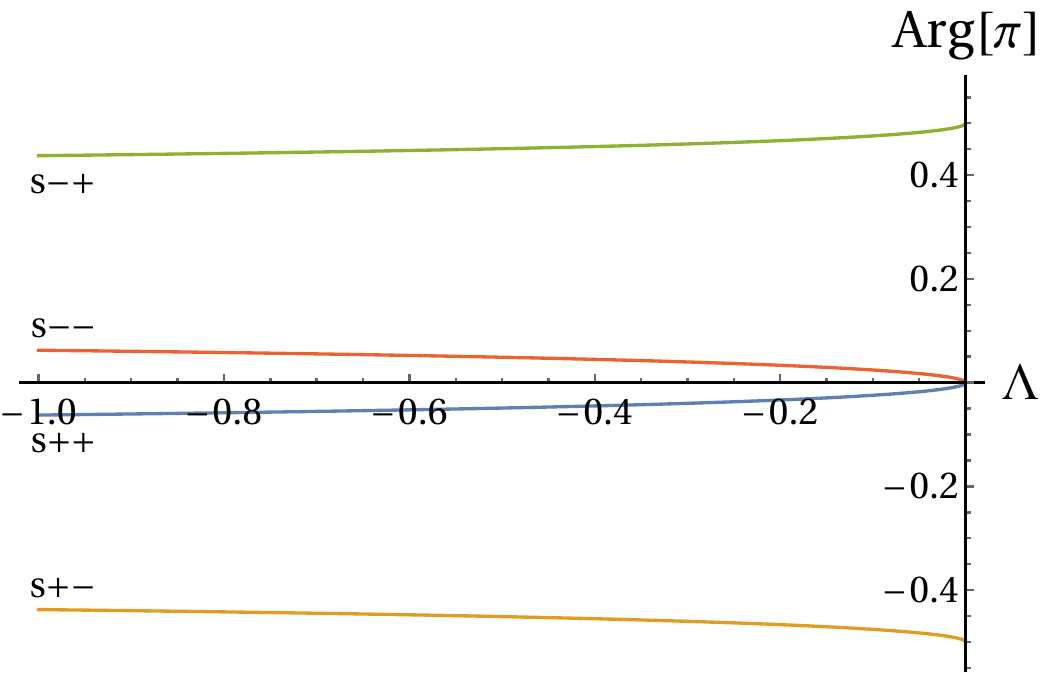}
    \caption{\label{TangentForLambdaNeg} The numerically evaluated {tangent angles} of the SDC at the saddles as a function of $\Lambda<0$ following from the expression for $\phi_{++}$ and the other three angles in the main text. The $y$-axis is in multiples of $\pi$.}
  \end{figure}

One can show that the two complex conjugate saddles, $s_{++}$ and $s_{+-}$, have the same value of $\Im(\psi)$ and hence lie on the same SDC. In addition, $\Re(\psi)$ is larger at $s_{+-}$, hence the steepest descent contour of $s_{+-}$ passes through $s_{++}$ {on descent.} 
We illustrate the SDC with a particular choice $\Lambda=-0.7$ in {Figures} \ref{SDCs++LambdaNeg}, \ref{SDCs+-LambdaNeg}, and \ref{SDCcontoursLambdaNeg}.

\begin{figure}
\centering
\begin{subfigure}{.495\textwidth}
   \centering
   \includegraphics[width=.9\linewidth]{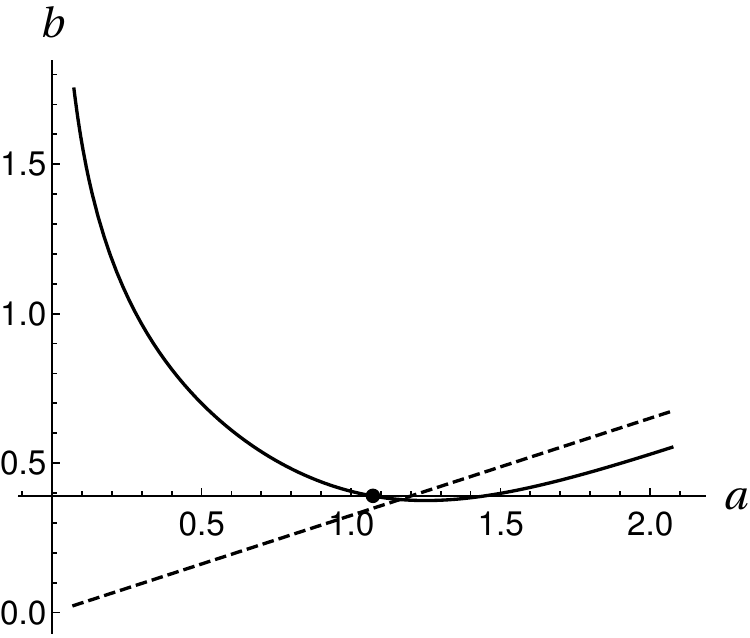}
   \caption{SDC in the complex plane for $s_{++}$.}
\end{subfigure}
\begin{subfigure}{.495\textwidth}
   \centering
   \includegraphics[width=.8\linewidth]{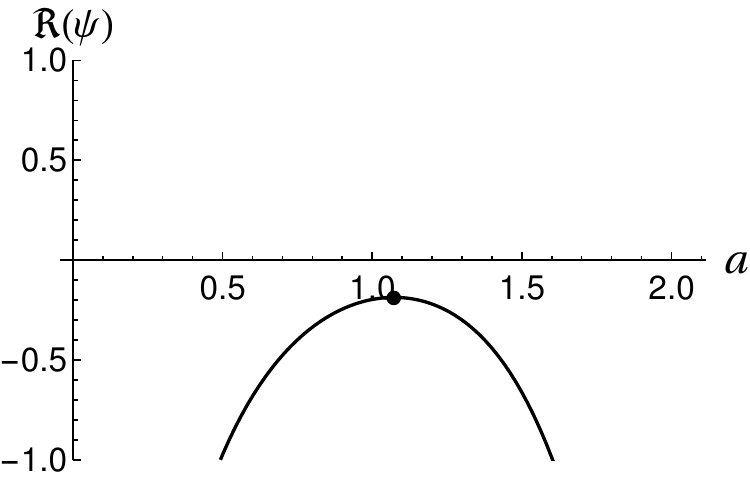}  
\caption{Real part of $\psi$ along the contour ($x$ axis being the parameter $t=a$ along the contour) confirming that it is the (steepest) descent curve.}
 \end{subfigure}
    \caption{\label{SDCs++LambdaNeg}Steepest descent contour for $\Lambda=-0.7$ passing through $s_{++}$. The dashed lines indicate the revealed asymptote at one end (being $\pi/10$), the other being $\pi/2$.}
\end{figure}

\begin{figure}
\centering
\begin{subfigure}{.495\textwidth}
   \centering
   \includegraphics[width=.5\linewidth]{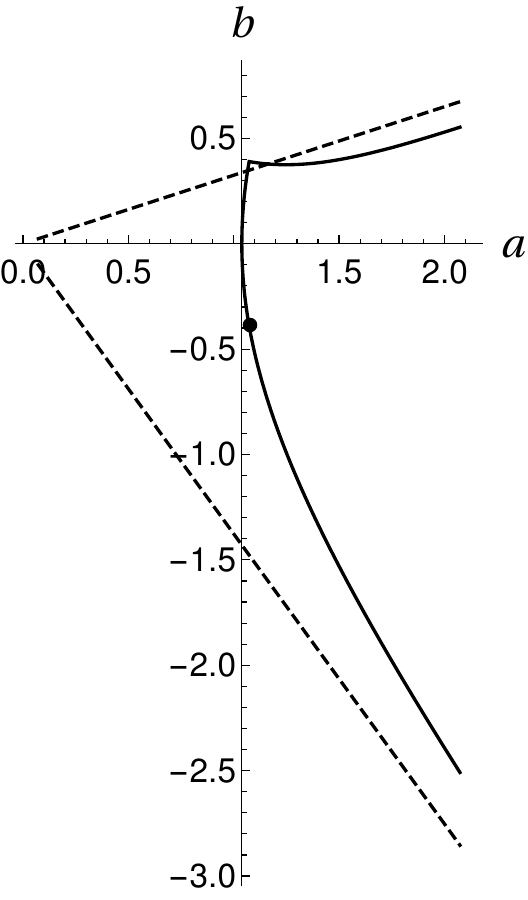}
   \caption{SDC in the complex plane.}
\end{subfigure}
\begin{subfigure}{.495\textwidth}
   \centering
   \includegraphics[width=.95\linewidth]{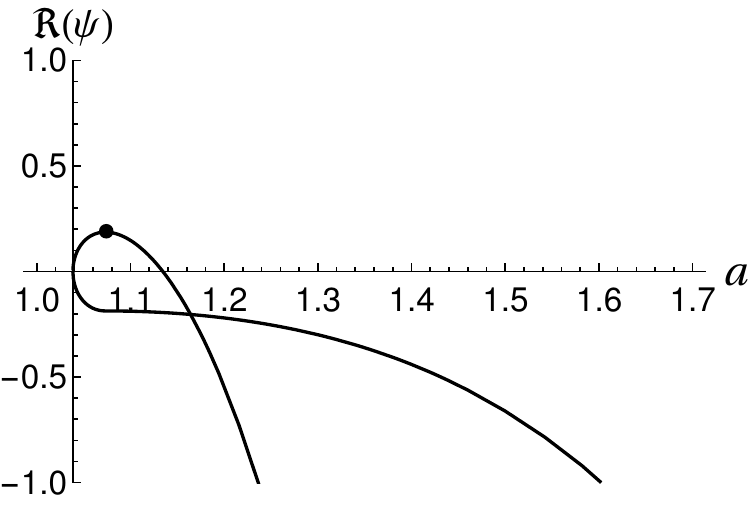}  
\caption{Real part of $\psi$ along the contour ($x$ axis being the parameter $t=a$ along the contour) confirming that it is the (steepest) descent curve.}
 \end{subfigure}
    \caption{\label{SDCs+-LambdaNeg}Steepest descent contour for $\Lambda=-0.7$ passing through $s_{+-}$ (the highlighted point) but also through $s_{++}$ being the complex conjugate of $s_{+-}$. The dashed lines indicate the revealed asymptotes (the new one being $17\pi/10$). Note that the real part of $\psi$ is greater at $s_{+-}$ than at $s_{++}$ and, as a result, the SDC corresponding to $s_{+-}$ continues along the SDC of $s_{++}$ once the second saddle is reached (arbitrarily choosing one of the two halves).}
\end{figure}

\begin{figure}
\centering
\begin{subfigure}{.495\textwidth}
   \centering
   \includegraphics[width=.95\linewidth]{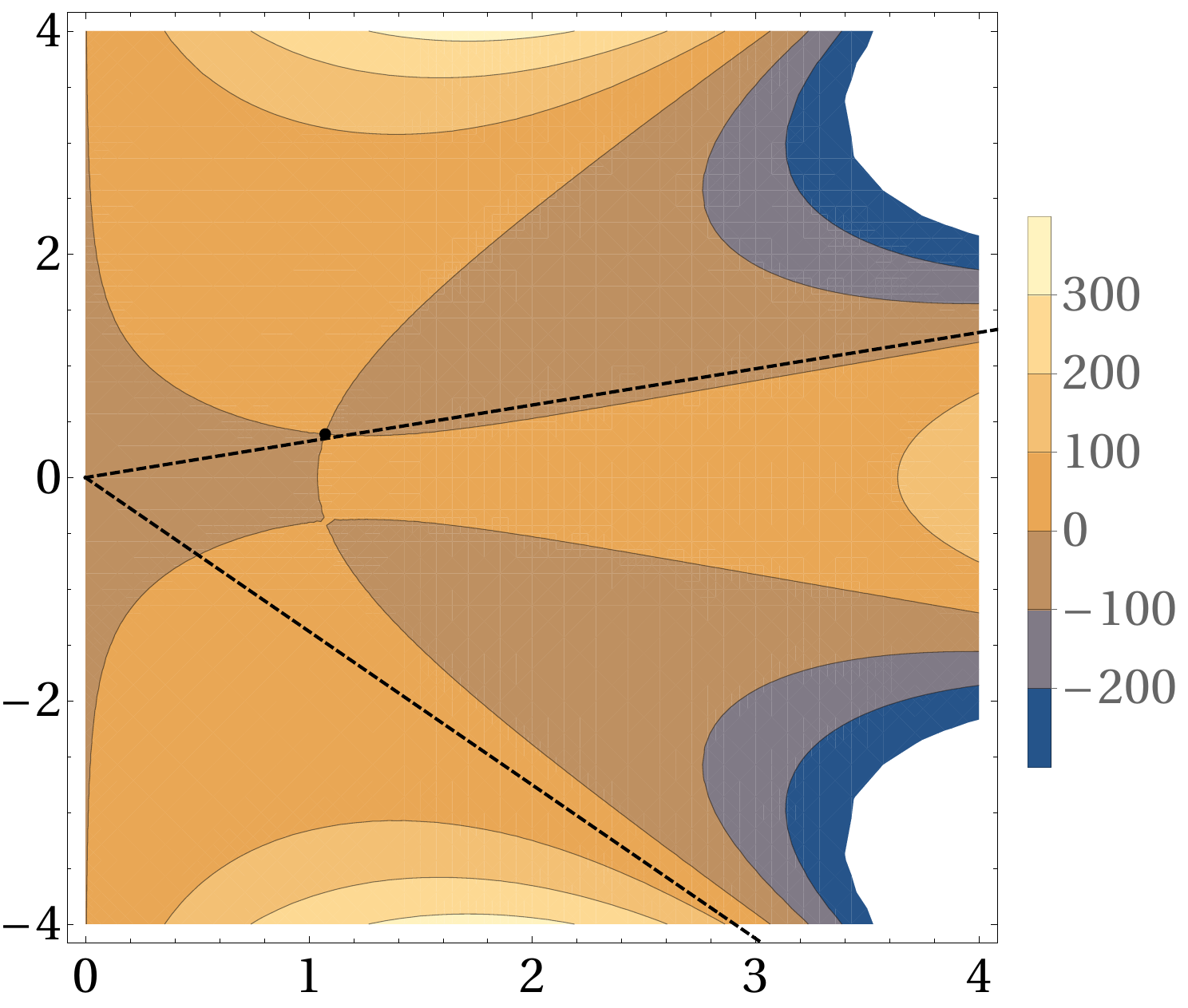}
   \caption{Contour lines of the imaginary part of $\psi$ in the complex plane.}
\end{subfigure}
\begin{subfigure}{.495\textwidth}
   \centering
   \includegraphics[width=.95\linewidth]{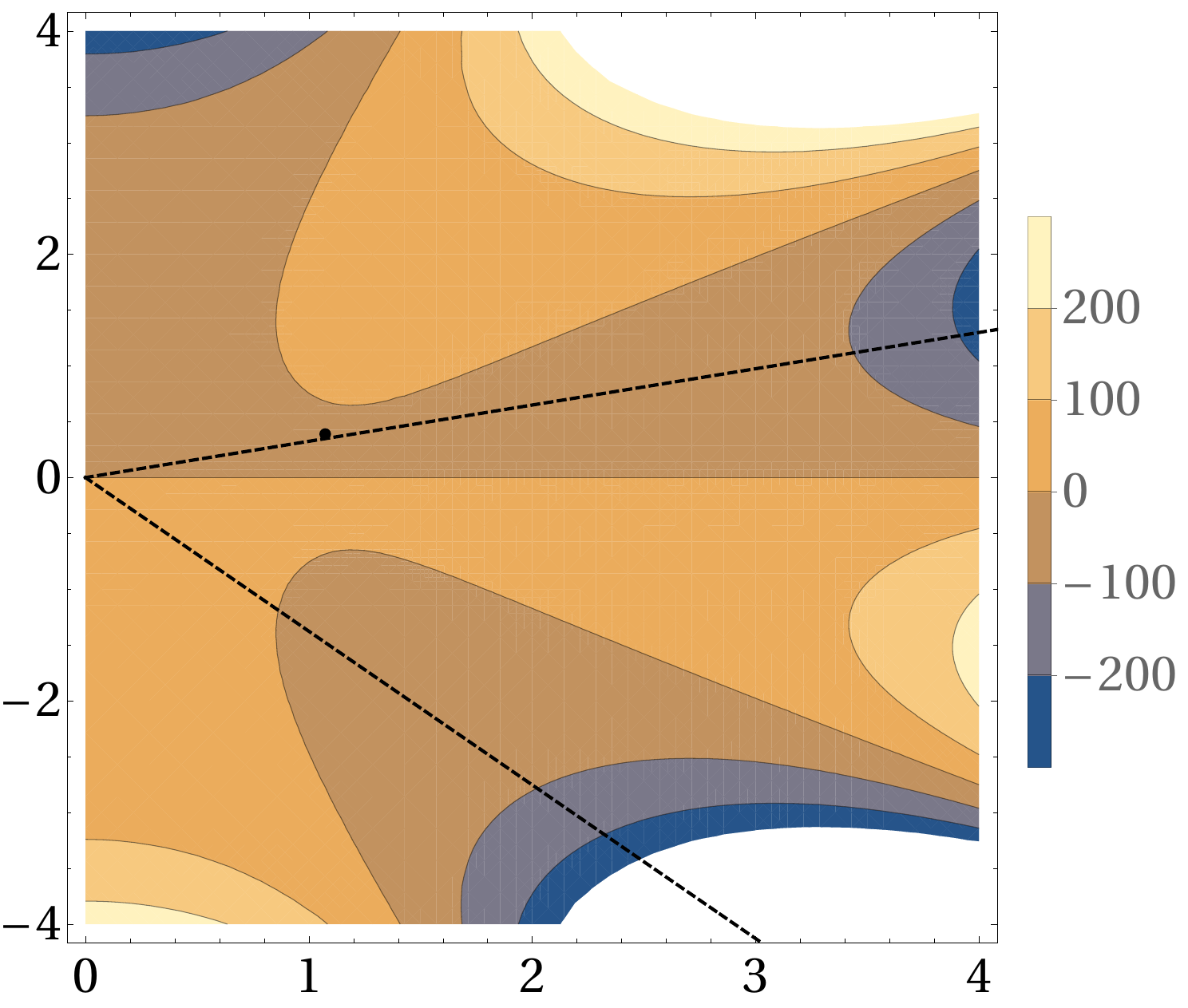}  
   \caption{Contour lines of the real part of $\psi$ in the complex plane.}
 \end{subfigure}
    \caption{\label{SDCcontoursLambdaNeg} The landscape of real and imaginary part of $\psi$ together with the highlighted location of the SDC passing through the saddle $s_{++}$ together with the calculated asymptotes and tangent at the saddle. Here, $\Lambda=-0.7$.}
\end{figure}

The approach to the contour integral approximation is the same as in the positive case.
We again have four independent solutions corresponding to saddles $s_{++},~s_{+-}$ and note that due to $\Re(\psi)|_{s_{+-}}>\Re(\psi)|_{s_{++}}$ the contribution of the neighbourhood of $s_{++}$ to the contour integral along the $s_{+-}$ contour (passing through $s_{++}$) is subleading.

We first evaluate $\psi$ and its second derivative at the saddle $s_{++}$. We have
\begin{align*}
  \psi_{++} &= -\left[184+56G-132\Lambda+120\frac{\Lambda}{G} \frac{1}{\sin^2(\varphi)}+15 \frac{\Lambda^2}{G^2}\frac{1}{\sin^4\varphi}\right] \\
  &\qquad \times \frac{(1-\Lambda)^{1/4}\sin(\varphi)}{240}+ i\left[\frac{4}{15} \left(1+G-3 \Lambda\right)G^{1/2} \cos(\varphi)\right],\\
   \psi_{++}'' &= 4 (1-\Lambda)^{-1/4} (i-\sqrt{-\Lambda})\left[\left(-1+G\right)\cos(\varphi)+i\left(1+G\right)\sin(\varphi)\right]\\
            &=4 G^{-1/2} \bigg[\sqrt{-\Lambda} \left(-1+G\right)\cos(\varphi)+\left(1+G\right)\sin(\varphi)\\
  &\qquad+i \left(\left(1-(1-\Lambda)^{1/2}\right)\cos(\varphi)+\sqrt{-\Lambda} \left(1+(1-\Lambda)^{1/2}\right)\sin(\varphi)\right)\bigg],
\end{align*}
where $G = (1-\Lambda)^{1/2}$ to slightly simplify the typesetting. Hence, the corresponding solution is approximately
\beq
  p_{++}(y) \sim \sqrt{2\eps^* \pi}  \cos\left(\frac{1}{\eps^*}\frac{4}{15} \left(1+(1-\Lambda)^{1/2}-3\Lambda\right)(1-\Lambda)^{1/4} \cos(\varphi)\right)\\
             \times \left(-\psi_{++}'' e^{2i{\rd \phi_{++}}}\right)^{-1/2}\exp\left[\left (-\frac{1}{\eps^*}\frac{1}{240} (1-\Lambda)^{1/4} \sin(\varphi)  \right)K\right],\label{eq:p++4LambdaNegAppendix}
\eeq
where 
\beq\label{Ugly_K_expression}
K = \left[184+56(1-\Lambda)^{1/2}-132\Lambda+120\frac{\Lambda}{(1-\Lambda)^{1/2}} \frac{1}{\sin^2(\varphi)}+15 \frac{\Lambda^2}{1-\Lambda}\frac{1}{\sin^4\varphi}\right]
\eeq
and $\phi_{++}$ is the angle of the SDC at $s_{++}$, see above. Note that there is another solution with $\sin$ instead of $\cos$ and $-\psi_{++}'' e^{2 i \phi_{++}}>0$ by the construction of the steepest descent curve.


Now, let us focus on the other saddle, $s_{+-}$. One can show that
\begin{align*}
  \psi_{+-} = -\psi_{++},\quad {\rd \phi_{+-}}=\frac{\pi}{2}-{\rd \phi_{++},} \quad -{\rd \phi_{+-}}''=-\overline{\psi_{++}''},
\end{align*}
and hence
\begin{align} 
  p_{+-}(y) &\sim \sqrt{2\eps^* \pi}  \cos\left(\frac{1}{\eps^*}\frac{4}{15} \left(1+(1-\Lambda)^{1/2}-3\Lambda\right)(1-\Lambda)^{1/4} \cos(\varphi)\right)\nonumber\\
            & \times \left(-\psi_{+-}'' e^{2i{\rd \phi_{+-}}}\right)^{-1/2}\exp\left[\left (\frac{1}{\eps^*}\frac{1}{240} (1-\Lambda)^{1/4} \sin(\varphi)  \right)K\right],\label{eq:p+-4LambdaNegAppendix}
\end{align}
with $K$ similarly defined in \cref{Ugly_K_expression}. There is, again, an additional solution when replacing $\cos$ with $\sin$.



\subsection{Approximation of the contour integral near the turning point}\label{appendix_approx_contours}

As stated in the main text, sufficiently close to the turning point, the approximations invoking {Laplace's method no longer work.} 
We shall take the advantage of the fact that each of the two pairs of the real saddles coalesce as $\Lambda\to 0^+$ and then separate out into two complex conjugate pairs
{
once $\Lambda$ has become negative, so that for $0<|\Lambda|\ll 1$ the saddles are  close to coalescence.}

We {thus} proceed here according to \cite[Chap{ter} 23]{temme2014asymptotic}, \cite[Chap{ter} 9]{olver1997asymptotics}, \cite{chester1957extension} where the main idea is to find and use a suitable change of variables so that one can use the known integral representation and asymptotics of Airy functions. Namely, it holds that 
\begin{multline} \label{eq:2Appendix}
  \frac{1}{2\pi i} \int_{C_{\Ai}}e^{\frac{1}{\eps}(\frac{1}{3} t^3 - \eta t)} f(t) \dd t \sim \eps^{1/3} \left({M(\eta)}+\Ord{\eps}\right) \Ai\left(\eta \eps^{-2/3}\right)\\-\eps^{2/3}\left({N(\eta)}+\Ord{\eps}\right) \Ai'\left(\eta \eps^{-2/3}\right),
\end{multline}
as $\eps\to 0^+$ {with $M(\eta)=\frac{1}{2}\left(f(\sqrt{\eta})+f(-\sqrt{\eta})\right)$, $N(\eta)=\frac{1}{2}\frac{1}{\sqrt{\eta}}\left(f(\sqrt{\eta})-f(-\sqrt{\eta})\right)$ and} where the contour $C_{\Ai}$ is one of the three Airy contours with the asymptotes of $(-1)^{1/3}$.

To this end, a cubic transformation is used such that the two coalescing saddles are mapped onto the two extrema of the cubic. In particular, we consider
\beq
  \Xi(t) = \frac{1}{3} t^3-\eta t+A,
\eeq
and we look for the values of parameters $A,~\eta$ such that the two saddles of the cubic, that is $t_\pm=\pm\sqrt{\eta}$, match the two coalescing saddles {$s_{+-},~s_{++}$, and below we have the latter matches $t_-$.}  Hence, we look for a transformation $s\to t$ such that
\beq
  \psi(s)=\Xi(t) \mbox{ at the two saddles.}
\eeq
That is
\beq
  i\left[(1-\Lambda)(1\pm\Lambda^{1/2})^{1/2}-\frac{2}{3}(1\pm\Lambda^{1/2})^{3/2}+\frac{1}{5}(1\pm\Lambda^{1/2})^{5/2} \right]=\mp\frac{2}{3} \eta^{3/2}+A.
\eeq
Hence $2A = \psi(s_{++})+\psi(s_{+-})$, resulting in
\begin{align}
  A&=-\frac{2}{15} i \left[-2(s_{+-}+s_{++})+\sqrt{\Lambda}(s_{++}-s_{+-})+3 \Lambda (s_{++}+s_{+-})\right]  \label{eq:A}\\
  &\sim i\left(\frac{8}{15}-\Lambda\right) \mbox{ as }{\Lambda\to0}.\nonumber
\end{align}

Further, due to the sign choices, the saddle $s_{++}$  maps onto $t_-$, and hence
\beq
  \frac{2}{3} \eta^{3/2} = \frac{\psi(s_{++})-\psi(s_{+-})}{2},
\eeq
and 
\begin{align}
    \eta^{3/2}& = \frac{1}{5} i \left[2(s_{++}-s_{+-})-\sqrt{\Lambda}(s_{++}+s_{+-})+3 \Lambda (s_{+-}-s_{++})\right]  \label{eq:eta}\\
  &\sim -i {\frac{1}{2}} \Lambda^{3/2} \mbox{ as }{\Lambda\to0}.\nonumber
\end{align}
This completes the specification of the cubic transformation $\Xi(t)$.

As our contour integral representation of the solution is of the form $\int_C \exp\left(\frac{1}{\eps^*} \psi\right)$, the function $f$ in \eqref{eq:2} follows from the transformation $s\to t$ as
\beq
  f(t)=\frac{\dd s}{\dd t}.
\eeq
As the dominant contribution comes from the saddles, we expand both $\Xi$ and $\psi$ around $s_{++}$  as
\begin{align*}
  \Xi(t)&=\psi(s)\\
        &=\psi(s_{++})+\psi''(s_{++}) \frac{1}{2}(s-s_{++})^2+\psi'''(s_{++})\frac{1}{6} (s-s_{++})^3 + \Ord{s-s_{++}}^4\\
  &=\Xi(t_-) + {\Xi''(t_-)} \frac{1}{2} (t-t_-)^2+\Xi'''(t_-)\frac{1}{6} (t-t_-)^3 + \Ord{t-{t_-}}^4,
\end{align*}
and hence we identify $s(t)$ near $t=t_-$ as  
\begin{align*}
  s(t) &= {s_{++} \pm  \left(\frac{\Xi''}{\psi''}\right)^{1/2} (t-t_-)\pm\frac{\Xi'''\mp\psi'''\left(\frac{\Xi''}{\psi''}\right)^{3/2}}{6 (\Xi'' \psi'')^{1/2}} (t-t_-)^2+\Ord{t-t_-}^3},
\end{align*}
{where we used $s(t_-)=s_{++}$. Hence, we may now identify the sought function $f=$d$s/$d$t$} {to sufficient {accuracy} for our requirements,} as
$$
{f(t)= \pm  \left(\frac{\Xi''}{\psi''}\right)^{1/2} \pm\frac{\Xi'''\mp\psi'''\left(\frac{\Xi''}{\psi''}\right)^{3/2}}{3 (\Xi'' \psi'')^{1/2}} (t-t_-)},$$
{and the coefficients of the approximation \eqref{eq:2Appendix} as
$$
{M_\pm} = \pm\left[\left(\frac{\Xi''}{\psi''}\right)^{1/2}+\sqrt{\eta}\frac{\Xi'''\mp\psi'''\left(\frac{\Xi''}{\psi''}\right)^{3/2}}{3 (\Xi'' \psi'')^{1/2}}\right], \quad {N_\pm} = \left[\frac{\Xi'''\mp\psi'''\left(\frac{\Xi''}{\psi''}\right)^{3/2}}{3 (\Xi'' \psi'')^{1/2}}\right],
$$}
where all derivatives of $\Xi$ are evaluated at $t_-$, while $\psi$ is evaluated at $s_{++}$.

Finally, note that the asymptotes of the SDC passing through $s_{++}$ {have tangent angles of} $\pi/10$ and $17\pi/10$ while $\psi(s)\sim i \frac{1}{5} s^5$ for large $s$. As  $\Xi(t)\sim \frac{1}{3} t^3$ for large $t$, we can see that the cubic transformation transforms the integration contour into a contour with asymptotes being $\left(e^{i\pi/2}(e^{i\pi/10})^5\right)^{1/3}$ and $\left(e^{i\pi/2}(e^{i17\pi/10})^5\right)^{1/3}$, that is $(-1)^{1/3}$. Therefore we have that
\beq
  \int_C e^{\frac{1}{\eps^*}\psi(s)} \dd s \sim \int_{\tilde{C}} e^{\frac{1}{\eps^*} \Xi(t)} \frac{\dd s}{\dd t} \dd t,
\eeq
with $\tilde{C}$ being one of the Airy contours.

As the largest contributions to the contour integral arise from the neighbourhood of the coalescing saddles, where $f(t)=\frac{\dd s}{\dd t}$ has been identified, we have
\beq
  \int_C e^{\frac{1}{\eps^*}\psi(s)} \dd s \sim 2 \pi i e^{\frac{A}{\eps^*}}\left[(\eps^*)^{1/3} \Ai\left(L\right) (M_\pm+\Ord{\eps^*})-(\eps^*)^{2/3}\Ai'\left(L\right) (N_\pm+\Ord{\eps^*})\right],
\eeq
with $L = \eta (\eps^*)^{-2/3}$ for ease of presentation.

The final check consists in verifying the assumed analyticity of the solution in $\eta$ (and thus in $\Lambda$, i.e.~in the spatial coordinate $y$). With the knowledge of $\eta(\Lambda)$, we may Taylor expand to reveal that  
\beq \label{Eq.A16}
  M_\pm = \Ord{1} \mbox{ as } \Lambda\to 0,
\eeq
but
\beq
  N_+ = \Ord{\Lambda^{-1/2}}, \quad N_-=\Ord{1} \mbox{ as } \Lambda\to 0.
\eeq

Therefore, we conclude that the contour integral representation of the solution near the turning point is the real or imaginary part of
\begin{equation}
  \label{eq:PacrossTurningPointAppendix}
  p(\xi) \sim 2 \pi i e^{\frac{1}{\eps^*} A} \left[(\eps^*)^{1/3} \Ai\left(\eta/(\eps^*)^{2/3}\right) M_- - (\eps^*)^{2/3} \Ai'\left(\eta/  (\eps^*)^{2/3}\right) N_-\right],
\end{equation}
with $A, \eta, M_-, N_-$ given above and being functions of $\Lambda(\xi)$.  

To explicitly see the behaviour across the turning point, we Taylor expand   and obtain a continuous function 
\begin{equation} \label{eq:PnearTP12}
  p\sim \frac{\pi (\eps^*)^{1/3}}{6^{2/3} \Gamma(2/3)} (-1)^{1/6} \left(2 i + (-1)^{2/3} \sqrt{\Lambda} + \frac{2}{\eps^*}\Lambda\right) e^{\frac{1}{\eps^*} \frac{8}{15} i}{\left( 1+o(1)\right)},  
\end{equation}
where {the $N_-$ contribution only generates terms within the $o(1)$ correction and we have} $\eps^*$ is small but fixed, while $\Lambda$ attains arbitrarily small values near the turning point, with $\Lambda/\eps^*= y$.  Recalling that the sought solution is the real or imaginary part of the contour integral and the presence of  $(-1)^{1/6}$ leads to six branches of potential solutions, we have 12 potential approximations at the turning points, see  
\cref{fig:12possibleApprox_num}.


To identify the correct approximation from all these potential solutions, a comparison with the numerical integration along the SDC contours is necessary.

\subsubsection*{Numerical evaluation of the contour integral solution along SDC}
Taking advantage of the explicit knowledge of the contour parametrisation for both $\Lambda>0$, $\Lambda<0$, we can numerically integrate the contour integral form of the solution for particular parameter values at fixed points {of}  $y$. We {compare these contour integral results for a range of values of $y$ with the analytic approximation of the contour integral near the turning point (the coalescing saddles case, \eqref{eq:PnearTP12}) and far from the turning point} (\cref{eq:p++4LambdaPosit}-\cref{eq:p+-4LambdaNeg}); see  \cref{fig:numerics,fig:12possibleApprox_num}.


\begin{figure}
\centering
\begin{subfigure}{.485\textwidth}
   \centering
   \includegraphics[width=.95\linewidth]{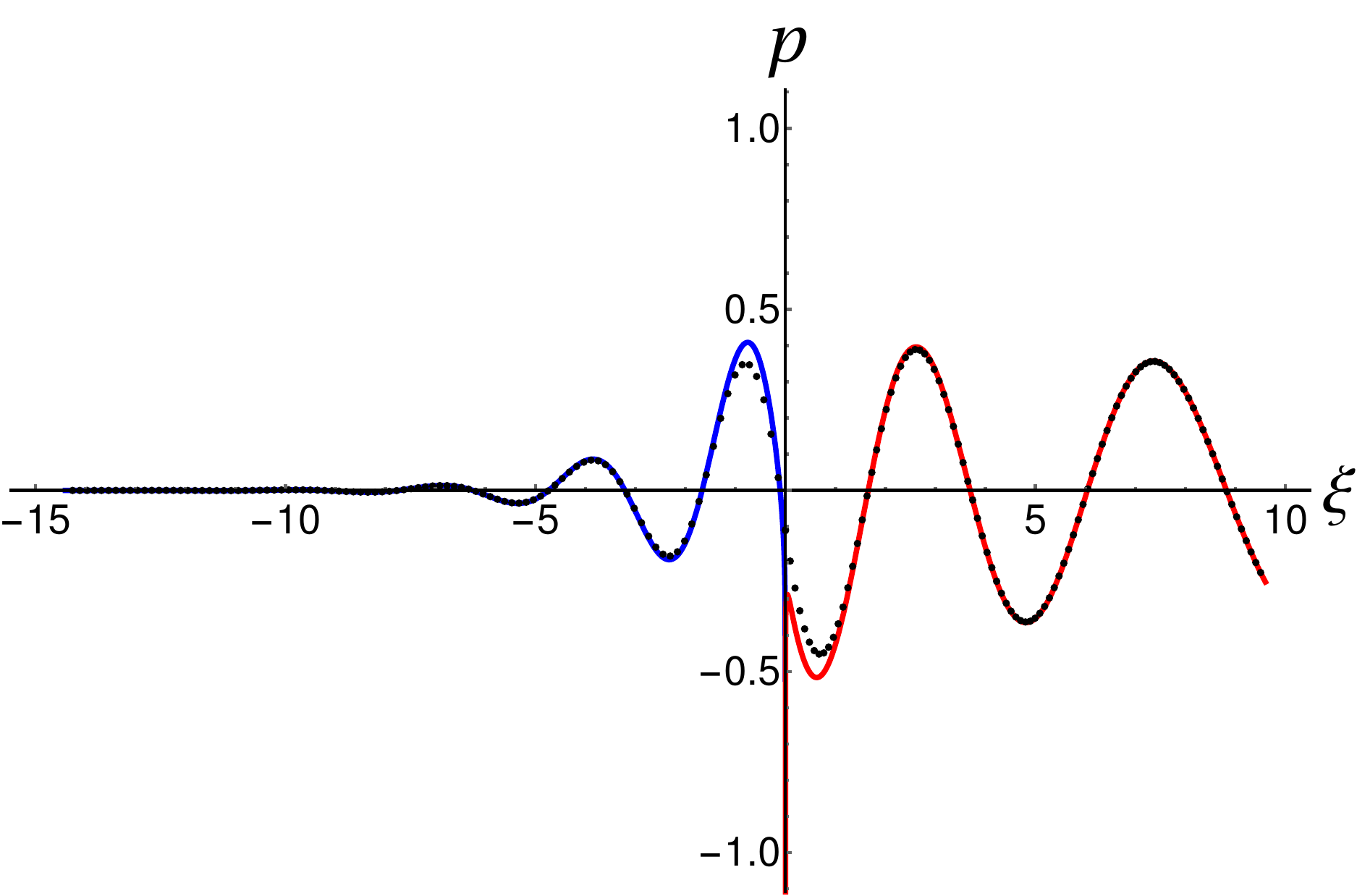}
\caption{The analytical solution {excluding the immediate vicinity of the turning point} (solid line) transitioning from an exponentially decaying solution to oscillatory is a combination of $p_{++}$ branch for $\Lambda<0$ (in blue), \cref{eq:p++4LambdaNeg}, and $p_{+-}$ branch for $\Lambda>0$ (in red), \cref{eq:p+-4LambdaPosit}.}
\end{subfigure}~~~~
\begin{subfigure}{.485\textwidth}
   \centering
   \includegraphics[width=.95\linewidth]{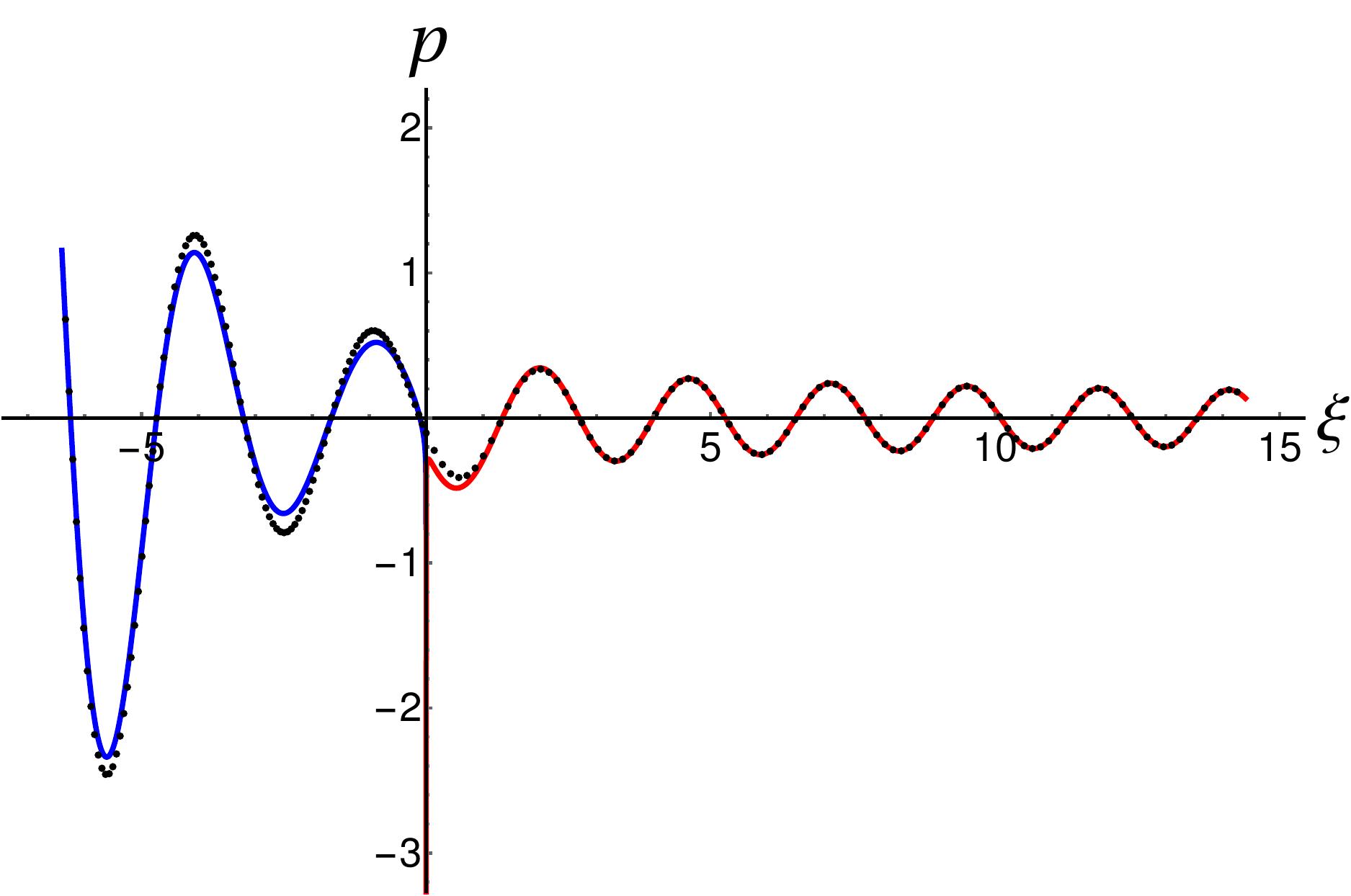}  
\caption{The analytical solution {excluding the immediate vicinity of the turning point} (solid line) transitioning from an exponentially growing solution to oscillatory is a combination of $p_{+-}$ branch for $\Lambda<0$ (in blue), \cref{eq:p+-4LambdaNeg}, and $p_{++}$ branch for $\Lambda>0$ (in red), \cref{eq:p++4LambdaPosit}.}
 \end{subfigure}
    \caption{\label{fig:numerics} {Results of the numerical  integration along the SDC contour shown as dots and the analytical estimates of the contour integral outside of the turning point as solid curves. Parameter values taken as $\rho=1/2^4,~{\eps}=1/2$ and hence, e.g., $\Lambda=0.9$ corresponds to $\xi=14.4$.}}
\end{figure}

\begin{figure}
\centering
   \includegraphics[width=.95\linewidth]{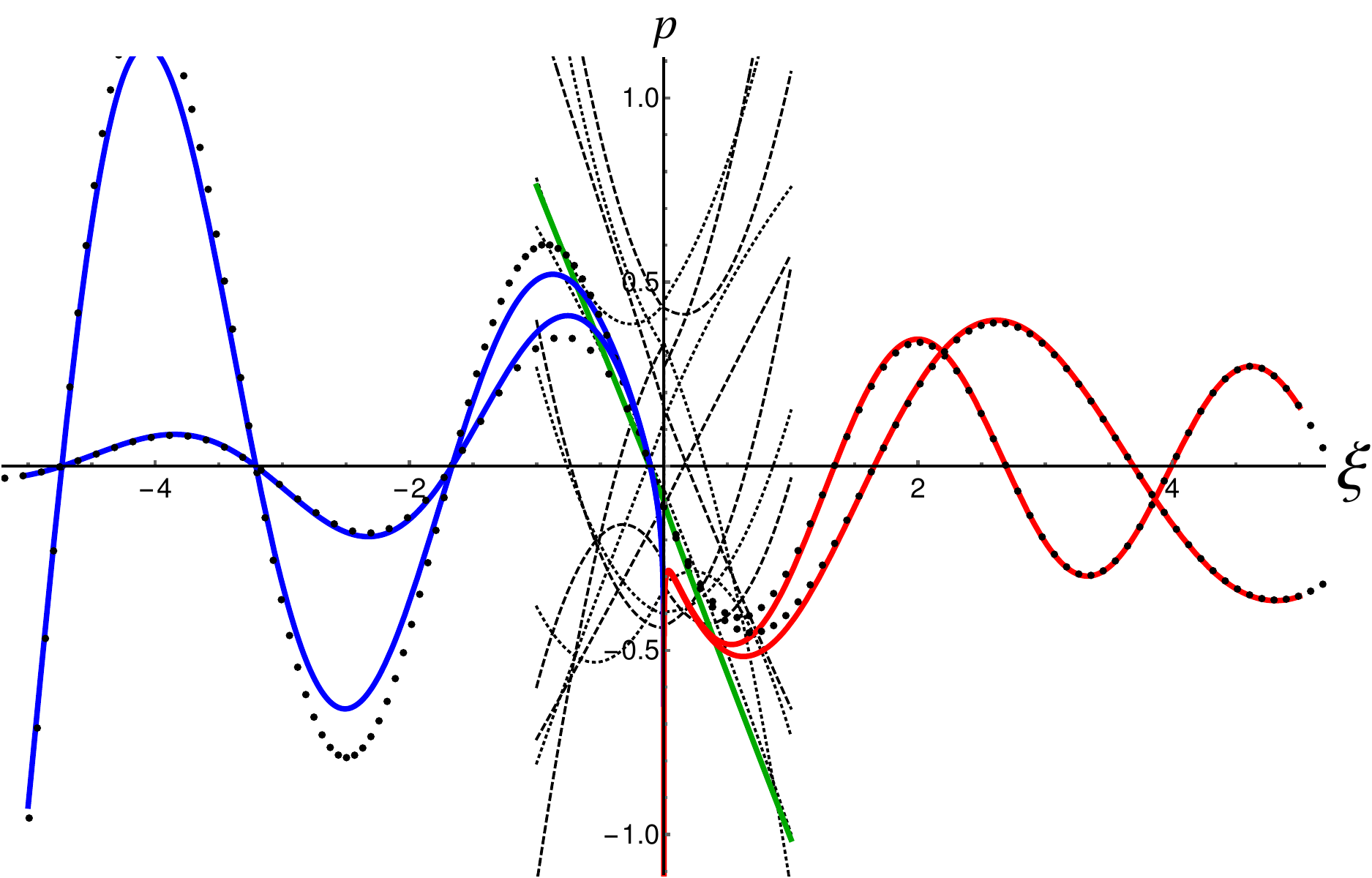}
   \caption{\label{fig:12possibleApprox_num}
     {Contour integral asymptotics for $\rho=1/2^4$, ${\eps}=1/2$ with the dots representing the numerically calculated contour integral along the SDC which allows us to calculate the behaviour even close to the turning point. Solid lines are the asymptotic {approximations for the multiple contour integral solutions}} away from the transition point, Eqs.~\cref{eq:p++4LambdaPosit}-\cref{eq:p+-4LambdaNeg} (note the blowup is outside of the range of validity). The black curves show all of the 12 potential approximations following from the analysis of coalescing saddles, \cref{eq:PnearTP12}.
     The match of the outer solution is remarkable except for a small neighbourhood of the turning point,   where one of the calculated linear approximation works well (one for all four solutions). The correctly behaving root corresponds to the fifth root of $(-1)^{1/6}$ and the real part of $p$. The explicit expression is in the text, given by \cref{eq:finalPnearTP}.}
 \end{figure}

 Solid lines are the asymptotics away from the transition point, Eqs.~\cref{eq:p++4LambdaPosit}-\cref{eq:p+-4LambdaNeg} (note the blowup is outside of the range of validity  {and is not shown in \cref{fig:numerics}}). The black lines {in \cref{fig:12possibleApprox_num}} show all the 12 potential approximations following from the analysis of coalescing saddles, \cref{eq:PnearTP12}.

 Note that the computations for $\Lambda<0$ and $\Lambda>0$ are quite different and so are the analytical expressions. Yet, the numerical solution seems to be smooth across the turning point $\Lambda=0$. In addition, the analytical results nicely match the numerics (and vice versa) with the notable exception being the vicinity of $\Lambda=0$.

\subsubsection*{Final form of solution near turning point} From the comparison of the numerical integration along the SDC contour, one can identify that the appropriate approximation (from the many contained in \cref{eq:PnearTP12}) corresponds to the real part of the fifth root of $(-1)^{1/6}$
\beq
  \label{eq:finalPnearTPAppendix}
  p&\sim \frac{\pi \left(r'(a)\right)^{1/3}}{6^{2/3} \Gamma(2/3)} \bigg[2\cos\left(\frac{8}{15 r'(a)}\right) +{r'(a)^{1/2}} \bigg(\Theta(y) \sin\left(\frac{8}{15 r'(a)}\right)\\ &-\Theta(-y) \cos\left(\frac{8 }{15 r'(a)}\right)\bigg) \sqrt{|y|} + 2  \sin\left(\frac{8 }{15 r'(a)}\right) y\bigg],
\eeq
where $\Theta$ stands for the Heaviside step function. We verified this choice of the root on other random parameter sets and it always led to a visually correct approximation of the behaviour near the turning point, see \cref{fig:VerifyingVia2cases}. Note that the {approximate} solution is continuous but has a discontinuity in the first derivative (due to the Heaviside step function).

Thus the WKB solution, \cref{eq:p++4LambdaPosit}-\cref{eq:p+-4LambdaNeg}, shows excellent agreement with numerical integration up to a proximity to the turning point, where we have a linear approximation \cref{eq:finalPnearTP}.

Note that this knowledge of the behaviour of the solutions reveals that the envelope is $\Ord{1}$ near the turning point; the leading order behaviour is actually a rescaled Airy function as follows from Eq \eqref{eq:PacrossTurningPoint}
\beq
  p \sim \underbrace{2 \pi i e^{i\frac{1}{\eps^*} \frac{8}{5}} (\eps^*)^{1/3} M_-}_{\mbox{constant}} \Ai\left((-i)^{3/2}  (\eps^*)^{2/3} \Lambda(\xi) \right)
\eeq
and hence the decay rates of the pattern tails correspond to the envelope behaviour of the Airy function matching those identified above in the outer WKB solution, \cref{eq:p++4LambdaPosit}-\cref{eq:p+-4LambdaNeg}.

\section{Numerical Methods}\label{appendix_numerics}

A complete copy of our numerical methods can be found in \href{https://github.com/AndrewLKrause/Heterogeneous-Localisation-Swift-Hohenberg}{this GitHub repository}\footnote{\url{https://github.com/AndrewLKrause/Heterogeneous-Localisation-Swift-Hohenberg}}. Briefly, we solve \cref{main_sh0} using a standard finite difference discretization of the Laplacian, leading to a five-point stencil for the operator $(1+\eps \nabla^2)^2$ in MATLAB. The resulting system of time-dependent ordinary differential equations is evolved in time using the function \verb|ode15s|, with a Jacobian sparsity pattern provided and absolute and relative tolerances set at $10^{-6}$. A minimum of $N=10,000$ grid points are used, though simulations for smaller $\eps$ were checked for convergence using more grid points and finer time stepping tolerances. Initial conditions were set as  independent and identically normally distributed random numbers for each grid point as $u(x,0) = \mathcal{N}(0,10^{-4})$. Simulations were run for $T = 20,000$ units of time, and checked that they had reached an approximate steady state by evaluating the difference of solutions at time $t=10,000$ from the final time.

To help a reader explore these dynamics without having to use the code above, we have also implemented the model using VisualPDE \cite{walker2023visualpde} at \href{https://visualpde.com/sim/?preset=Heterogeneous-Swift-Hohenberg}{this simulation link}\footnote{\url{https://visualpde.com/sim/?preset=Heterogeneous-Swift-Hohenberg}}. This website provides a crude, yet rapid and interactive way to vary the parameters in the model and immediately observe the dynamics. The solution $u(x,t)$ is plotted in colour starting from small random initial data as described above, and the function $r(x)$ is plotted as a fixed black curve. The model implemented is of the form,
\beq
    \pd{u}{t} = r(x)u-\left(1+\eps^2 \pdd{}{x}\right)^2u + au^2+bu^3+cu^5, \quad \quad \quad x\in[0,1],
\eeq
so one can easily observe {locally} supercritical dynamics by setting all of $a$,$b$, and $c$ to be non-positive, and {locally} subcritical dynamics by, e.g., setting $b>0$ and $c<0$ (as long as at least one higher-order nonlinearity is negative to ensure bounded solutions). Subcriticality can also be observed for $a>0$ for some values of $b$; see \cite{burke2006localized} for details in the case $c=0$. One can also modify the heterogeneity $r(x)$, and the value of $\eps$. The default ranges provided should work without needing to change the time or space steps; modifying {the system to be} outside of these ranges, or using a different nonlinearity, are possible, but may require {resolving}  the time and space step sizes to {obtain} well-behaved solutions. We would advise using this website only to get a rough picture of the dynamics, and to use the codes shared above via GitHub for a more accurate numerical treatment.

Finally, we remark that \cref{fig:bifurcation_diagrams} was produced using the numerical continuation code BifurcationKit in the Julia language \cite{veltz:hal-02902346}. These BifurcationKit continuations were also used to check the direct method of continuation and computation of the energy $E(u^*)$ which was done in MATLAB.

\section{On the restriction Im($\varphi_\pm(x)$)=0}\label{appc}

Our objective in this appendix is to motivate and justify the neglect of the WKB solutions associated with asymptotically fast exponential growth or decay, characterised by the terms of expression \eqref{etas}. 

\subsection{Regions of validity of the WKB solutions}\label{appendix_validity}
We first of all consider when any of the WKB solutions are valid. The inner problem for $x=a$ such that $r(a)=\lambda$ is 
$$0 = \eps^4 p^{(4)} + 2   \eps^2 p'' +p  - {\rho (x-a) }  p ,$$ where $\rho = {r'(a)}$. 
We use the more general WKB ansatz,
\begin{align*}
p \sim {\exp}\left({\sum_{n=0} }\frac{1}{\eps} \eps^n S_n\right).
    \end{align*}
By a dominant balance argument, we obtain
    \beq
        S_0 \sim i {(x-a)},
        \quad
        S_1 \sim - \frac{1}{4} \ln  {(x-a)},
        \quad
        S_2 \sim  \frac{ {(x-a)}^{-3/2}}{\sqrt{|\rho}|}.
    \eeq
As the WKB  approximation is valid for ${|S_0|}/\eps \gg {|S_1|} \gg \eps {|S_2|}$, we get the region of validity as
    \begin{align}\label{val1}
      { | x - a| } \gg \eps^{2/3} \rho^{-1/3} .
    \end{align}
    Similarly, considering another turning point where $1=\sqrt{r(a)-\lambda}$, the region of validity of the local WKB approximation is again
    \begin{align*}
        { | x - a| } \gg \eps^{2/3} \rho^{-1/3}.
    \end{align*}

\subsection{Constructing WKB solutions with Im($\varphi_\pm(x))\neq 0$}

We first consider a scale for the rate of growth or decay of the solutions, in particular away from a simple root of $r(x)-\lambda.$ This root will be located by $x=a$ below and the region where the asymptotically fast exponential growth or decay occurs will, without loss, be taken to be  on the left of $x=a$, which requires $r(x)-\lambda<0$ in this region. 
Then, with $\mu$ and $\theta$ defined via 
$$ \mu (x)=\tan\theta(x) = (\lambda-r(x))^{1/2} \geq 0 , $$ 
for $x\leq a$, and approaching the simple  root of $r(x)-\lambda=0$ at $x=a$ with $x<a$, $r(x)<\lambda$  we have  $\mu^2 \approx r'(a) (a-x)$ for $x$ close to $a$ with $r'(a)>0$ and 
  $$ (1+\mathrm{i}\mu) = (1+\mu^2)^{1/2}  \mathrm{e}^{\mathrm{i}\theta}.$$

Recall we have the definition 
  \begin{align}
         \varphi_\pm(x) = \pm \int_y^x \sqrt{1 \pm \sqrt{r(\alpha) - \lambda}} \, \mathrm{d}  \alpha, \label{varphi1}
    \end{align}
where the lower limit $y$ has been taken to be $a$ without loss. As $r(x)<\lambda$ for $x<a$, this expression entails that $\varphi_\pm$ has non-zero imaginary real part for  $x<a$. In particular, using the above results, 
\begin{eqnarray*} \mbox{Im}(\varphi_\pm(x)) &=&  \pm \int_{a}^x (1+\mu^2(\alpha))^{1/2}  \sin\left(\frac 1 2 \theta(\alpha) \right) \mathrm{d}\alpha \\ &=&  \mp \frac 1 2  \int_{x}^a \left((1+\mu^2(\alpha))^{1/2} \pm 1 \right) \mathrm{d}\alpha \approx \mp (a-x),~~~ \mp \frac{r'(a)}8
(a-x)^2
\end{eqnarray*} 
where the final approximations are for $|a-x|$ sufficiently small, and dropping higher order powers of  $|a-x|$. 
Thus, the imaginary part of \cref{varphi1} generates an amplitude term of the form 
\begin{equation}\label{evsl}  \exp\left[\pm\frac 1{\varepsilon} (a-x)\right],~~~ \exp\left[\pm\frac 1{8\varepsilon} r'(a) (a-x)^2\right]
\end{equation} 
for $x<a$, $|a-x|$ sufficiently small.

The asymptotically fast exponential decay terms generate evanescent solutions, that have decayed once 
\begin{equation}\label{rng}a-x\gg \left(\frac{8}{r'(a)}\right)^{1/2} \varepsilon^{1/2}.
\end{equation} 
This is true for the limiting case of $\lambda=0$ in particular, so we expect a bleed of pattern on a lengthscale of $\eps^{1/2}$ beyond the boundary of $r(x)=0$, which the leading order theory in the main text predicts to be a boundary of the patterning.

Strictly, in reaching these conclusions, we also need $\varepsilon^{1/2} \gg \varepsilon^{2/3}$, where the latter characterises the distance away from the root of $r(x)-\lambda$ for the WKB approximation to be valid; see \cref{val1}. We thus require  $1\gg \varepsilon^{1/6}$, and thus very small $\varepsilon$ for these insights  to be valid. They are nonetheless asymptotically consistent for sufficiently small  $\varepsilon$ and informative even for less extreme values.

\subsection{On the relevance of  WKB solutions with Im($\varphi_\pm(x))\neq 0$ }

 For typical cases, once  $a-x\gg \varepsilon^{1/2}$, the evanescent case simply decays away. This may be observed in Fig.~\ref{fig:homog_het_comparison}b for example, where $\eps=0.01$ but the bleed of the pattern to the left of the red-dash at $x=0.25$, due to a decaying evanescent solution, persists for a length of about $\eps^{1/2}=0.1$.

More generally, this correction does not contribute to the leading order pattern localisation prediction  developed  in the main text, where zero is used as an asymptotically correct, but crude, approximation for any evanescent solution.
In addition,  the bleed of the evanescent solutions is anticipated to induce  an associated correction to the wavenumber constraints of \cref{wvc}, and \cref{cn2}, since the extent of patterning is altered, albeit by an asymptotically small amount. Again such corrections are neglected in the main text. 

One way in which to consider prospective non-trivial contributions from  evanescent solutions, despite their asymptotically rapid decay, is to rescale the solution. This will increase the  global maximum of the solution so that the magnitude of the evanescent solution remains significant. However, then the global maximum of the solution is typically not $O(1)$, which is not consistent with the linearisation required for the linear Turing instability analysis in the main text. More generally, this motivates why only WKB solutions that are $O(1)$ throughout their domain of definition  are  considered in this paper in the context of the ({locally} supercritical) Turing instability study of \cref{sec_Localisation}, ruling out numerous solutions characterised by the terms of \cref{etas0}.

Another possible exception to obtain a prospective non-trivial contribution from  evanescent WKB solutions is to scale them so that they are $O(1)$ at their maximum and thus non-trivial but only in highly isolated regions due to the asymptotically rapid decay away from these regions.  We anticipate such patterns would require highly localised initial conditions to force pattern formation in a highly localised region, supported by our numerical observations. We use 
 initial conditions that represent a noisy forcing across the whole domain that is homogeneous on average, if not on localised patches,  so that we consider the impact of the heterogeneity in $r(x)$ alone in the linear stability analysis, and not  the interaction of heterogeneity in both $r(x)$ and heterogeneity in the initial conditions that is effectively already patterned. With the initial conditions we use there is no numerical evidence of such highly localised perturbations, as illustrated by plots (b)-(d) in Fig.~\ref{fig:homog_het_comparison} and plots 
 (a),(b) in Fig.~\ref{fig:super_sub_comparisons} where the  {locally} supercritical Turing bifurcation has taken place to induce patterning. 
Hence, while the evanescent WKB solutions can potentially be used to construct  extremely localised solutions, their relevance is not indicated for the scope of this study -- a study of the the impact of the heterogeneity in $r(x)$ rather than a coupling  of different sources of heterogeneity. Hence such possibilities are not considered in the main text.  

A further possibility is that the location roots of $r(x)-\lambda$ are within $O(\varepsilon^{1/2})$ of each other or a boundary, so that there is not space for the evanescent solution to decay. There is no reason why such solutions may not manifest, though they have been excluded from analytical consideration because of the \emph{a posteriori} observations that often such solutions have not impacted  pattern localisation via a  {locally} supercritical Turing bifurcation, and also that they will not impact the predictions for sufficiently small $\eps$. This may be explicitly seen in  
panels~(b)-(d) of Fig.~\ref{fig:homog_het_comparison} and plot~(b) in Fig.~\ref{fig:super_sub_comparisons}.
However, a case with a loss of predictive ability of the theory presented in the main text presents in panel (a) of Fig.~\ref{fig:super_sub_comparisons}.
Here,  pattern bleeding  prevents localisation for larger values of $x$. We can estimate when this loss of predictive ability for the localisation of pattern will occur. For a sinusoidal function of the form $r(x) = \pm \cos(2 \pi \zeta f(x))$, with $\zeta $ constant and $f(x)$ possessing smooth O(1) derivatives, the gap between two zeros is approximately 
$ 1/[2\zeta |f'|]$ providing 
\begin{equation} \label{zeta}  \Delta |f| \approx|f'|\Delta x,
\end{equation}
where $ \Delta |f|$  the difference in $|f(x)| $ between the two zeros of $r(x)$, $f'$ is the derivative at one of the zeros 
and $\Delta x$ is the gap. With this holding, the gap between the two zeros can be written as 
 $\pi/r'$, where $r'$ is the value of $|r'(x)|$ at the one of  the zeros. 
This gap must be much greater than twice the decay lengthscale, as by given by  
$ 2 \left(8 \eps/r'\right)^{1/2}$ from \cref{rng}. 
Thus we require 
\begin{equation} \label{c4}  \frac \pi {r'} \gg 2 \left(\frac{ 8 \eps }{r'}\right)^{1/2}, ~~~~\mbox{which reduces to}~~~~~~~  \eps\ll \frac{\pi^2} {32r'} . \end{equation}
For the heterogeneity of Fig.~\ref{fig:super_sub_comparisons}(a),   $r(x) =-\cos(20\pi x^2)$, one can trivially confirm that 
\cref{zeta} holds for larger $x$, as the zeros of $r(x)$ are close since  $\zeta=10\gg 1$ but $f'(x)\sim O(1)$. Hence, \cref{c4} gives a condition for pattern separation. 
 Substituting an estimate for $r'$ for $x\approx 1$ then reduces condition \cref{c4}  to $\eps\ll  0.0025$. However, for panel (a) of Fig.~\ref{fig:super_sub_comparisons} we have  $\eps = 0.002,$ so that the conditions for the separation of patterning is not satisfied within this plot for larger $x$, and the emergence of pattern separation failure is numerically observed. In contrast  the separation condition \cref{c4} holds for panel (b) of the same figure, where pattern separation is clearly evident in the numerical observations.

The same reasoning transfers with little deviation to the fast exponentially growing solutions. To keep the maximum amplitude to be $O(1)$, such solutions must either be highly localised or restricted to regions sandwiched by roots of $r(x)-\lambda$ that are within  $O(\varepsilon^{1/2})$ of each other or a boundary. The reasons  for typically not considering these solutions are then inherited from those from the evanescent solutions. 

Thus, in summary, we do not consider WKB solutions involving terms of the form of \cref{etas0} for a combination of \emph{a priori} reasons, such as consistency with a linearised Turing instability analysis, and \emph{a posteriori} reasons, namely that the impact of any remaining  solutions of this form are not seen in numerical observations, especially once $\eps\ll 1$ is sufficiently small. Thus the main text proceeds from \cref{etas0} excluding these solutions unless explicitly stated otherwise, which in turn entails only considering solutions with Im$(\phi_\pm(x))=0$. In general, the theory compares well with numerical observation. However,  Fig.~\ref{fig:super_sub_comparisons}(a) also  highlights that the future study of evanescent and possibly exponentially growing solutions may  be a worthwhile future refinement when the heterogeneity $r(x)$ changes sufficiently rapidly, so as to violate \cref{c4}.

\end{document}